\documentclass[fleqn,usenatbib]{mnras}

\usepackage{newtxtext,newtxmath}

\usepackage[T1]{fontenc}

\DeclareRobustCommand{\VAN}[3]{#2}
\let\VANthebibliography\thebibliography
\def\thebibliography{\DeclareRobustCommand{\VAN}[3]{##3}\VANthebibliography}

\usepackage{graphicx}	
\usepackage{amsmath}	
\usepackage{bbm}
\usepackage{gensymb}
\usepackage{subcaption}
\usepackage{booktabs}
\usepackage{multirow}
\usepackage{siunitx}
\usepackage{mathtools}
\usepackage{xcolor}
\usepackage{ulem}

\definecolor{J1}{HTML}{ffffd4}
\definecolor{J2}{HTML}{fed98e}
\definecolor{J3}{HTML}{fe9929}
\definecolor{J4}{HTML}{cc4c02}
\RequirePackage{color}

\title[S-PLUS jellyfish candidates]{Systematic analysis of jellyfish galaxy candidates in Fornax, Antlia, and Hydra from the S-PLUS survey: A self-supervised visual identification aid}

\author[Gondhalekar, Chies-Santos, de Souza et al.]{Yash Gondhalekar$^{1}$\thanks{yashgondhalekar567@gmail.com},
Ana L. Chies-Santos$^{2}$ \thanks{ana.chies@ufrgs.br},
Rafael S. de Souza$^{3}$\thanks{r.da-silva-de-souza@herts.ac.uk},
Carolina Queiroz$^{4}$, 
\newauthor
Amanda R. Lopes$^{5}$,
Fabricio Ferrari$^{6}$,
Gabriel M. Azevedo$^{2}$,
Hellen Monteiro-Pereira$^{2}$,
\newauthor
Roderik Overzier$^{7,8}$,
Analía V. Smith Castelli$^{5,9}$,Yara L. Jaffé$^{10}$, Rodrigo F. Haack$^{5,9}$, 
P.T. Rahna$^{11}$,
\newauthor 
Shiyin Shen$^{12}$,
Zihao Mu$^{12}$,
Ciria Lima-Dias$^{13}$, Carlos E. Barbosa$^{14}$, Gustavo B. Oliveira Schwarz$^{14,15}$,
\newauthor
Rogério Riffel$^{2,16}$,
Yolanda Jimenez-Teja$^{17}$,
Marco Grossi$^{18}$,
Claudia L. Mendes de Oliveira$^{14}$,
\newauthor
William Schoenell$^{19}$,
Thiago Ribeiro$^{20}$,
Antonio Kanaan$^{21}$\\
\\
$^{1}$Department of CSIS, BITS Pilani K.K Birla Goa Campus, Goa, 403726, Goa, India\\
$^{2}$Instituto de Física, Universidade Federal do Rio Grande do Sul (UFRGS), Av. Bento Gonçalves, 9500, Porto Alegre, RS, Brazil\\
$^{3}$Centre for Astrophysics Research, University of Hertfordshire, College Lane, Hatfield, AL10~9AB, UK\\
$^{4}$Departamento de F\'isica Matem\'atica, Instituto de F\'{\i}sica, Universidade de S\~ao Paulo, Rua do Mat\~ao, 1371, CEP 05508-090, S\~ao Paulo, Brazil\\
$^{5}$Instituto de Astrofísica de La Plata, UNLP-CONICET, Paseo del Bosque s/n, B1900FWA, Argentina\\
$^{6}$Instituto de Matemática Estatística e Física, Universidade Federal do Rio Grande, Rio Grande, RS, Brazil\\
$^{7}$Observatório Nacional, Rua General José Cristino, 77, São Cristóvão, 20921-400 Rio de Janeiro, RJ, Brazil\\
$^{8}$Leiden Observatory, Leiden University, Niels Bohrweg 2, 2333 CA, Leiden, The Netherlands\\
$^{9}$Facultad de Ciencias Astrónomicas y Geofísicas, Universidad Nacional de La Plata, Paseo del Bosque s/n, B1900FWA, Argentina\\
$^{10}$Departamento de Física, Universidad Técnica Federico Santa María, Avenida España 1600, Valparaíso, Chile\\
$^{11}$ Centro de Estudios de Física del Cosmos de Aragón (CEFCA), Plaza San Juan, 1, 44001 Teruel, Spain.\\
$^{12}$Shanghai Astronomical Observatory, Chinese Academy of Sciences, Shanghai, China\\
$^{13}$Instituto Multidisciplinario de Investigaci\'on y Postgrado, Universidad de La Serena, Ra\'ul Bitr\'an 1305, La Serena, Chile\\
$^{14}$Universidade de São Paulo, IAG, Rua do Mato 1225, Sao Paulo, SP, Brazil\\
$^{15}$Universidade Presbiteriana Mackenzie, R. da Consolação, 930 - Consolação, São Paulo, Brazil\\
$^{16}$Instituto de Astrof\'\i sica de Canarias, Calle V\'\i a L\'actea s/n, E-38205 La Laguna, Tenerife, Spain\\
$^{17}$Instituto de Astrofísica de Andalucía–CSIC, Glorieta de la Astronomía s/n, E–18008 Granada, Spain\\
$^{18}$ Observatório do Valongo, Universidade Federal do Rio de Janeiro, Rio de Janeiro, RJ, Brazil\\
$^{19}$GMTO Corporation 465 N. Halstead Street, Suite 250 Pasadena, CA 91107, USA \\
$^{20}$Rubin Observatory Project Office, 950 N. Cherry Ave., Tucson, AZ 85719, USA \\
$^{21}$Departamento de Física - CFM - Universidade Federal de Santa Catarina, PO BOx 476, 88040-900, Florianópolis, SC, Brazil\\
}

\date{Accepted XXX. Received YYY; in original form ZZZ}

\pubyear{2023}

\begin{document}
\label{firstpage}
\pagerange{\pageref{firstpage}--\pageref{lastpage}}
\maketitle
\begin{abstract}
We study 51 jellyfish galaxy candidates in the Fornax, Antlia, and Hydra clusters. These candidates are identified using the JClass scheme based on the visual classification of wide-field, twelve-band optical images obtained from the Southern Photometric Local Universe Survey. A comprehensive astrophysical analysis of the jellyfish (JClass > 0), non-jellyfish (JClass = 0), and independently organized control samples is undertaken.
We develop a semi-automated pipeline using self-supervised learning  and similarity search to detect jellyfish galaxies. The proposed framework is designed to assist visual classifiers by providing more reliable JClasses for galaxies.
We find that jellyfish candidates exhibit a lower Gini coefficient, higher entropy, and a lower 2D Sérsic index as the jellyfish features in these galaxies become more pronounced. 
Jellyfish candidates show elevated star formation rates (including contributions from the main body and tails) by $\sim$1.75 dex, suggesting a significant increase in the SFR caused by the ram-pressure stripping phenomenon.
Galaxies in the Antlia and Fornax clusters preferentially fall towards the cluster's centre, whereas only a mild preference is observed for Hydra galaxies. 
Our self-supervised pipeline, applied in visually challenging cases, offers two main advantages: it reduces human visual biases and scales effectively for large datasets. This versatile framework promises substantial enhancements in morphology studies for future galaxy image surveys.
\end{abstract}




\section{Introduction}

The distribution of galaxies of different morphological types is not uniform through space. Most galaxies are in groups and clusters, while a smaller fraction are isolated in the field and voids. The density of the environment influences immensely the morphological types that are dominant in that region of the Universe. The morphology-density relation shows that the fractions of ellipticals and lenticular galaxies increase with environmental density, while the fractions of spirals and irregular decrease \citep{dressler80, goto03, houghton15, pfeffer23}.

Galaxies in dense environments are more subjected to environmental interaction, both gravitational (with neighbouring galaxies or the cluster gravitational potential) and hydrodynamical (with the intracluster gas). Such interactions may end up suppressing the star formation of late-type galaxies and changing their morphology, turning spirals and irregulars into ellipticals and S0s. The primary hydrodynamical process that takes place in clusters and groups is the ram pressure stripping \citep[RPS,][]{Gunn&Gott72}, which strips out the interstellar gas from the galaxies' disks and may form a unique type of galaxy called jellyfish.

Jellyfish galaxies, distinguished by their tentacle-like features composed of ionised gas and star-forming regions, represent a distinctive category of galaxies undergoing transformation \citep[see][and references therein]{review_jellyfish}. These galaxies are subject to RPS, which significantly affects their morphology and may enhance their star formation \citep{enhancedSFR, roman2019,Azevedo2023}. RPS involves the removal of the galaxy's cold interstellar gas by the hot intracluster medium, generally opposing the galaxy's movement \citep[e.g.,][]{rps99}. Although RPS is more prevalent in spiral galaxies \citep{stripping_spiral_virgo, Poggianti16, virgo, roman2019, LoTSS-I, LoTSS-II}, it can also occur in elliptical \citep{elipticalfireball}, dwarf \citep{dwarf_fireball}, and ring galaxies \citep{gaspV}. Consequently, studying jellyfish galaxies and their formation provides essential insights into galaxy interactions, their environmental effects, and overall evolution.

RPS galaxies were first observed several decades ago \citep{first_obs_RPS}. However, recent advances in observational surveys and cosmological simulations have enabled more comprehensive and detailed investigations into these objects. Using the high-resolution \texttt{TNG100} (i.e., box size of $100\,h^{-1}$Mpc) simulations, \citet{Yun18} identified satellite galaxies in massive groups and clusters exhibiting asymmetric gas distributions and tails, characteristics indicative of ram pressure stripping. Their findings suggest that approximately 13 per cent of cluster satellites at redshifts $z < 0.6$ bear the signatures of ram pressure stripping and associated gaseous tails. When the analysis was confined to gas-rich galaxies, this proportion escalated to 31 per cent. Additionally, \citet{Yun18} pointed out that these estimates could be considered conservative lower limits, as potential jellyfish candidates could be overlooked due to their random orientation, possibly missing approximately 30 per cent of them. Recently, \citet{2023arXiv230409202Z} extended \citet{Yun18}'s study to incorporate TNG50 with TNG100 simulations to present a richer sample of jellyfish candidates residing in hosts at the lower mass end, outskirts of groups or clusters, and at the higher redshift regime.\citet{2023MNRAS.524.3502R,2023arXiv230409199G} further investigated their evolution and loss of cold gas. They find that while jellyfish candidates undergo dominating star formation in their main bodies (i.e., discs), no significant overall enhancement was observed in their star formation rates compared to the control sample consisting of satellite and field galaxies with similar properties known to affect star formation (redshift, stellar mass, host mass, gas content).

From an observational perspective, galaxies undergoing ram pressure stripping have been scrutinised using photometry and integral field spectroscopy (IFS) over a wide spectral range, extending from the ultraviolet to radio frequencies \citep{Jaffe15, gaspI, virgo, UVITrps, roman2019, LoTSS-I}. These studies have led to the detection of significant amounts of ionised, atomic, and molecular gas in the tails and discs of these galaxies \citep{Jaffe15, gaspI, virgo, gaspXIII, roman2019, HI_JO206, HI_JO206eJO201, HI_JO204, molecular_JW100}. Many dedicated works have been performed in the past decade, focusing specifically on these galaxies and probing them in diverse environments at different redshifts. Such efforts have resulted in the discovery of dozens to hundreds of jellyfish galaxy candidates in both low-redshift ($z \lesssim 0.1$) and medium-redshift ($0.2 < z < 0.9$) clusters and groups \citep{wings, gaspI, LoTSS-I, LoTSS-II, durret21, durret22}. Notably, over 70 jellyfish candidates have been found within the A901/2 multi-cluster system alone \citep{roman2019, roman2021, ruggiero19}.

A defining feature of jellyfish galaxies is their enhanced star formation activity. These galaxies have been observed to possess higher star formation rates (SFRs) compared to other star-forming galaxies within clusters, with SFRs often exceeding even those of starburst galaxies \citep{Merluzzi13, enhancedSFR, roman2019, LoTSS-I}, with a notable enhancement within their `tentacle' structures \citep{SFR_tails}. This intensified activity is believed to result from compression and shock waves generated as the galaxy traverses through the surrounding  intracluster medium \citep{gaspXXX}. However, the enhancement of SFRs for jellyfish candidates belonging to galaxy groups is yet to be fully understood since some studies find an enhancement \citep[e.g.,][]{Tutku22} while some do not \citep[e.g.,][]{oman21, LoTSS-II}. This elevated star formation rate in cluster jellyfish candidates points to a phase of active evolution in these galaxies, shedding light on the potential mechanisms driving galaxy evolution.
However, the ultimate fate of these dynamically evolving galaxies remains uncertain. One possibility is that RPS could transform spiral and irregular galaxies into lenticular and elliptical galaxies, as removing gas could eventually lead to quenching \citep{disktoS0}. Additionally, spirals may undergo a process termed `diffusion', culminating in their transformation into dwarf galaxies \citep{roman2021}. Another intriguing possibility is that the observed ultra-compact dwarfs (UCDs) and intracluster globular clusters (GCs) in low-redshift clusters may originate from \ion{H}{ii} regions formed in the tails of jellyfish candidates, given the observed similarities in their mass \citep{gaspXIII, giunchi23}.

Traditionally, jellyfish candidates have been identified through visual inspection in optical wavelengths, which has resulted in a classification scheme based on observed stripping signatures in the optical bands, known as JClass \citep{Poggianti16}. This scheme encompasses a spectrum of cases ranging from the most extreme (JClass 5) to progressively milder (JClass 1) instances. For example, Fig.~1 of \citet{roman2019} and Figs.~1--3 of \citet{Poggianti16} show visual examples of different JClass candidates. IFS data can be used to categorise jellyfish candidates into various stages of stripping to complement this idea by contrasting H$\alpha$ emission images with those of continuum emission \citep{gaspI,gaspIX,Azevedo2023}. Nonetheless, this approach is also reliant on visual criteria.

Despite its popularity, human visual inspection possesses a few drawbacks. It is time-consuming and can be susceptible to errors due to biases introduced by disturbed morphology, bright knots of star formation, and debris tails. Given the importance of jellyfish classification in understanding their astrophysical properties and evolution, it is important to inspect alternative approaches to visual classification. Machine learning techniques present a complementary strategy to identify these objects and mitigate these challenges. The application of machine learning has gained prominence in recent years as a powerful tool to automate image classification in astronomy \citep[e.g.,][]{Moore2006,Selim2017,Goddard2020,Teimoorinia2020,Vega2021,Quanfeng2023}.

In the realm of machine learning methods, self-supervised learning (SSL) representation has recently gained significant attention due to its ability to learn generalisable and semantically meaningful data representations without manual labelling  \citep[e.g.]{Liu_2021,e24040551,Ericsson_2022}. SSL does not necessarily require large datasets  to perform well, which makes it beneficial for scenarios where only a small sample of objects is known \citep{Nouby2021}. Various SSL approaches have been proposed, including Momentum Contrast \citep[MoCo;][]{moco}, Bootstrap Your Own Latent \citep[BYOL;][]{grill2020bootstrap}, and Augmented Multiscale Deep InfoMax \citep[AMDIM;][]{amdim}. 

\citet{Hayat_2021} applied SSL to multi-band galaxy images from the Sloan Digital Sky Survey (SDSS), demonstrating that it could achieve performance comparable to or better than supervised learning with half or fewer labels for galaxy morphology classification and redshift estimation tasks. \citet{Sarmiento_2021} found that SSL representations were more resilient to non-physical properties, such as instrumental effects, and more closely tied to physical properties than Principal Component Analysis (PCA) representations. Detailed astrophysical studies revealed that SSL representations closely relate to galaxies' physical properties, such as velocity dispersion, stellar mass, and metallicity. Public-access tools developed by \citet{Stein2021}, as well as work by \citet{Stein_2022}, have further illustrated how SSL can be employed for large-scale similarity searches to identify rare astronomical objects, explicitly showcasing its utility in detecting strong gravitationally lensed galaxies.

In this study, we use the S-PLUS multi-band survey data (\citealt{SPLUS}) to identify instances of RPS. Specifically, we employ the narrow-band filter $J0660$ to detect H$\alpha$ emitters within three nearby galaxy clusters: Fornax, Antlia, and Hydra.
\cite{logrono19} has shown that H$\alpha$ will fall within the $J0660$ filter for sources up to $z \leq 0.015$ for the J-PLUS survey \citep{Cenarro19}, which has an identical filter set to S-PLUS.  
The S-PLUS survey offers a suitable dataset because of its extensive coverage of these three nearby clusters. This work uses broad-band optical combined with the $J0660$  classifications, which can better view RPS than just optical images (e.g., \cite{Poggianti16, McPartland16}. We visually classify these RPS candidates based on their stripping strength (JClass) and subsequently develop a semi-automated detection pipeline using SSL, demonstrating it as a concept validation. For the pipeline, we learn representations of the galaxy images using SSL and perform a similarity search on these representations to yield the most similar galaxies to a given `query' galaxy to assist visual inspection. We use two downstream tasks, query by example and supervised classification using the SSL representations, to evaluate the SSL representation quality. This work primarily focuses on applying SSL methods in computer vision, particularly for galaxy images that are widely accessible yet often require further labelling. Distinguishing this work from prior studies, we apply these techniques to a relatively small dataset of approximately 200 images. This approach holds significant interest due to the frequent underperformance of supervised learning methods in the context of limited data.

This paper is organised as follows. Section~\ref{sec:data} outlines the S-PLUS data employed in this work, providing details on the selection criteria and data preprocessing. Our methodology, discussing the visual inspection of H$\alpha$ emitters and the SSL training details for classification, is detailed in Section~\ref{sec:methodology}. In Sections \ref{sec:results} and \ref{sec:astro_results}, we validate our semi-automated detection approach, present the astrophysical properties of the jellyfish candidates, and discuss the implications of our findings, respectively. We then summarise our main findings in Section~\ref{sec:discussion}, leading to our concluding remarks and potential future work in Section~\ref{sec:conclusions}. All magnitudes presented in this paper are in the AB system.

\section{Data}
\label{sec:data}

The Southern-Photometric Local Universe Survey (S-PLUS; \citealt{SPLUS}) has already observed approximately 3,200 deg$^{2}$ of the southern hemisphere. Its goal is to map an extensive area exceeding 9,000 deg$^{2}$ using an optimised photometric system (\citealt{Cenarro19}). This system incorporates five broad-band (BB) filters ($g, r, i, z$ being SDSS-like and $u$ being Javalambre) and seven narrow-band (NB) filters, covering a wide spectral range from 3700 to 9000 \AA. The NB filters offer unparalleled insights into nearby galaxies because of their ability to detect prominent stellar features such as [OII], Ca H+K, H$\delta$, H$\alpha$, Mgb, and Ca triplets.
Furthermore, S-PLUS reaches about one magnitude deeper than the SDSS (\citealt{SDSSDR12}), providing strong constraints on the star formation histories and photometric redshifts of galaxies. Observations for the project are made using a 2 deg$^{2}$ field of view camera fitted with a 9k $\times$ 9k CCD at a 0.55 arcsec/pixel scale. This equipment is mounted on a fully robotic 0.8-m diameter telescope (T80-South) located at Cerro Tololo, Chile.

The photometric data from S-PLUS are calibrated according to the methodology described by \citet{AlmeidaFernandes2022}. The calibrated magnitudes for all 12 bands are measured in six distinct apertures in addition to the astrometry and other photometric parameters. The entire catalogue of images and data is available through the S-PLUS web portal\footnote{\url{https://splus.cloud}}, which provides various tools for querying and visualising the data. For this study, we have chosen to focus our analysis on the photometric data 

obtained using \texttt{SExtractor} \citep{SExtractor},
employing the so-called ``dual-mode'' and selecting \texttt{AUTO} magnitudes.

\subsection{Data selection}
\label{subsec:data_selection}

To ensure that the H$\alpha$ emission from our candidates falls within the $J0660$ filter, a visual inspection was performed of all galaxies from three nearby clusters (at redshift $z\leq 0.015$) included in the S-PLUS Data Release 1 \citep[DR1;][]{SPLUS}. Galaxies that exhibited an excess in H$\alpha$ emission were explicitly sought. Six S-PLUS fields were analysed in Antlia, 23 in Fornax from DR1 and iDR3, and four in Hydra. An additional twenty fields on the outskirts of Fornax were also inspected as they became available at the time of the start of the visual inspection.
We obtained a sample of candidate H$\alpha$-excess objects by subtracting the $r_\mathrm{SDSS}$ band image from the narrow H$\alpha$ filter. Before subtraction, the  $r_\mathrm{SDSS}$ band was scaled to match the global count rates between the objects in common on the two images.
Following this selection process, we identified 158 H$\alpha$ emitting candidate galaxies, with 38, 47, and 73 originating from Antlia, Fornax, and Hydra, respectively. This set constitutes our primary sample.

\begin{figure}
\includegraphics[width=\linewidth]{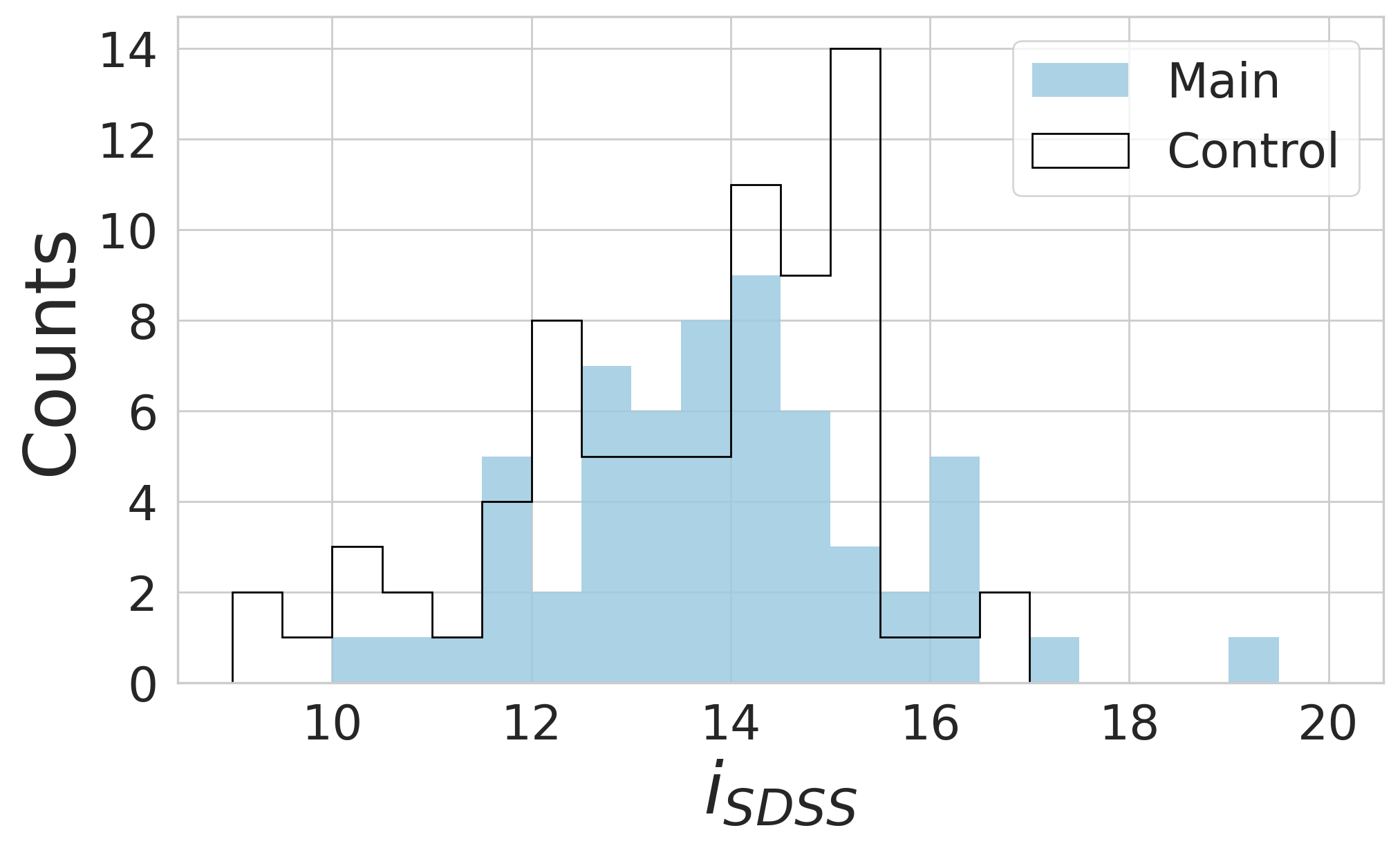}
\caption{Comparison of $i_{SDSS}$-band magnitudes between the main and control samples used for visual inspection. The solid blue histogram represents the main sample, while the black curve represents the control sample.}
\label{fig:hist_control_sample}
\end{figure}

We assembled a control sample of 75 additional galaxies from the Fornax cluster for visual inspection. These galaxies were not previously identified as exhibiting H$\alpha$ emission excess and exhibited a magnitude distribution in the $r_{SDSS}$ band similar to that of our selected candidates. In Fig.~\ref{fig:hist_control_sample}, we present the magnitude distributions in the $i_{SDSS}$ band, which does not contain H$\alpha$ emission excess, for both the primary and control samples.

\section{Methodology}
\label{sec:methodology}

Having discussed the potential of self-supervised learning, we now discuss the implementation details. Our approach consists of two key stages: first, we pre-select jellyfish candidates based on a visual inspection of S-PLUS multi-band images. Subsequently, we benchmark SSL as a means to assist visual classification in future applications.

\subsection{Visual inspection}
\label{subsec:visual_inspection}

Stripping galaxy candidates are typically classified through visual inspection based on the evidence of stripping signatures in optical bands, ranging from the most extreme (JClass 5) to the weakest (JClass 1) cases \citep{Poggianti16, roman2019}. These classifications are determined by the formation of tentacles of ripped gas, which occur because of the interaction between the galaxy's interstellar medium and the intracluster medium (ICM).

In this study, we performed the visual classification of the main and control samples internally in a private project using the Zooniverse platform\footnote{\url{https://www.zooniverse.org/}}. This task was accomplished by six classifiers \footnote{Carolina Queiroz, Ana L. Chies Santos, Yash Gondhalekar, Yara Jaffé, Rahna P.T., and Mu Zihao.}, who categorised galaxies with no visual stripping evidence as JClass 0 and flagged galaxies with merger evidence. The assignment of the final JClass to galaxies disregards JClasses from visual classifiers who flagged the galaxy as a merger. 
To aid the classification process, we provided a composite image consisting of three panels: an image displaying only H$\alpha$ emission (J0660 narrow-band), an RGI image generated using \texttt{Trilogy} \citep{Coe12} with $r_{SDSS}$, $g_{SDSS}$, and $i_{SDSS}$, and a colour image (RGI) combined with H$\alpha$ emission to accentuate the star-forming regions, depicted in pink.

These star-forming regions, typically bright in H$\alpha$, resemble irregularly distributed star-formation clumps, often called debris. Occasionally, no image was available for one or more broad bands. In such cases, neighbouring bands were selected to compose the images (e.g., $u_{JAVA}$ or $z_{SDSS}$). However, this did not rectify the problem for a few galaxies, in which case we either obtained a dark or no image. We then replaced the $\rm RGI$ + H$\alpha$ image with the full 12-band image or omitted this frame.

\begin{figure}
    \centering
  \includegraphics[width=0.95\linewidth]{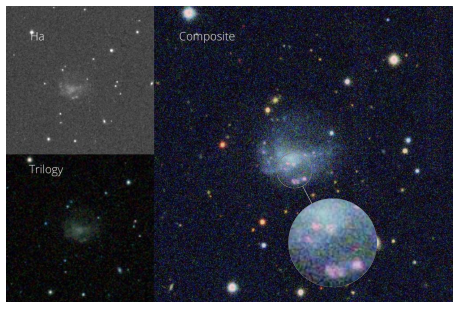}
    \caption{Example of composite image inspected in Zooniverse. Upper left panel: $J0660$ narrow-band image. Bottom left panel: $RGI$ coloured image. Right panel: $RGI$ + H$\alpha$ image. The central galaxy is NGC1437A from the Fornax cluster, classified as JClass 3.  The zoomed inset highlights the pink clumps denoting star-forming regions for visual clarity.}
    \label{fig:zooniverse}
\end{figure}

\begin{figure}
    \includegraphics[width=0.9\linewidth]{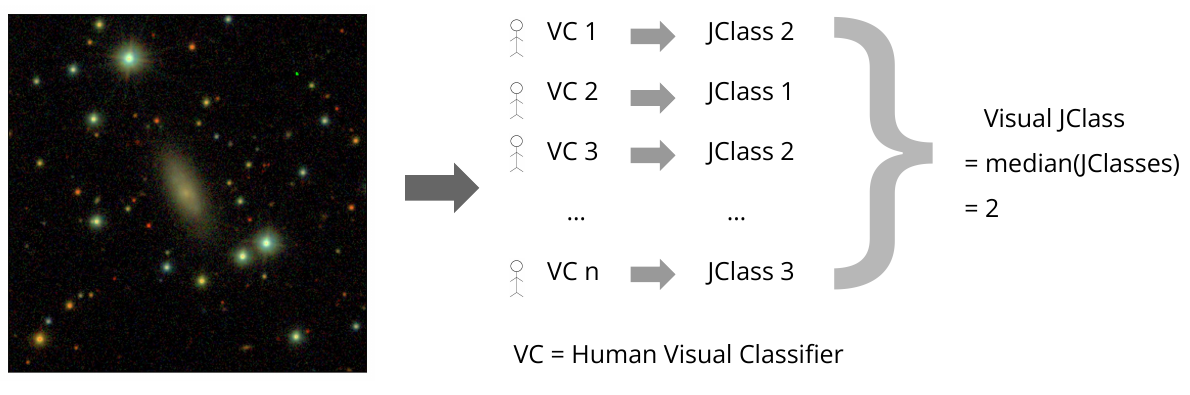}
    \caption{Workflow of visual inspection consisting of $n$ human classifiers for assigning a JClass to a galaxy.}
    \label{fig:visualWorkflow}
\end{figure}

If more than one galaxy was present in the field, the classifiers were instructed to classify only the galaxy positioned at the centre of the frames. The final JClass designation corresponds to the median of all classifications. The visual inspection and classification workflow is described in Fig.~\ref{fig:visualWorkflow}. In Fig.~\ref{fig:zooniverse}, we present an example of a composite image for the galaxy NGC1437A \citep{serra23} from the Fornax cluster, as inspected in Zooniverse. Note the pink clumps of star formation in the right panel; this galaxy was classified as JClass 3.

\begin{figure}
    \includegraphics[width=\linewidth]{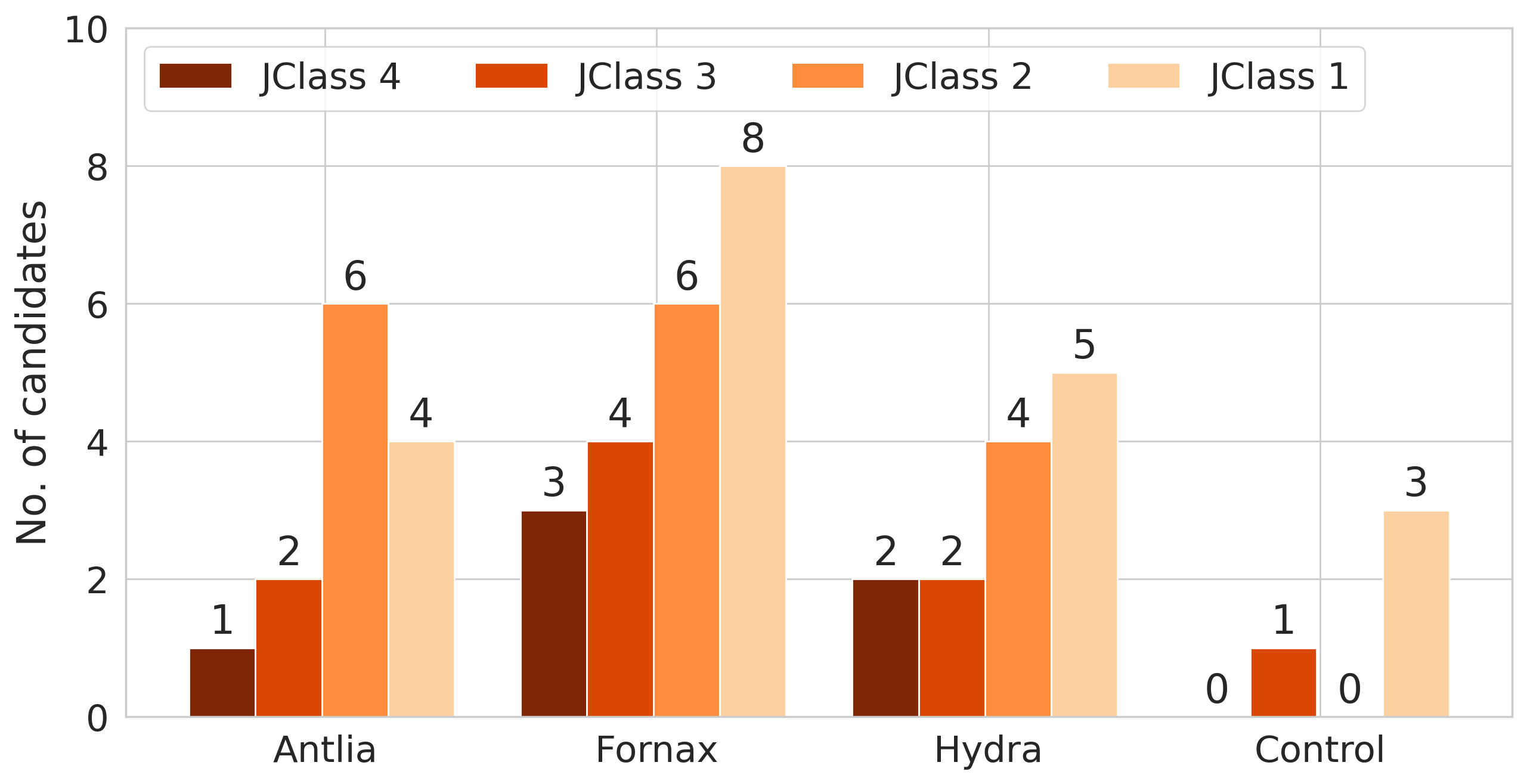}
    \caption{Frequency of jellyfish candidates based on their respective JClass rankings for each galaxy cluster and the control group, arranged in descending order from the strongest to the weakest.}
    \label{tab:visual_class}
\end{figure}

Following the visual inspection, we identified 51 jellyfish candidates with JClass ranging from 1 to 4 (no example with JClass 5 was found in the dataset). These include 13 galaxies from Antlia, 25 from Fornax, and 13 from Hydra. Notably, four of the 25 galaxies from Fornax belong to the control sample. Four jellyfish candidates are included in the control sample because the data selection, which relied only on H$\alpha$ emission as discussed in Sect.~\ref{subsec:data_selection}, is independent of visual inspection that identified jellyfish candidates. The distribution of JClass across each cluster is presented in Fig.~\ref{tab:visual_class}.

Thus, $\sim$30\% of the H$\alpha$ emitters are jellyfish candidates (excluding the four jellyfish candidates from the control sample). \cite{Yun19} analysed 2600 satellites in the IllustrisTNG simulation, selecting galaxies with some gas, stellar masses higher than $10^{9.5}\,M_{\odot}$ and in clusters, and massive groups with halo masses $10^{13} < M_{200c}/M_{\odot} < 10^{14.6}$. They found that $\sim$31\% of the galaxies were jellyfish at $z<0.6$. Observationally, \cite{roman2019} finds $\sim$ 16\% of the star-forming galaxies in A901/2 at $z=0.0165$ to be jellyfish candidates. \cite{Vulcani22} studied a sample of late-type, blue, and bright ($B < 18.2$) galaxies in clusters from the {\sc WINGS} and OmegaWINGS surveys ($0.04 < z < 0.07$) within two virial radii. Their study found $\sim$15\% of the sample as stripping candidates and $\sim$20\% of galaxies with ``unwinding arms", which could be attributed to jellyfish seen face-on. Although the sample selection criteria and type of data are different from previous works, our fraction of stripping candidates falls in a similar range of values compared to recent literature values.

\subsection{Image pre-processing for machine learning}
\label{subsec:image_preprocess}

Although convolutional neural networks (CNNs) possess the ability to pinpoint objects within an image, deliberately steering their attention towards the foreground object can significantly enhance their performance \citep{Cao2021}. Observational datasets comprise several point sources and extended sources with sizes akin to galaxies, which could potentially confuse the network. In Appendix~\ref{appn:galmaskMotivation}, we use the Grad-CAM visualisation method to illustrate that SSL representations may dominantly encode information about the background sources instead of the galaxy, which is undesirable. Thus, we preprocessed the S-PLUS images in our study to mitigate any prospective bias during the learning procedure by removing background sources from our galaxy image dataset.

\begin{figure}
    \centering
    \includegraphics[width=\linewidth]{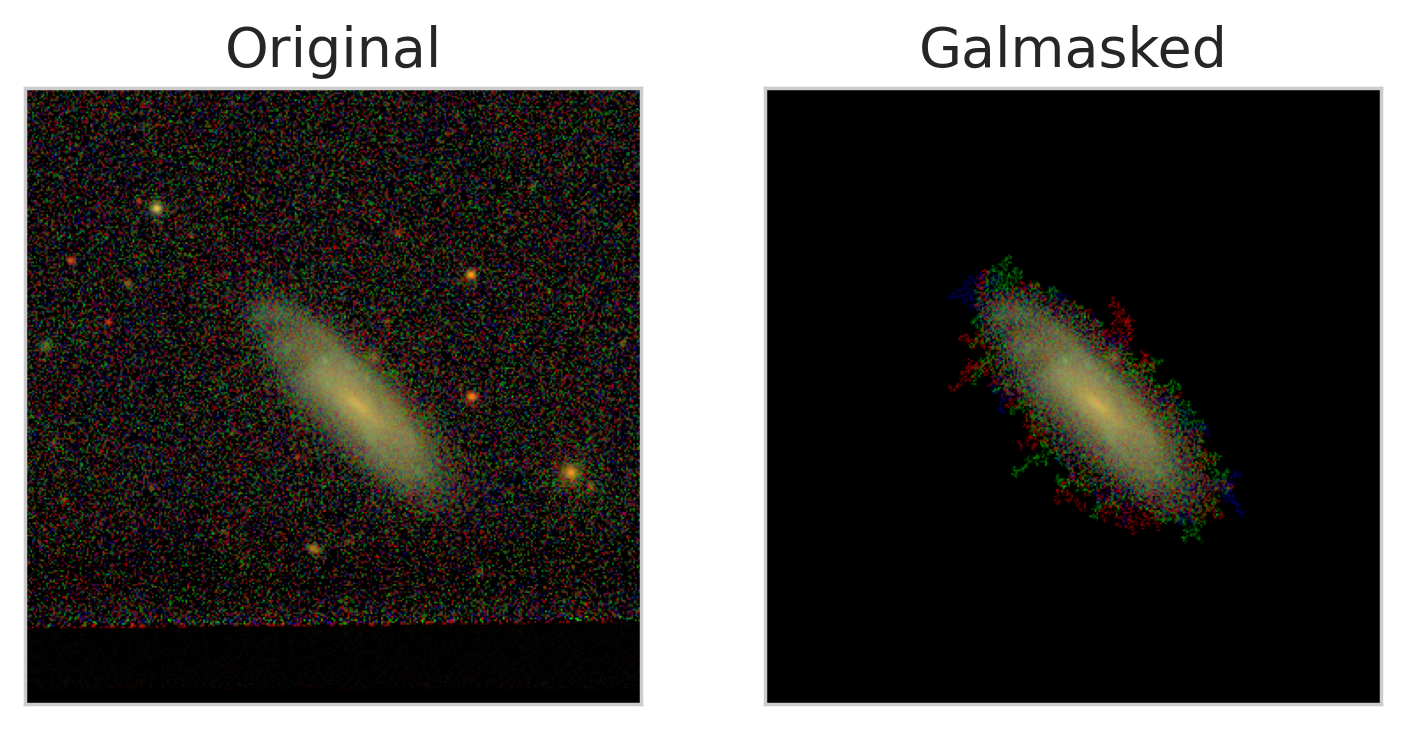}
    \caption{An example application of \texttt{galmask} on the jellyfish candidate IC1885 having JClass = 2. \texttt{galmask} is used during preprocessing to remove unwanted background sources.}
    \label{fig:galmaskExample}
\end{figure}

For this purpose, we employ the \texttt{galmask} (v0.2.0) Python package \citep{Gondhalekar_2022}, designed to eliminate background sources from our images. \texttt{galmask} is applied independently to each band. Before its utilisation for background source removal, we first generate a segmentation map using \texttt{NoiseChisel} \citep{noisechisel_segment_2019}, followed by the \texttt{Segment} program from Gnuastro \citep{gnuastro}. We found \texttt{NoiseChisel} to be particularly well suited to detect the dim, dispersed tails characteristically seen in jellyfish signatures, which has proved challenging for traditional signal-based threshold methods. To avoid inadvertently eliminating the peripheries of galaxies, such as the extended tails, we opted for a slightly conservative set of parameters within \texttt{galmask}. Segmentation is followed by optional deblending and connected-component labelling, which selects the connected component closest to the centre to isolate the central galaxy region from background sources. An example visualisation of the galaxy image before and after the application of \texttt{galmask} is shown in Fig.~\ref{fig:galmaskExample}, which demonstrates that background sources present in the original image are masked.

In our dataset of 51 jellyfish and 183 non-jellyfish images, \texttt{galmask} yielded successful outputs for 46 and 171 images, respectively. The failure of \texttt{galmask} to process a handful of images is attributed to the extreme faintness of the galaxies (some of which were undetectable by \texttt{NoiseChisel}) or difficulties encountered during the extraction of the galaxy cutouts. As \texttt{galmask} operates separately on each band, we discarded any galaxy images that did not yield a successful output across all 12 bands. After this, we manually inspected all \texttt{galmask} outputs to identify any failures in the masking process. Any images showing portions of the galaxy that were erroneously removed during the masking stage within \texttt{galmask} were also discarded. This results in 43 jellyfish and 140 non-jellyfish images to carry forward in our analysis. We also estimate the background level in the central galaxy and subtract it from the outputs of \texttt{galmask}. In addition, we apply an arcsinh transformation to all images to enhance contrast.
Although this procedure results in a reduced dataset size in our already small dataset, SSL is data-rich in that it can learn meaningful representations even with less data. Thus, we opt for a `stricter' selection of galaxies to include in our final dataset. Additionally, our data set has a high imbalance, with jellyfish examples constituting only approximately 22\% of the total (this fraction of jellyfish candidates is similar to the values found in the literature; see Sect.~\ref{subsec:visual_inspection}).

\subsection{Self-supervised Learning using \texttt{SimCLR}}
\label{sec:ssl}

Our self-supervised approach is based on the SimCLR framework \citep[e.g.,][]{Cheng2020Simclr}, which is a contrastive method and offers an elegant, end-to-end solution for learning generalised feature representations from unlabelled data. We have modified certain aspects of SimCLR, such as the base encoder network and data augmentation, to suit our requirements better (details of our modification in the architecture and hyperparameters are described in Appendix~\ref{sec:moreTrainingDetails}). A batch of $N$ images from the training dataset is sampled at each training iteration. For each sampled image $\boldsymbol{x}$, two independent augmentation functions are applied to produce $\tilde{\boldsymbol{x}}_i$ and $\tilde{\boldsymbol{x}}_j$. This doubles the batch size to $2N$ images.

Both $\tilde{\boldsymbol{x}}_i$ and $\tilde{\boldsymbol{x}}_j$ are passed through the base encoder network, denoted as $f(\cdot)$ (often a convolutional network for image data), which extracts a one-dimensional representation vector, $\boldsymbol{h}_i = f(\tilde{\boldsymbol{x_i}})$. Subsequently, a projection head network, usually a single hidden layer multilayer perceptron (MLP), denoted as $g(\cdot)$, projects the representation vector onto a space where a contrastive loss function is applied, i.e., $z_i = g(\boldsymbol{h}_i)$. Instead of directly applying the contrastive loss function to the representations, using a projection head during training promotes learning more potent representations \citep{Cheng2020Simclr}.

A `positive' pair, denoted as ($\tilde{\boldsymbol{x}}_i, \tilde{\boldsymbol{x}}_j$), consists of two distinct augmented views derived from the same original image $\boldsymbol{x}$. Conversely, a `negative' pair comprises two images not derived from the same original image. In contrastive learning, these negative pairs are essential to learning differentiable feature representations by enforcing the model to learn distinct representations for different images.
The contrastive loss function, or NT-Xent loss, is formulated as follows:
\begin{equation}\label{eqn:contrastiveLoss}
l_{i, j} = -\log{\dfrac{\exp({\textrm{sim}(z_i, z_j) / \tau})}{\sum_{k=1}^{2N} \mathbbm{1}_{[k \neq i]}\exp{(\mathrm{sim}(z_i, z_k) / \tau})}}
\end{equation}
where $\textrm{sim}(\boldsymbol{p}, \boldsymbol{q}) = \dfrac{\boldsymbol{p} \cdot \boldsymbol{q}}{\Vert \boldsymbol{p} \Vert \Vert \boldsymbol{q} \Vert}$ is the similarity function, $\tau$ is a temperature hyperparameter controlling the sensitivity of the loss function (\citealt{zhang2021temperature}), and $\mathbbm{1}_{[k \neq i]} \in {0, 1}$ is the indicator function that equals one only if $k \neq i$. The loss is averaged over all positive pairs in the sampled mini-batch, and the weights of networks $f$ and $g$ are adjusted to minimise it.

One defining feature of SimCLR is its use of large batch sizes (as high as 8192) to keep track of negative examples, bypassing the need for more complex structures like memory banks \citep{Wu2018,moco}. It harnesses the robustness of data augmentations and allows multiple negative examples for each positive pair instead of the traditional single negative example per positive pair \citep[see, e.g.][]{Liu_2021}. Such an approach enhances the effectiveness of the contrastive loss function and improves the quality of the learnt representations.

Data augmentation is vital for learning valuable representations. A good data augmentation pipeline is particularly important, given the small size of our dataset. They must maintain the semantic meaning of the images \citep{Hayat_2021}, compelling the model to learn features that persist through transformations. Ultimately, this results in learnt representations invariant to these transformations \citep{tian2020makes,Xiao2020,Wang2021}, enhancing the generalisability of these representations.
Our data augmentation pipeline encompasses the following procedures:
\begin{itemize}
    
    \item {\bf Centre crop}: Each image is centre-cropped to a size of $200 \times 200$ pixels ($\sim9 \times 10^{-3}$ pc). As the galaxies often reside near the centre of the image, centre-cropping can be beneficial.

    \item {\bf Random-crop-and-resize}: After centre-cropping, we randomly crop a section from the image and resize it to $72 \times 72$ pixels. This is done for a few reasons. First, in contrastive learning, we only need to determine if two image patches are from the same image, not necessarily requiring the whole image. Second, smaller images not only expedite training but have also been shown to improve performance in SSL scenarios \citep{Cao2021}.

   \item {\bf Random horizontal and vertical flip}: We apply each flip with a 0.5 probability. Horizontal flips help the model to be invariant to the galaxy's horizontal mirror-image transformation. Although less common, vertical flips are also useful, as we want the learnt representation to be unaffected even if the galaxy appears inverted.

     \item {\bf Custom Colour Jitter}: As this study does not deal with RGB images, the conventional colour jitter technique (randomly adjusting the brightness, contrast, saturation, and hue of an image) cannot be used. To introduce colour jitter into our multi-channel images, we multiply pixel values by a uniformly sampled value in the $[0.8, 1.2]$ range, keeping it fixed for a particular channel \citep{rs13112181}. This effect scales the channel-wise mean and standard deviation by the sampled value, introducing an element of randomness. This transformation is randomly applied with a probability of 0.8.

   \item {\bf Random rotation}: Each image is rotated by a random angle sampled from $[0\degree, 360\degree]$ to ensure invariance with the spatial orientation of the galaxy in the image, as done in \citet{Hayat_2021}. Although this rotation includes horizontal and vertical flips as a special case, random rotation provides more flexibility.

   \item {\bf Gaussian Blur}: With a probability of 0.5, we blur an image with a Gaussian kernel of size $9 \times 9$ pixels, selecting the standard deviation, $\sigma$, uniformly at random in the $[0.1, 2.0]$ range, akin to \citet{Cheng2020Simclr}. This step enables the representations to remain largely unaffected by varied levels of image smoothing. To some degree, this also serves as a way to achieve invariance with the Point Spread Function (PSF), even though the standard deviation of the Gaussian blurring kernel is not explicitly scaled using the PSF Full Width at Half Maximum \citep{Hayat_2021,Stein2021}.
   
\end{itemize}
The augmentation techniques we employ enhance those used in the initial SimCLR model, tailored to offset the limitations posed by our small training dataset. Introducing variety into the training images can induce the model to learn more robust and invariant features. We argue that applying a centre-cropping operation before a random-resize-and-crop, instead of a standalone random-resize-and-crop, is more advantageous for our dataset. This assertion stems from the preprocessing step, which substantially reduces background objects. By prioritising a centre-cropping operation, we are tilting the odds in our favour to extract random crops from the central galaxy instead of from the background areas. An ablation study that examines the significance of these augmentations is presented in Appendix~\ref{appn:dataAugMoreDetails}.

\subsection{Model implementation}
\label{subsec:training} 

First, the encoder is pretrained. Following the pretraining phase, the projection head is discarded as the representations of the encoder are considered more meaningful than the projection head since the latter is found to lose critical information necessary for downstream tasks \citep{Cheng2020Simclr}. As a result, the pretrained encoder is used as a fixed-feature extractor to obtain image representations. Since the augmentations described in Section~\ref{sec:ssl} were explicitly designed only for contrastive learning, these augmentations were discarded for deriving representations from the fixed feature extractor \citep{Cheng2020Simclr}. The images are standardised before feeding into the model using the channel-wise mean and standard deviation calculated across the training dataset. Instead of pretraining the encoder on a large dataset and then fine-tuning it on our small dataset, the pretraining is performed directly on the target dataset, a strategy that has also recently shown promise in the low-data regime (\citealt{Nouby2021}; \citealt{Cao2021}). The extremely small size of our dataset is used to test whether (a) SSL can learn meaningful representations of galaxies and (b) SSL representations can encode important features for downstream tasks such as classification, which was previously unexplored in an astronomical context for small data.

Contrastive learning approaches often benefit from a longer training duration and larger batch sizes, as they expose the model to more negative examples \citep{Cheng2020Simclr}. Given RAM limitations, we select a batch size of 128, the maximum feasible size for our application. Furthermore, smaller resolution images pave the way for larger batch sizes, as noted in \citep{Cao2021}. The model undergoes training with the contrastive loss function for 1000 epochs, with optimisation using the Adam with decoupled weight decay method \citep[AdamW;][]{loshchilov2019decoupled}, featuring a weight decay of $10^{-4}$ and a learning rate, $lr = 10^{-4}$. A large number of epochs ensures better convergence on our small dataset. We adopt a cosine annealing schedule to regulate the learning rate, with the minimum learning rate set to $lr / 50$ and without restarts \citep{loshchilov2016sgdr}. The maximum number of epochs for the scheduler is set to the number of epochs in our training run, i.e., 1000. The temperature parameter, $\tau$, is set to 0.05. Hyperparameter tuning was performed using $K$-Fold cross-validation, with more details provided in Appendix~\ref{appn:hyperTuning-ss}. Weights \& Biases \citep{wandb} was used for tracking mode training and validation experiments.

\section{Self-supervised learning results} 
\label{sec:results}

In this section, we discuss the results of the SSL for similarity search and provide an example application to assist and improve subjective visual classification. Due to limited examples in our dataset, we combine the training and testing sets for the analysis in this section.

\subsection{Query by example}
\label{sec:query}

We conduct a query by example (or similarity search), similar to the works of \citet{Stein2021,Hayat_2021} to inspect what types of galaxies are clustered closely in the self-supervised representation space. To conduct this experiment, cosine similarities are calculated between the representation vectors of a chosen query galaxy image and all other galaxy images in the dataset, and the similarities are ordered in decreasing order to select the four closest representation vectors (and the corresponding images) to the representation of the query galaxy. The query galaxy image is then visually compared for any morphological similarities with the selected closest images to gain insights into the clustering in the self-supervised representation space.

\begin{figure*}
\begin{subfigure}{.99\textwidth}
    \centering
    \includegraphics[keepaspectratio,width=\textwidth]{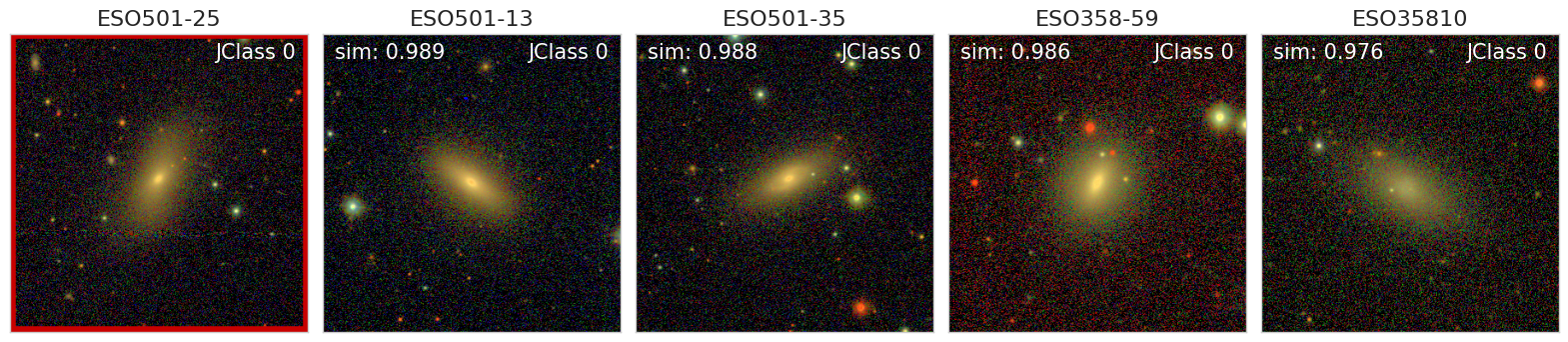}
\end{subfigure}
\begin{subfigure}{.99\textwidth}
    \centering
    \includegraphics[keepaspectratio,width=\textwidth]{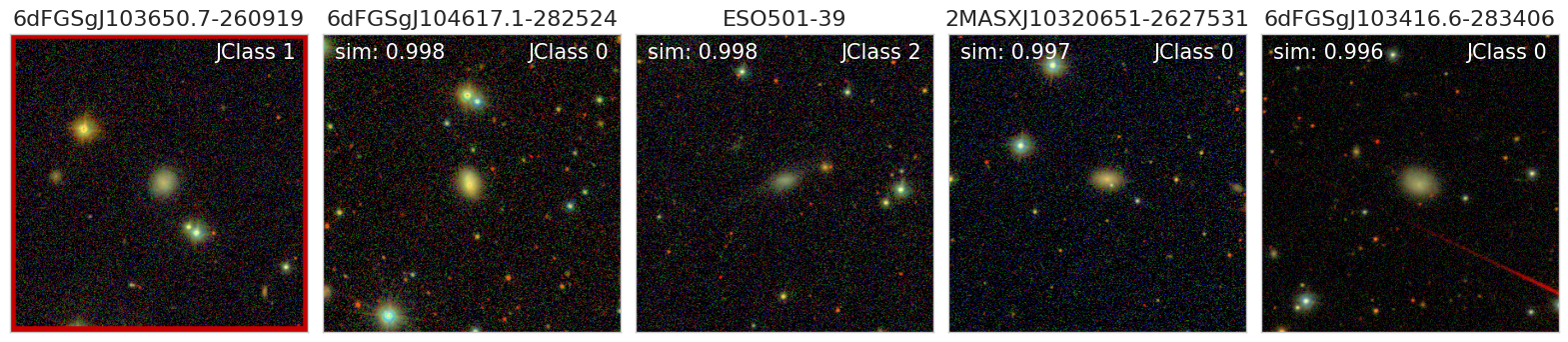}
\end{subfigure}
\begin{subfigure}{.99\textwidth}
    \centering
    \includegraphics[keepaspectratio,width=\textwidth]{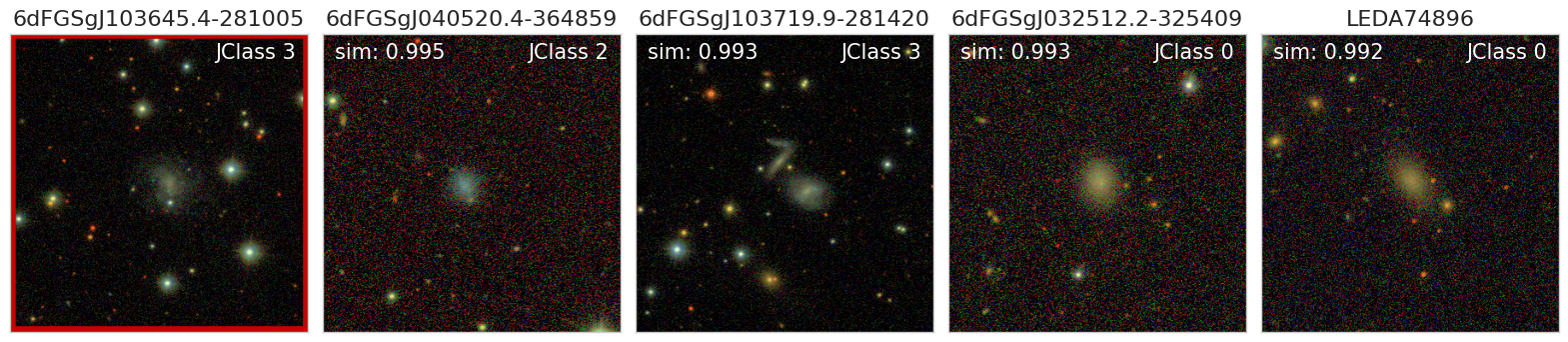}
\end{subfigure}
\begin{subfigure}{.99\textwidth}
    \centering
    \includegraphics[keepaspectratio,width=\textwidth]{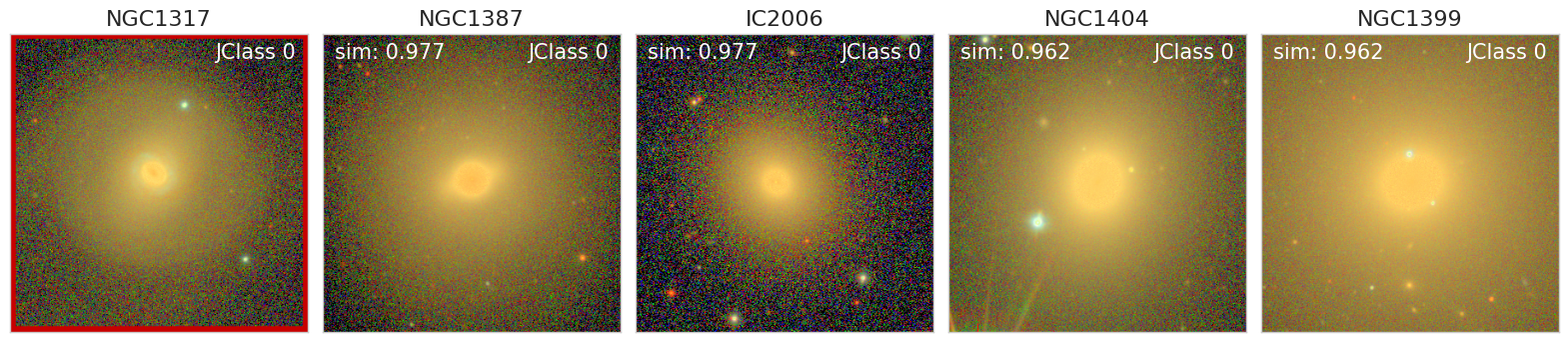}
\end{subfigure}
\begin{subfigure}{.99\textwidth}
    \centering
    \includegraphics[keepaspectratio,width=\textwidth]{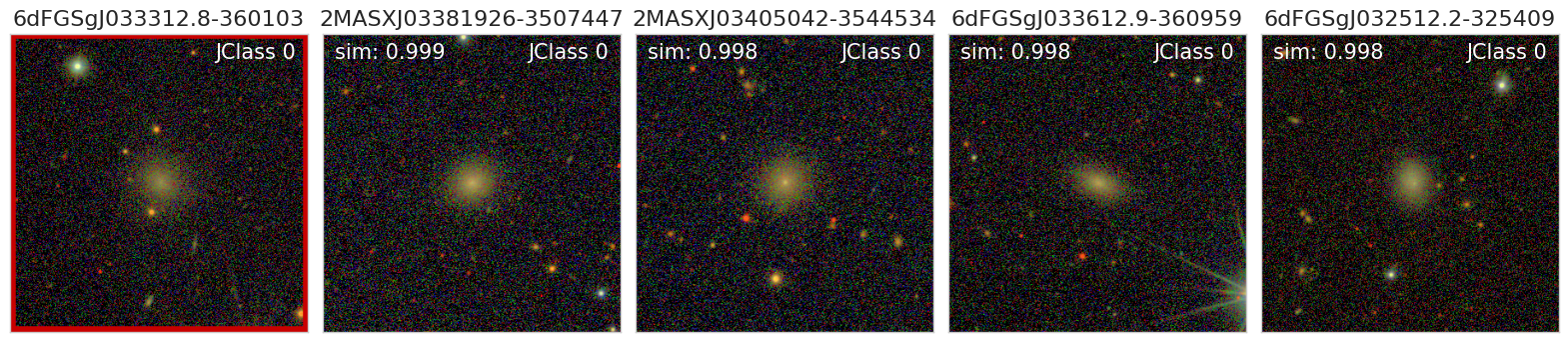}
\end{subfigure}
\begin{subfigure}{.99\textwidth}
    \centering
    \includegraphics[keepaspectratio,width=\textwidth]{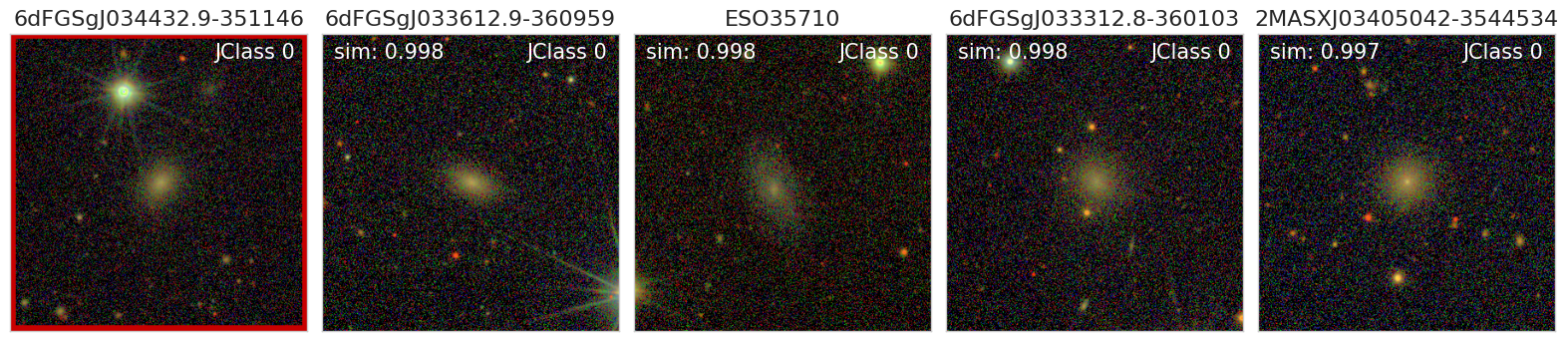}
\end{subfigure}
\caption{Illustration of the query by example returning the four closest images to the query image (outlined in red) as obtained by the similarity search. The JClass obtained from visual classification, and the cosine similarity values are marked on the images. The galaxy's name is shown on top of each image. Various cases are shown, such as non-jellyfish (JClass 0) query images or jellyfish (JClass 1, 3) query images. While we use \texttt{galmask} during preprocessing, the images shown here are without the use of \texttt{galmask}.}
\label{fig:simSearch}
\end{figure*}

\begin{figure*}\ContinuedFloat
\begin{subfigure}{.99\textwidth}
    \centering
    \includegraphics[keepaspectratio,width=\textwidth]{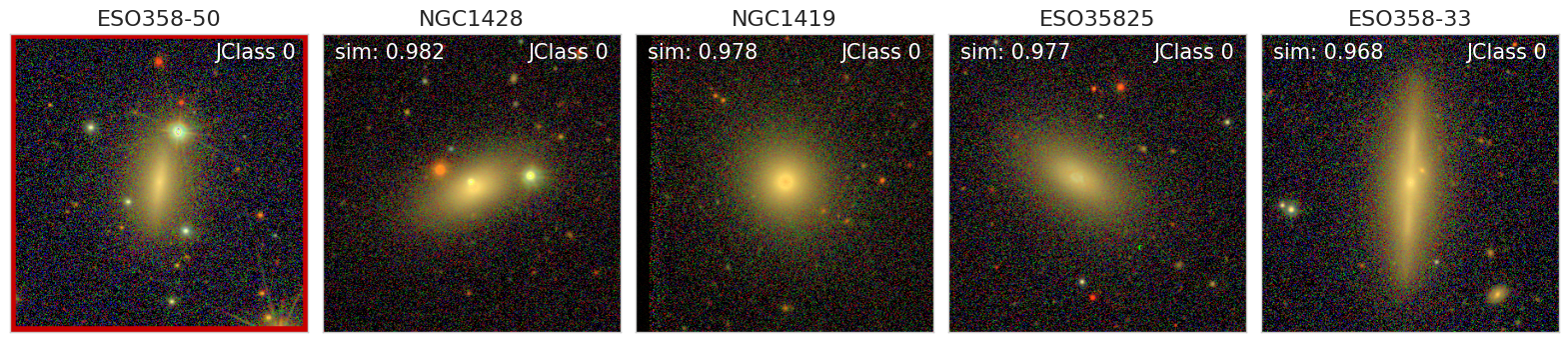}
\end{subfigure}
\begin{subfigure}{.99\textwidth}
    \centering
    \includegraphics[keepaspectratio,width=\textwidth]{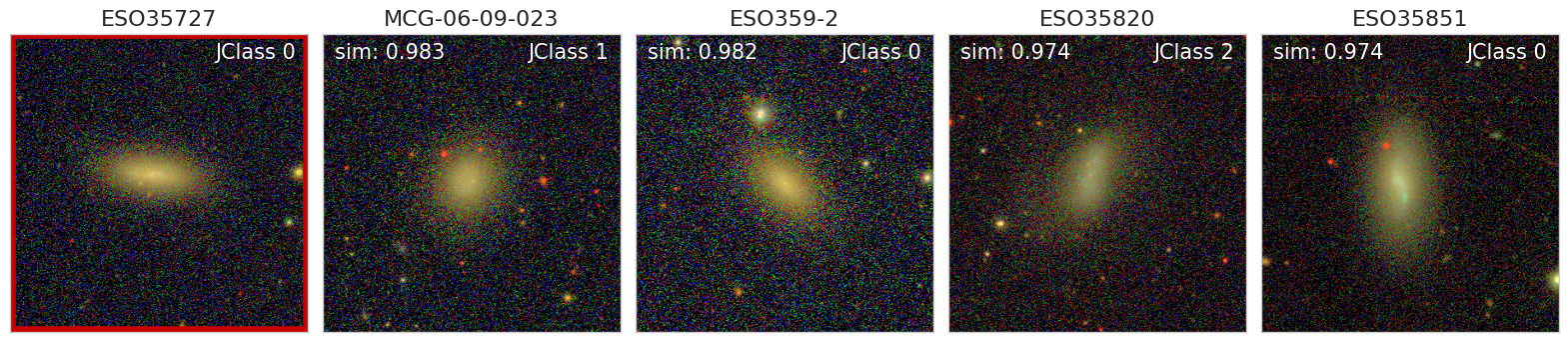}
\end{subfigure}
\begin{subfigure}{.99\textwidth}
    \centering
    \includegraphics[keepaspectratio,width=\textwidth]{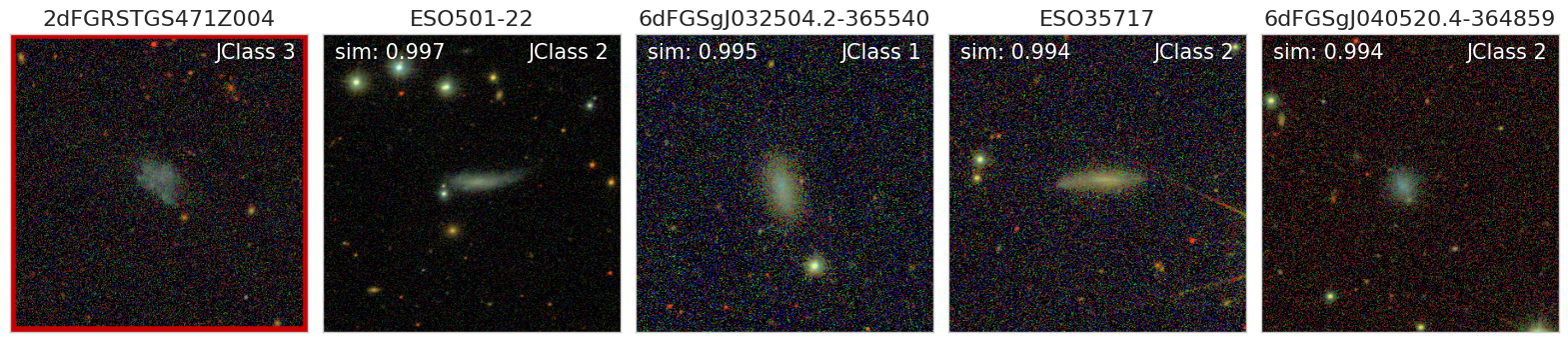}
\end{subfigure}
\begin{subfigure}{.99\textwidth}
    \centering
    \includegraphics[keepaspectratio,width=\textwidth]{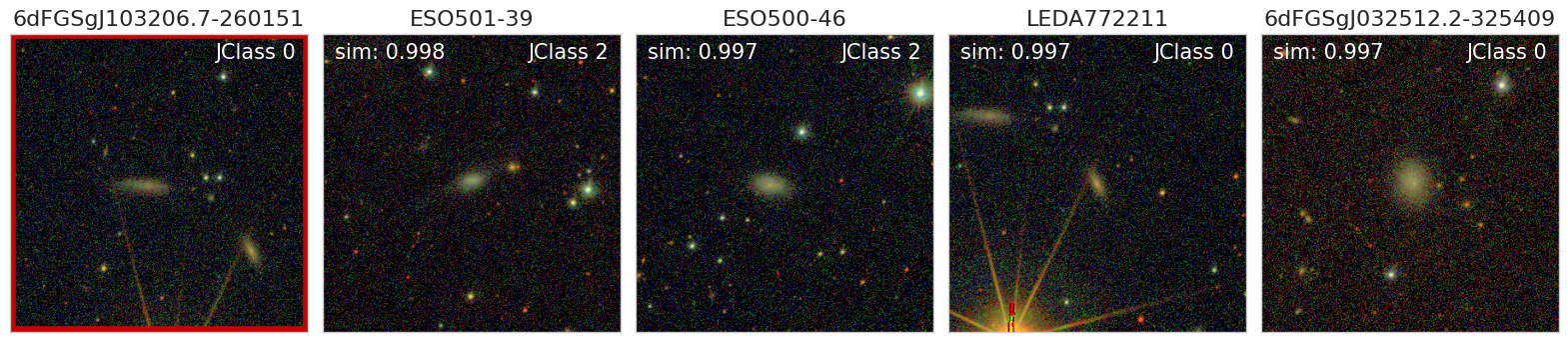}
\end{subfigure}
\caption{{\it (continued)}}
\end{figure*}

Fig.~\ref{fig:simSearch} shows the results of our similarity search. It can be observed that the similarity search returns semantically similar images to the query image. In particular, the query search returns galaxies with similar colours and visual morphological characteristics. The query search is unaffected by the rotation of galaxies by any arbitrary angle and robust to the number and shape of background sources within the images. The former is likely because of the random rotation data augmentation used during self-supervised pretraining. We hypothesise the latter is mainly due to the use of \texttt{galmask} in our internal preprocessing pipeline, which conveniently removes many unwanted sources from the background. Slight correspondence is observed between the JClass of the query galaxy and the JClass of the most similar images to the query galaxy. Jellyfish (non-jellyfish) query galaxies tend to have jellyfish (non-jellyfish) galaxies as the most similar galaxies to the query (jellyfish: $\textrm{JClass} > 0$; non-jellyfish: $\textrm{JClass} = 0$). However, since these JClasses are based on visual classification rather than SSL, it is challenging to interpret these correlations. Overall, we conclude that the self-supervised representations encode important information about the galaxies, which allows the clustering of the galaxies in a morphologically meaningful manner. 
\subsection{Re-calibrating visual classification}
\label{sec:fixBad}

\begin{figure}
    \centering
    \includegraphics[width=0.85\linewidth]{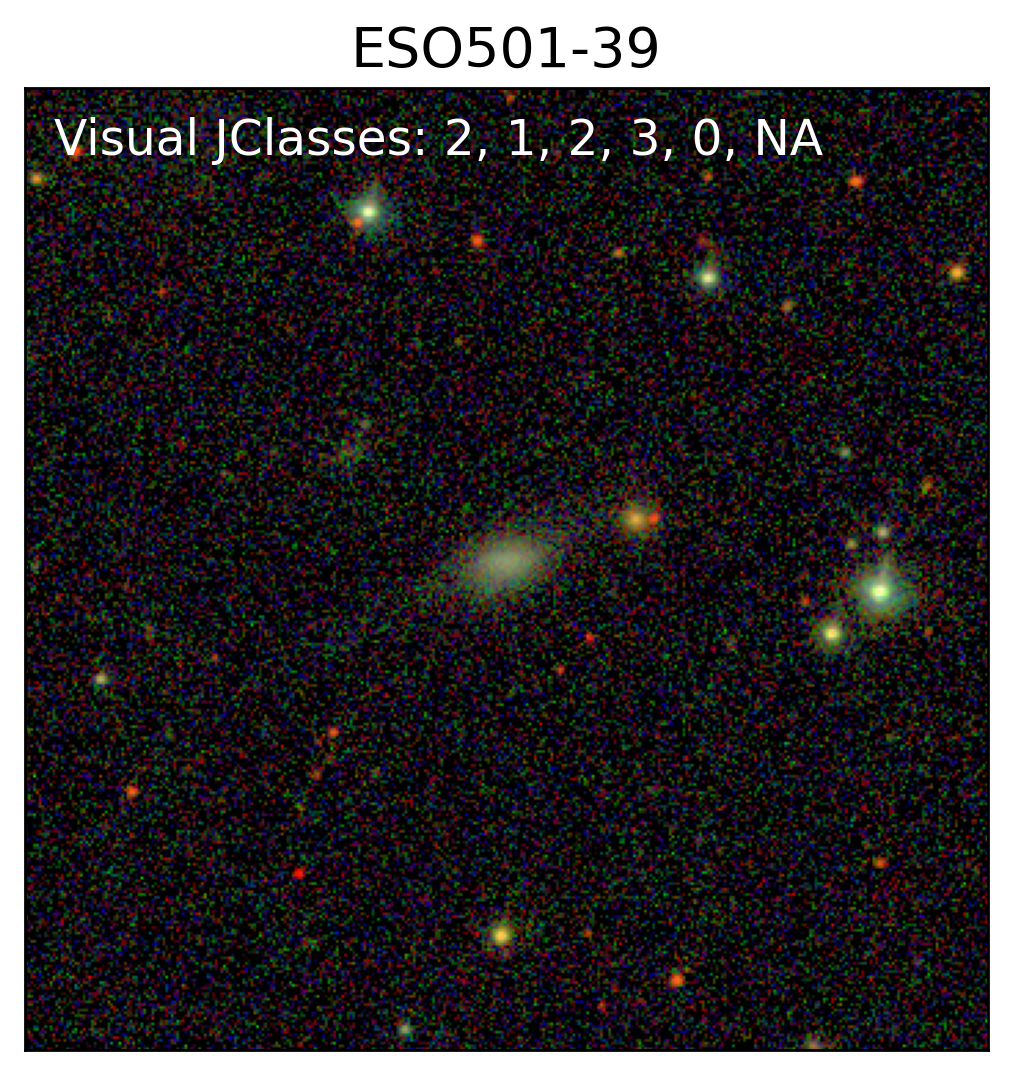}
    \caption{Illustration of the subjectiveness of visual classification. As denoted in the image, five of the six visual classifiers assigned a JClass to this galaxy, while one classifier did not assign any JClass (denoted by ``NA"). Out of the five classifiers, one assigned a JClass 3, two assigned a JClass 2, one assigned a JClass 1, and the other classifier assigned a JClass 0. Thus, considerable uncertainty prevails in the visual analysis since different visual classifiers assigned various JClasses. As described in Sect.~\ref{subsec:visual_inspection}, the final visual JClass is the median across all the JClasses, i.e., JClass 2, which might not be reliable due to the significant uncertainty across visual classifiers.}
    \label{fig:motivateRecalibration}
\end{figure}

During visual inspection, each visual classifier individually assigns labels to an image, with no measurable boundaries between different JClasses. Such a methodology largely hinges on the classifier's prior domain knowledge and the guidelines provided before the image assessment. Fig.~\ref{fig:motivateRecalibration} demonstrates this behaviour. Even though taking the median of visual classifications of different classifiers attempts to minimize individual human biases, it would be affected if there is a large disagreement among the visual classifiers. Therefore, strategies that counteract human biases can improve this subjective classification procedure. As discussed in Sect.~\ref{sec:query}, self-supervised representations present a morphologically significant structure, with analogous galaxies closely clustered. This finding serves as the basis for our proof-of-concept application, which shows how SSL can increase the quality of visual classification in a data-efficient way. This section thus examines the potential of using self-supervised representations to refine visually labelled JClasses.

Since the JClass assigned by visual inspection is based on a subjective assessment of jellyfish-ness, a linear evaluation protocol (see Appendix.~\ref{sec:classificationCompare}) in which a  supervised logistic regression classifier is employed on the self-supervised representations using the JClasses as ground-truth labels will be affected by the quality of these labels. The linear evaluation will not yield a precise disturbance strength estimate, either, since it is only a binary classification (jellyfish vs. non-jellyfish). A multi-label classification (with ground-truth categories $\textrm{JClass} = 0, 1, 2, 3, 4$) will degrade due to the increased severity of the class imbalance. Thus, a supervised regressor trained on self-supervised representations is not ideal for improving JClass. To mitigate these issues, we develop a new downstream task to assign JClass to galaxies leveraging self-supervised representations to assist visual classifiers in their classification.

Since we propose not to rely solely on visual inspection and aim to improve estimates of visual JClasses, this section focuses only on galaxies with high uncertainty in the JClass among the visual classifiers. Hence, our initial step involves identifying galaxies that present a visual classification challenge. If a galaxy receives more than $\lceil N / 2 \rceil$ unique visually assigned JClasses from $N$ visual classifiers (where $\lceil \rceil$ denotes the ceiling function), it signifies a considerable classification uncertainty, which renders the galaxy visually complex. We refer to this set of galaxies as our ``target'' sample. Here, $N = 6$ (see Sect.~\ref{subsec:visual_inspection}). Notably, around 85\% of the galaxies in our target sample are jellyfish candidates, suggesting a more considerable disparity among visual classifiers in assigning JClass to jellyfish than non-jellyfish galaxies. Since our goal is to yield precise JClass estimates for identifying novel instances of galaxies exhibiting jellyfish characteristics, we limit our self-supervised application to the target sample.

We predict JClass using the self-supervised representations as follows: given a target galaxy, $K$-nearest neighbours to it are found in the representation space, and the mean of JClasses of the nearby galaxies, weighted by their cosine similarities, is assigned as the JClass of the target galaxy. The nearby galaxies are chosen such that they are not already in the target sample. We have chosen to use $K = 4$. The JClass assigned using SSL is determined by the following relationship:
\begin{equation}
\textrm{JClass}_{ss} = \dfrac{\sum_{i=1}^{K} s_i \, \textrm{JClass}_{v_i}}{\sum_{i=1}^{K} s_i}
\end{equation}
Here, $\textrm{JClass}_{v}$ and $\textrm{JClass}_{ss}$ represent the visually assigned and the self-supervised predicted JClass, respectively. $s_i$ denotes the cosine similarity between the query image and the $i^{th}$ image similar to the query. Such a weighted scheme enables assigning more weightage to more similar galaxies. We emphasise that we do not train a $k$-nearest neighbour classifier on the representations  to predict the JClass since the self-supervised model is already trained to encode relevant information about the galaxies.

\begin{figure*}
    \centering
    \includegraphics[keepaspectratio,width=0.75\textwidth]{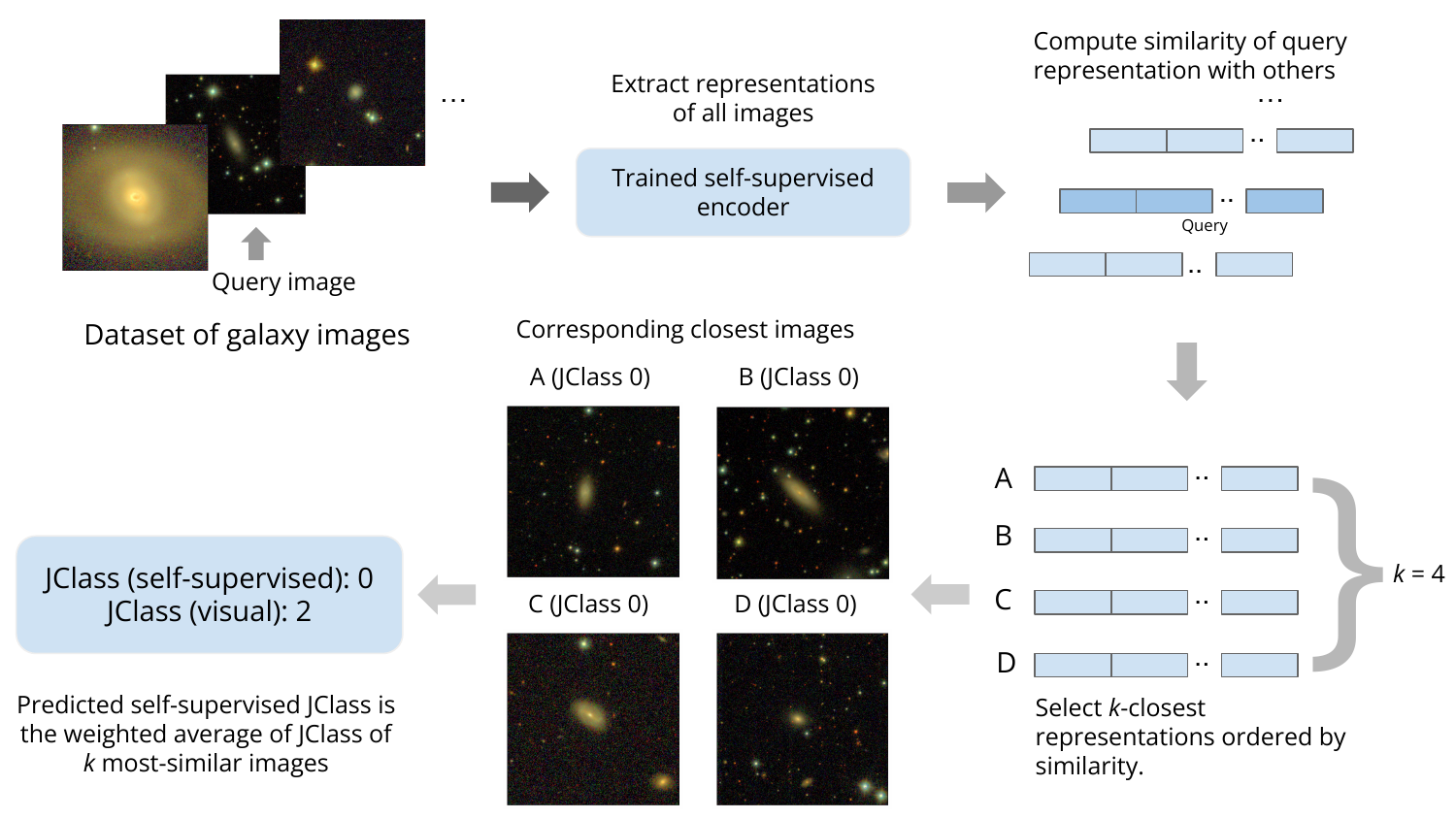}
    \caption{Workflow demonstrating the use of SSL to assign JClass to a galaxy solely based on the similarity search on the extracted representations of galaxies. Before training the self-supervised encoder, we find it crucial to account for background sources to prevent affecting the similarity search, as discussed in Sect.~\ref{subsec:image_preprocess} and demonstrated in Appendix.~\ref{appn:galmaskMotivation}.}
    \label{fig:SSframework}
\end{figure*}

A similarity search is then conducted using the target galaxy as the query, similar to Sect.~\ref{sec:query}. Fig.~\ref{fig:SSframework} illustrates our framework to assign JClass to galaxies based on the similarity search on the representations of galaxies. Although our proposed approach uses visual JClasses for the final prediction after the self-supervised learning is performed, these visual JClasses are not used as ground-truth labels in training our self-supervised model. This means that our self-supervised approach learns patterns in the galaxy images based on the observed data alone and does not use the visual JClasses to learn to distinguish between jellyfish and non-jellyfish galaxies. This characteristic feature of our self-supervised approach alleviates human biases and thus provides benefits over approaches such as training a supervised CNN on the galaxy images or training a supervised classifier on the self-supervised representations.

\begin{figure*}
\centering
\begin{subfigure}{.99\textwidth}
    \centering
    \includegraphics[keepaspectratio,width=\textwidth]{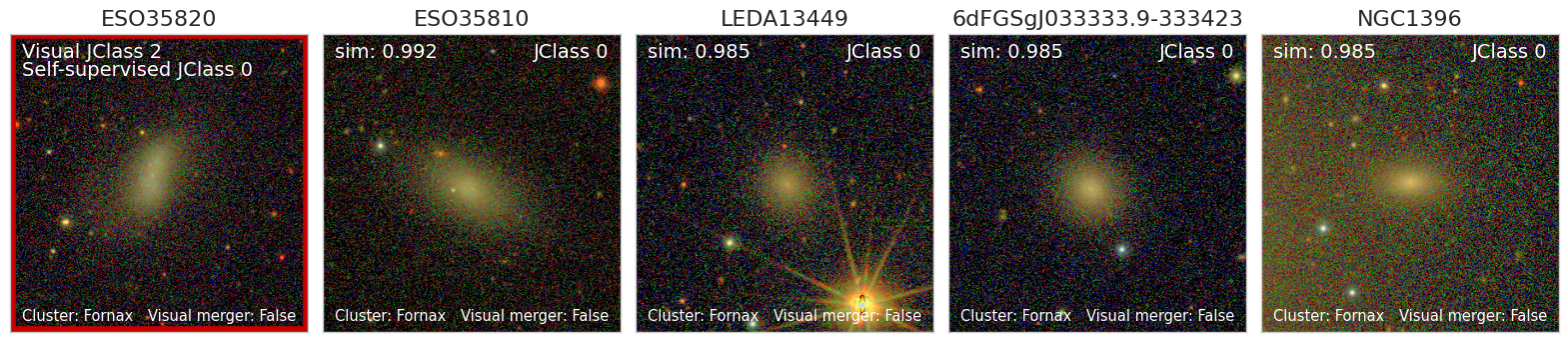}
\end{subfigure}
\begin{subfigure}{.99\textwidth}
    \centering
    \includegraphics[keepaspectratio,width=\textwidth]{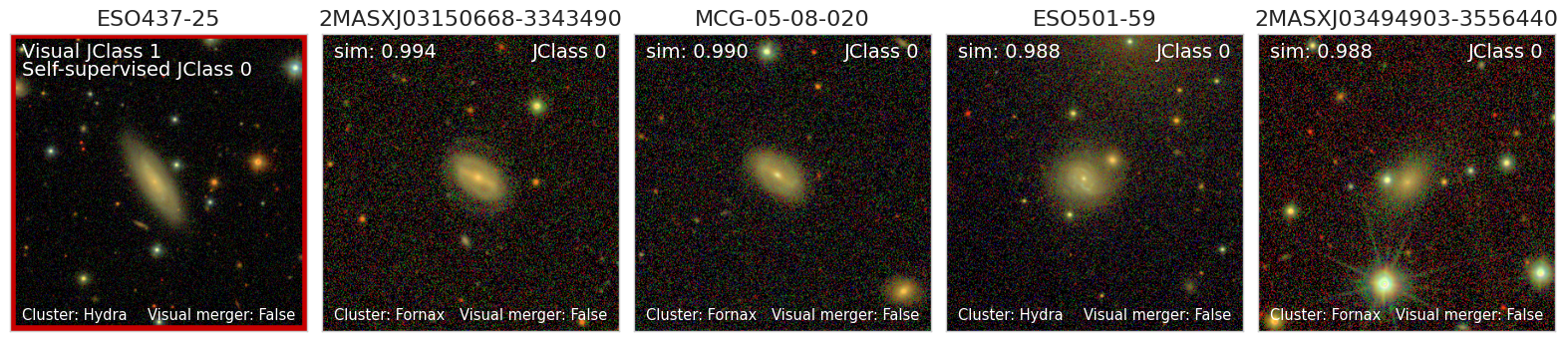}
\end{subfigure}
\begin{subfigure}{.99\textwidth}
    \centering
    \includegraphics[keepaspectratio,width=\textwidth]{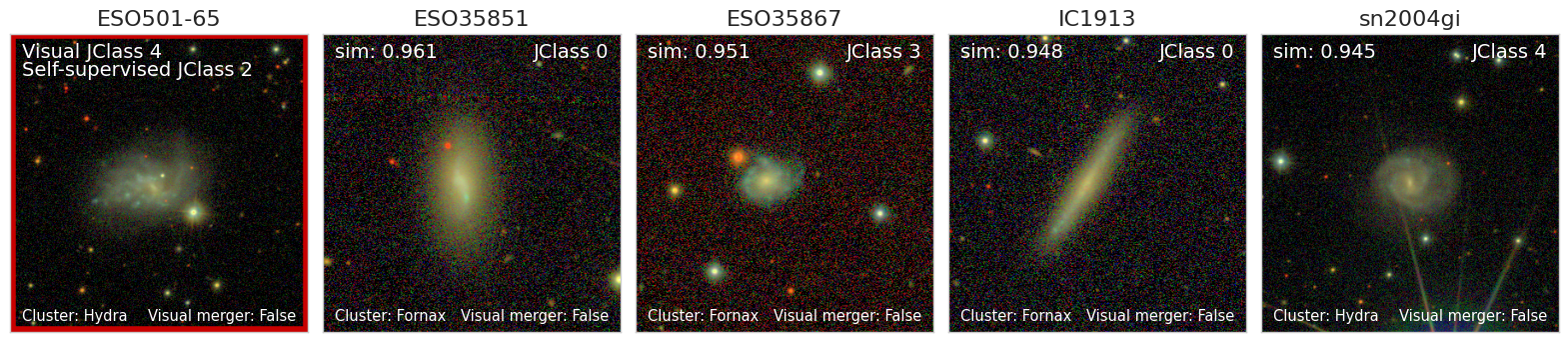}
\end{subfigure}
\begin{subfigure}{.99\textwidth}
    \centering
    \includegraphics[keepaspectratio,width=\textwidth]{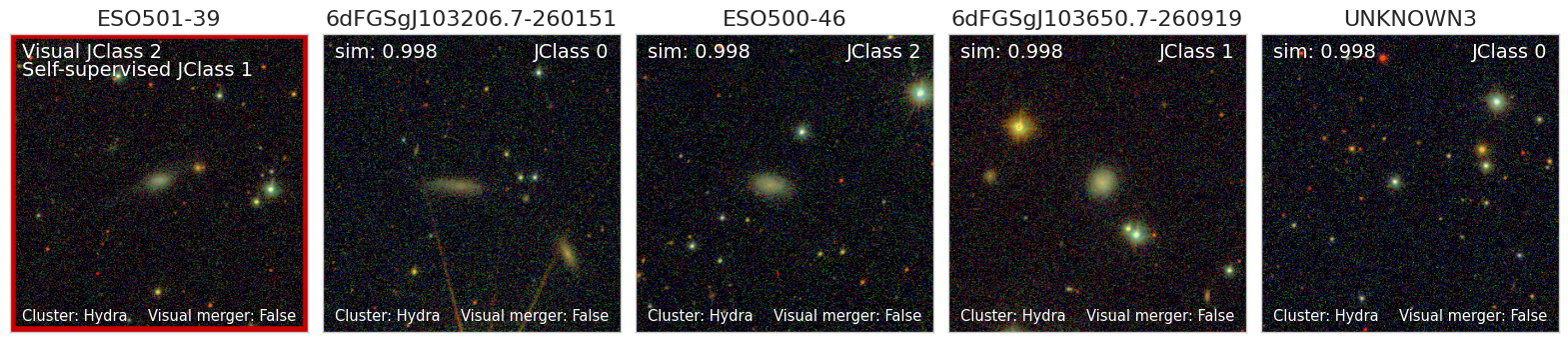}
\end{subfigure}
\begin{subfigure}{.99\textwidth}
    \centering
    \includegraphics[keepaspectratio,width=\textwidth]{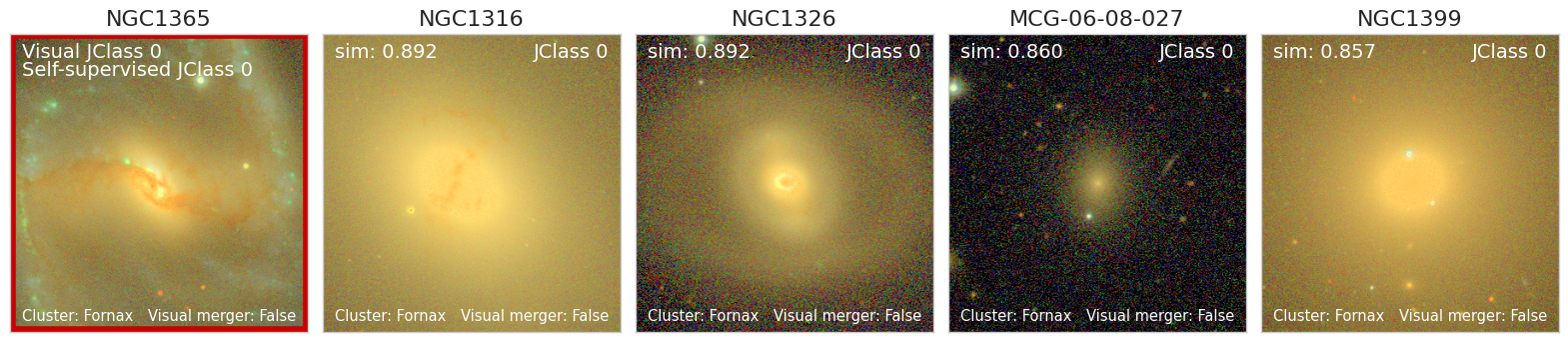}
\end{subfigure}
\begin{subfigure}{.99\textwidth}
    \centering
    \includegraphics[keepaspectratio,width=\textwidth]{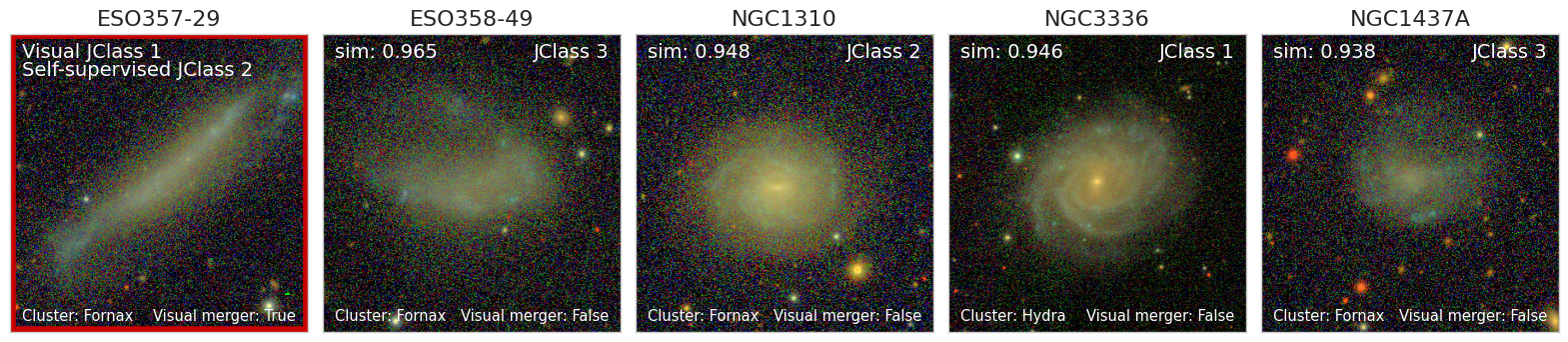}
\end{subfigure}
\caption{An example application of our framework to assign a JClass to galaxies that are visually confusing to classify. The images outlined in red are the query images and have significant deviations in their visual JClass (see text for details). The four images closest to each query image are shown as obtained by SSL. The JClass predicted by SSL (see text for details) is also shown for each query image. The JClasses mentioned in images not outlined in red are obtained from visual classification. Cosine similarity values are indicated. Visual classification also predicts whether a given galaxy shows merger signatures, shown in each image's bottom-right. A galaxy is considered a merger only when more than half of the visual classifiers voted in favour of a merger.}
\label{fig:fix-bad-classify-ss}
\end{figure*}

An example application of our framework is provided in Fig.~\ref{fig:fix-bad-classify-ss}. The top two rows show two weak jellyfish query images (JClass 2 and 1, respectively), whereas the self-supervised approach predicted it to be a non-jellyfish galaxy. This occurs because galaxies most similar to the query had a JClass 0. The third and fourth rows show two cases where the self-supervised approach predicted a milder jellyfish signature, i.e., a lower JClass, than visual classification (JClass 2 instead of JClass 4 and JClass 1 instead of JClass 2). We assume that the visual JClasses of all non-query images (not outlined in red) are fairly accurate since we have only selected cases where the majority of the visual classifiers agreed on a common JClass. Hence, for the third and fourth rows, the fact that similar galaxies to the query image contain a mix of jellyfish and non-jellyfish galaxies suggests that the corresponding query image likely contained some features similar to non-jellyfish galaxies and some features similar to jellyfish candidates. As a result, cases where similar images to the query contain both jellyfish and non-jellyfish galaxies might be the most complex to classify visually. JClasses predicted using SSL could be the most beneficial for visual inspection for such cases.

In the second-last row, visual and self-supervised approaches match their JClass predictions -- such cases are relatively less complex for visual classification. In the last row, the self-supervised approach predicted the query galaxy to be a stronger jellyfish candidate, resulting from all similar galaxies also being jellyfish. In this case, the visual similarity of morphological signatures between the query and the similar images is not entirely apparent. Despite jellyfish candidates being rare in the dataset, all four similar galaxies are jellyfish, which could strongly indicate jellyfish signatures in the query. However, we note that the query was identified as a merger by visual classification\footnote{ A galaxy is classified as a merger if more than half of the classifiers visually classified it as a merger.}. Hence, it is possible that the self-supervised model could not distinguish well between jellyfish and merger galaxies; instead, it predicted a higher JClass. 

As part of a complementary validation test, we assessed the agreement between the self-supervised and visual JClasses for cases with confident visual classification (i.e., those with a maximum of two distinct visual JClasses across all visual classifiers). This criterion identified 34 non-jellyfish (JClass 0) galaxies\footnote{Applying this criterion yielded only one JClass = 1 example. Although not utilised here, this example further underscores the uncertainty in visually classifying jellyfish candidates.}. Among these, 33 galaxies identified visually as JClass 0 were also predicted as JClass 0 using the self-supervised approach, while the remaining galaxy was classified as JClass 1 by the self-supervised method. Consequently, a high level of agreement is observed between the visual and self-supervised JClasses for these cases. This experiment suggests that the self-supervised model effectively captures abstract features that align with human inspection in classifying a galaxy as a non-jellyfish. However, the limited number of jellyfish examples in this test restricts our ability to fully assess the model's accuracy in identifying jellyfish galaxies, indicating a need for further investigation with a more balanced dataset.

The experiments in this section show that SSL can help visual inspectors classify visually complex cases to improve the JClass prediction. We call this an ``improvement'' since the JClasses predicted based on the nearest neighbour search in the self-supervised representation space alleviates human-level uncertainties associated with visual-only inspection, primarily because learning to distinguish jellyfish from non-jellyfish galaxies does not use labels but is majorly learnt from the data itself. A practical application of our method is to train a self-supervised model on a larger galaxy dataset containing secure jellyfish candidates based on visual inspection and use it for fast JClass assignment on any new galaxy. Unlike pure visual inspection, such an approach may scale to large astronomical datasets better. See Sect.~\ref{sec:discussion} for more discussion.

\section{Astrophysical analysis}
\label{sec:astro_results}

This section explores the astrophysical properties of the jellyfish candidates as labelled by the visual inspection. We present the morphological features of the jellyfish candidates and their spatial distribution around the three cluster systems. We also estimate their star formation rates and stellar masses and analyse their distributions on the phase-space diagrams.  It is important to note that although candidates in JClass 1 and 2 are considered jellyfish galaxy candidates in the study, they are weak examples of jellyfish galaxies. Therefore, they may represent 'disturbed' morphologies rather than exhibiting true jellyfish signatures.

\subsection{Morphological properties via {\sc Morfometryka}}
\label{sec:morpho}

We perform a morphometrical analysis of the galaxies to look for possible correlations in the JClass--morphology space, using the {\sc Morfometryka} code \citep{Ferrari_2015}. We select two non-parametric and one parametric morphological indicator: i) The normalised information entropy $H$, which summarises the distribution of pixel values in the galaxy region in the images -- smooth (clumpy) galaxies have low (high) $H$; 
ii) the Gini coefficient $G$ as used by \citet{Lotz2004}, which is another measure of the flux distribution across pixels; iii) the best-fit S\'ersic index from the two-dimensional S\'ersic fit to the galaxy images, \texttt{nFit2D}, which quantifies the curvature of the radially centred light distribution (which would be the brightness profile in 1D case). We rely solely on $r$-band statistics due to their comparatively high signal-to-noise ratio (S/N). All measurements are performed considering pixels inside the Petrosian region \citep[see][for details]{Ferrari_2015}

\begin{figure}
\centering
\includegraphics[keepaspectratio,width=0.85\linewidth]{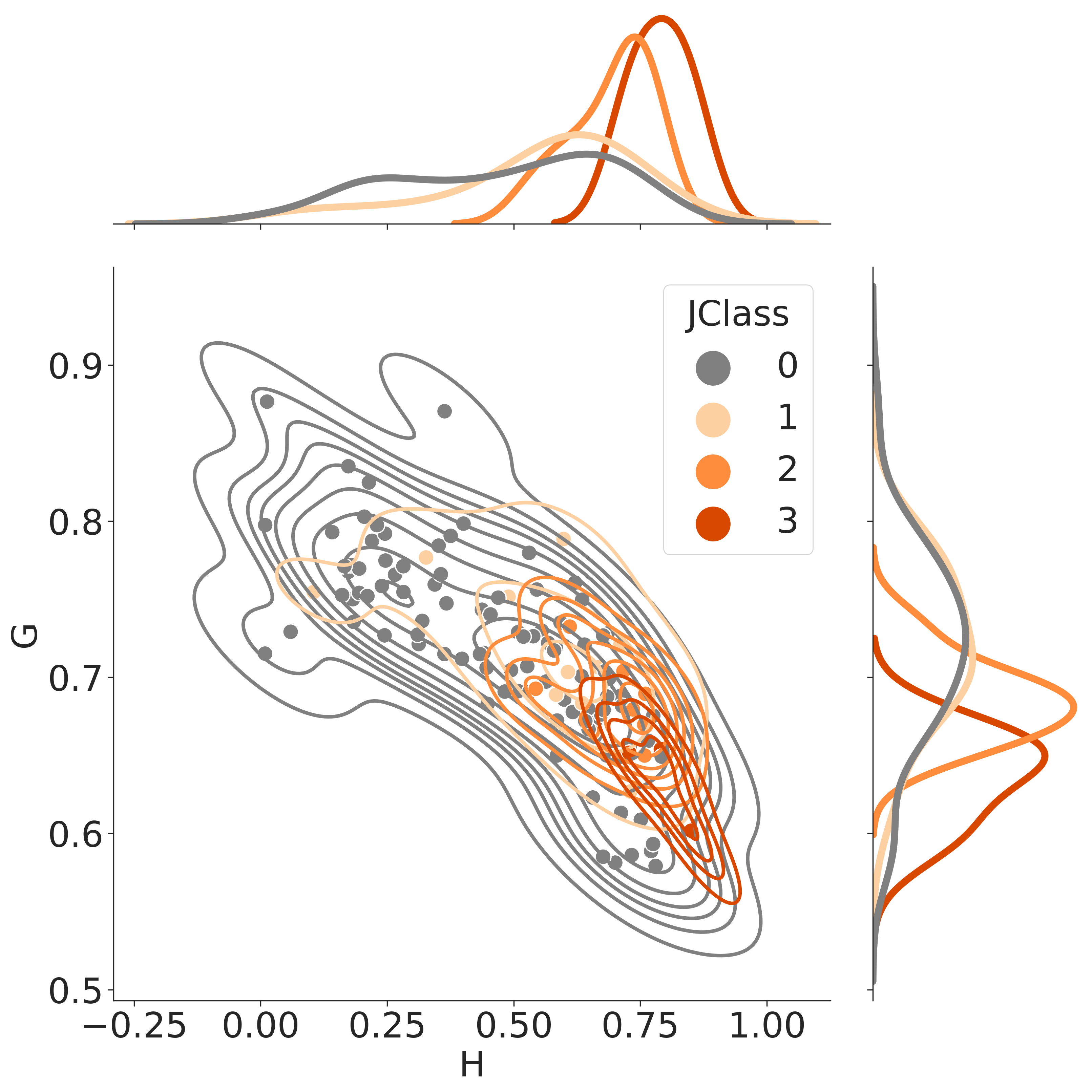}
\includegraphics[keepaspectratio,width=0.85\linewidth]{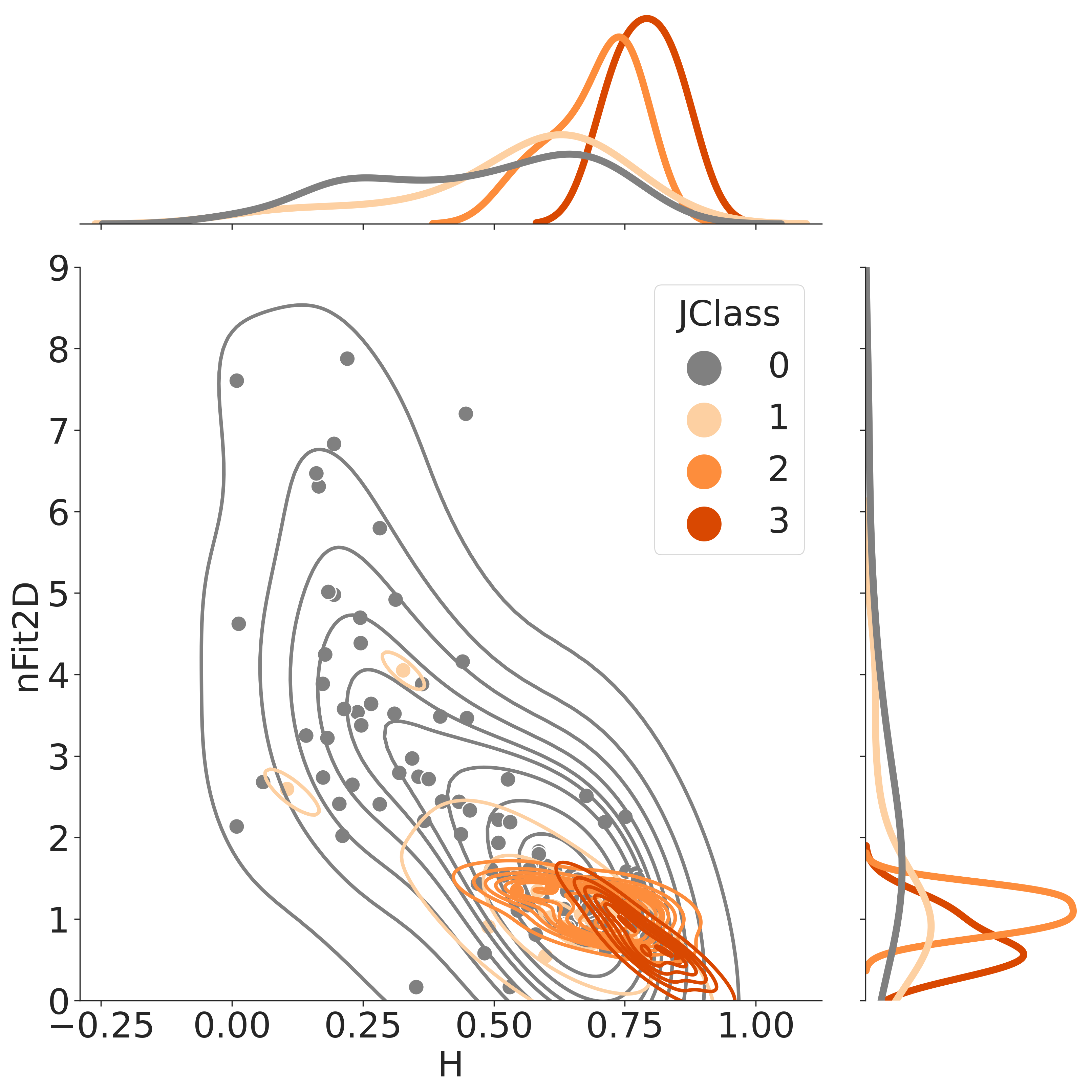}
\caption{Distribution of the morphological parameters considered in this study ($G$, $H$, and \texttt{nFit2D}), coloured by the JClass. The data points and the bivariate KDE curves are shown in the main plot, whereas the univariate KDE curves are shown on the top and right of the figures. The univariate KDEs are normalized independently of each other. Cases with unusual conditions during the morphometry calculations were excluded, as indicated by the quality flag (see Table~3 of \citet{Ferrari_2015} for more details).}
\label{fig:morphJoint}
\end{figure}

Fig.~\ref{fig:morphJoint} shows the joint distributions of $H-G$ and $H-\textrm{nFit2D}$, along with the marginal distributions of each morphological parameter, coloured by the JClass. The most extreme jellyfish candidates in our dataset (with $\textrm{JClass} = 4$) were not considered due to insufficient examples for their kernel density estimate (KDE) calculation.  We note that JClass 1 candidates are the weakest examples of RPS, so they nearly overlap with JClass 0 in the distribution of the three morphological parameters considered. The $H-G$ plot shows a non-trivial correlation between the position in the $H-G$ space and the JClass  of the increasingly stronger JClasses (JClass = 2, 3). In particular, JClass 2 and 3 galaxies have higher entropy than other galaxies, suggesting that galaxies with strong RPS evidence are clumpier than galaxies with weak or no evidence of RPS. A similar pattern is observed with $G$, where JClass 2 and 3 candidates have lower $G$, suggesting that the flux in such galaxies is spread across more pixels than in galaxies with weak or no RPS evidence, which is in contrast to \citet{Bellhouse_2022} who found $G$ alone to be an insufficient indicator to separate ram-pressure stripping galaxies from the general population of galaxies. The $H-\textrm{nFit2D}$ plot additionally shows that JClass 3 candidates have lower S\'ersic indices than lower JClass candidates in the plot, which indicates the former are more disc-like than galaxies with weaker or no RPS evidence (JClass 0, 1, and 2). From an astrophysical perspective, this observation is expected as jellyfish galaxies generally possess a disc component \citep[e.g.,][]{Poggianti16}.

Although the small sample size limits the conclusions from our analysis here, we find hints that the clumpiness of galaxies increases, and their  ``disc-ness" increases as the JClass increases,  as suggested by the marginal distribution of the morphological parameters. Repeating this study on larger datasets would allow deducing more meaningful conclusions. We also note that mergers may share a morphology parameter space similar to jellyfish candidates, which is observed in several studies \citep[e.g.,][]{McPartland16,Bellhouse_2022,krabbe24}, but this is not discussed here. 

\subsection{SFR vs. mass} 
\label{subsec:sSFR}

The star formation rates (SFR) of the jellyfish candidates are derived from the H$\alpha$ fluxes. Such flux measurements are obtained using the Three Filter Method \citep[3FM;][]{Pascual2007, VilellaRojo2015} applied to $r$, $J660$ and $i$-band images. This approach creates emission line maps by assuming that the two broad-band filters can trace the continuum of the galaxy within the narrow-band filter, where the emission line is located. The 3FM is based on colour relations, so the images must be calibrated and PSF-corrected. Moreover, a low pass (Butterworth) filter is applied to all images to decrease noise. A Voronoi binning is performed to reach an S/N of 20 in the $J0660$ image. The S/N limit was chosen after several tests to ascertain artefacts or bad pixels are excluded. Details of the pre-processing procedures will be discussed in Lopes et al. ({\it in prep})\footnote{Examples of H$\alpha$ maps and the main code used to derive the SFR can be found at \url{https://github.com/amanda-lopes/Halpha-SPLUS-Jelly}}. We integrated the resulting H$\alpha$ map to estimate the H$\alpha$ flux within a radius encompassing 90\% of the flux ($r_{90}$) in the $r$-band for each galaxy. The choice of $r_{90}$ aims to maximise the inclusion of the emission structure in the analysis. From the H$\alpha$ flux, we derived the H$\alpha$ luminosity, which is converted to SFR following the relation given by \cite{Kennicutt1998}. The SFRs are corrected for dust and [NII] following the relation proposed by \cite{Kouroumpatzakis2021}. This procedure computes the total integrated star formation rates for the entire galaxy region. SFR errors are derived by a simple propagation of errors.

The stellar masses ($M_{\star}$) are obtained by fitting the galaxy spectral energy distributions (SEDs) with the \texttt{CIGALE} code (\citealt{Cigale}; version 2020.0). SED modelling was performed for only 84 galaxies from the main and control samples (15, 46, and 23 from Antlia, Fornax, and Hydra, respectively), which had their photometry measured in the S-PLUS filters (\citealt{haack24}; \citealt{smith24}) using \texttt{SExtractor}. The spectroscopic redshifts and distances were obtained from the NASA/IPAC Extragalactic Database (NED)\footnote{The NASA/IPAC Extragalactic Database (NED) is operated by the Jet Propulsion Laboratory, California Institute of Technology, under contract with the National Aeronautics and Space Administration.}.  

\begin{figure}
    \includegraphics[width=0.9\linewidth]{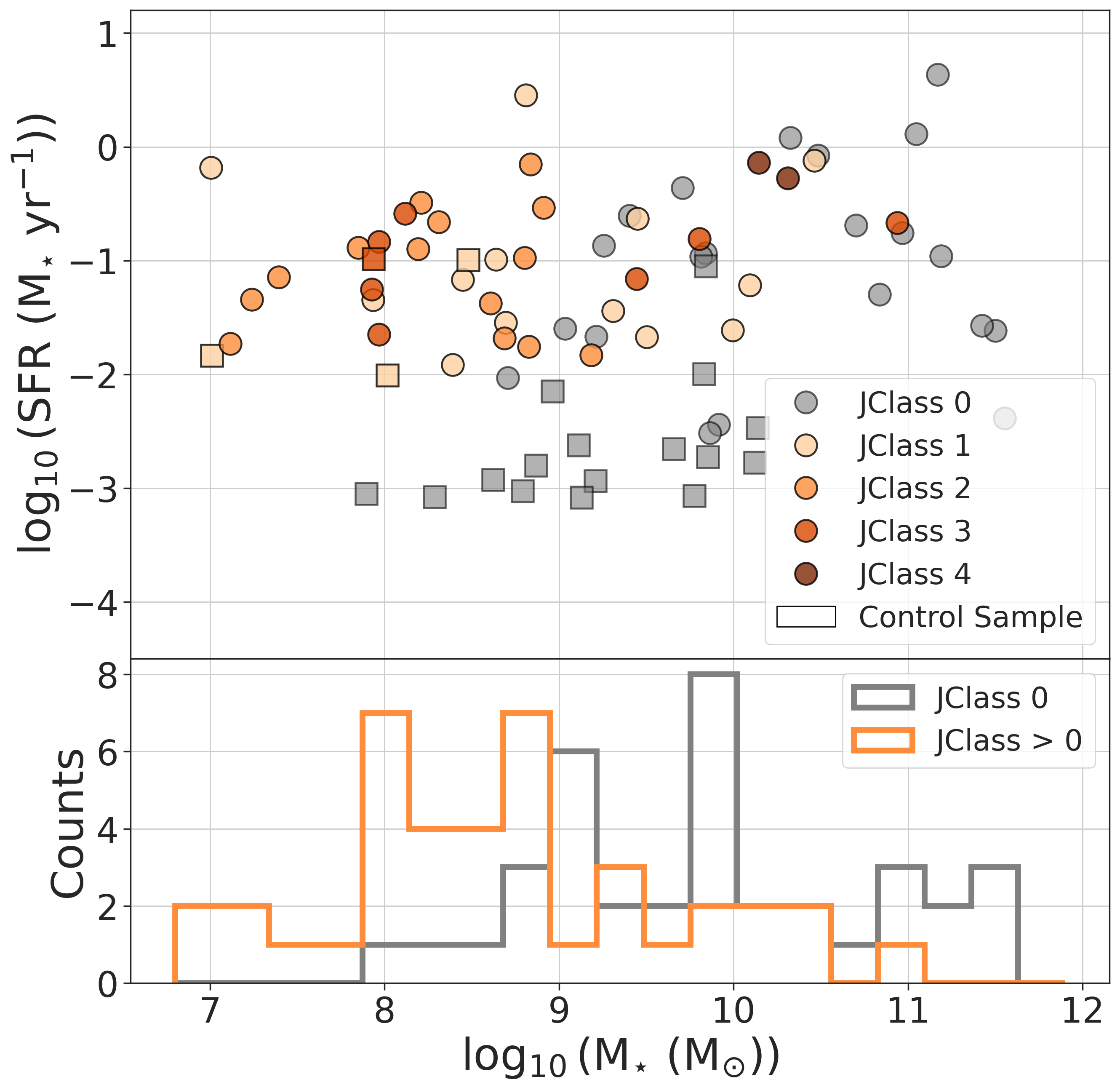}
    \caption{Star formation rate versus stellar mass for varying JClass candidates. Dots (squares) indicate galaxies from the main (control) sample. The lower panel shows the distribution of the stellar masses for non-jellyfish and jellyfish candidates. Examples with significant errors in the SFR calculation $\Big(\dfrac{\textrm{SFR}_{\textrm{error}}}{\textrm{SFR}} > 50\%\Big)$ are excluded. Typical errors in the SFR are $\lesssim30\%$.}
    \label{fig:SFR_mass}
\end{figure}

In Fig.~\ref{fig:SFR_mass}, we compare the star formation rate (SFR) as a function of $M_{\star}$ of all jellyfish candidates (JClass > 0) vs. normal (i.e., non-jellyfish; JClass 0) or star-forming galaxies, combined from the three clusters. Comparison between each JClass is not performed due to the low number of examples for each JClass. It can be observed that no significant trends are observed in the SFR vs. $M_{\star}$ relation for the jellyfish candidates. The SFRs of the jellyfish candidates generally tend to possess elevated star formation compared to non-jellyfish candidates; however, this elevation is not apparent at the high stellar mass end. The jellyfish candidates are skewed towards lower stellar masses, as seen in the lower panel. Thus, one possible reason for the non-elevation could be the rarity of jellyfish candidates at higher stellar masses. Possible implications of these observations are discussed in Sect.~\ref{sec:discussion}.  

We perform a two-sample Kolmogorov-Smirnov (KS) test to statistically quantify the SFR comparison, similar to \citet{roman2019}. The p-values for the SFR comparison of jellyfish vs. non-jellyfish galaxies and the main vs. control sample galaxies are lower than $10^{-4}$. For any reasonable confidence level (e.g., 95\%, 99\%), we thus reject the null hypothesis that the SFRs of the two samples are drawn from the same distribution for the jellyfish vs. non-jellyfish and the main vs. control comparison. These findings are qualitatively in line with the results of other studies \citep[e.g.,][]{Poggianti16, enhancedSFR, roman2019}, which found increased star formation in jellyfish candidates compared to other normal or star-forming galaxies.

\subsection{Direction of infalling} \label{subsec:trail_vector}

Jellyfish galaxies leave a trail of RPS gas in the opposite direction of motion \citep{LoTSS-II, gaspI, Fumagalli2014}. Thus, the trail direction provides information about the most probable direction of the galaxy's motion in the cluster system \citep[e.g.,][]{McPartland16, Smith2010,smith22}. We discuss the projected motion of the jellyfish candidates in each cluster: Antlia, Hydra, and Fornax, shown by trail vectors \citep[e.g.,][]{roman2019, roman2021}.

\citet{roman2021} found that the shift between the peak and the centre of light of galaxies (calculated using {\sc Morfometryka}) is a better proxy for the motion direction than visual inspection, especially for disturbed morphologies. The motivation for using such an approach lies in the fact that while the centre of light of the galaxy is sensitive, the peak of light is resilient to perturbations in the galaxy morphology due to ram pressure stripping so that the difference between them can be used as a tracer of the galaxy's motion. As a result, we derive the trail vectors using {\sc Morfometryka} measurements, where the direction is based on the following relation: $(x, y)_{\rm peak} - (x, y)_{\rm col}$. Similar to Sect.~\ref{sec:morpho}, calculations are performed only on $r$-band images.

\begin{figure*}
\begin{subfigure}{.33\linewidth}
\centering
\includegraphics[width=\textwidth]{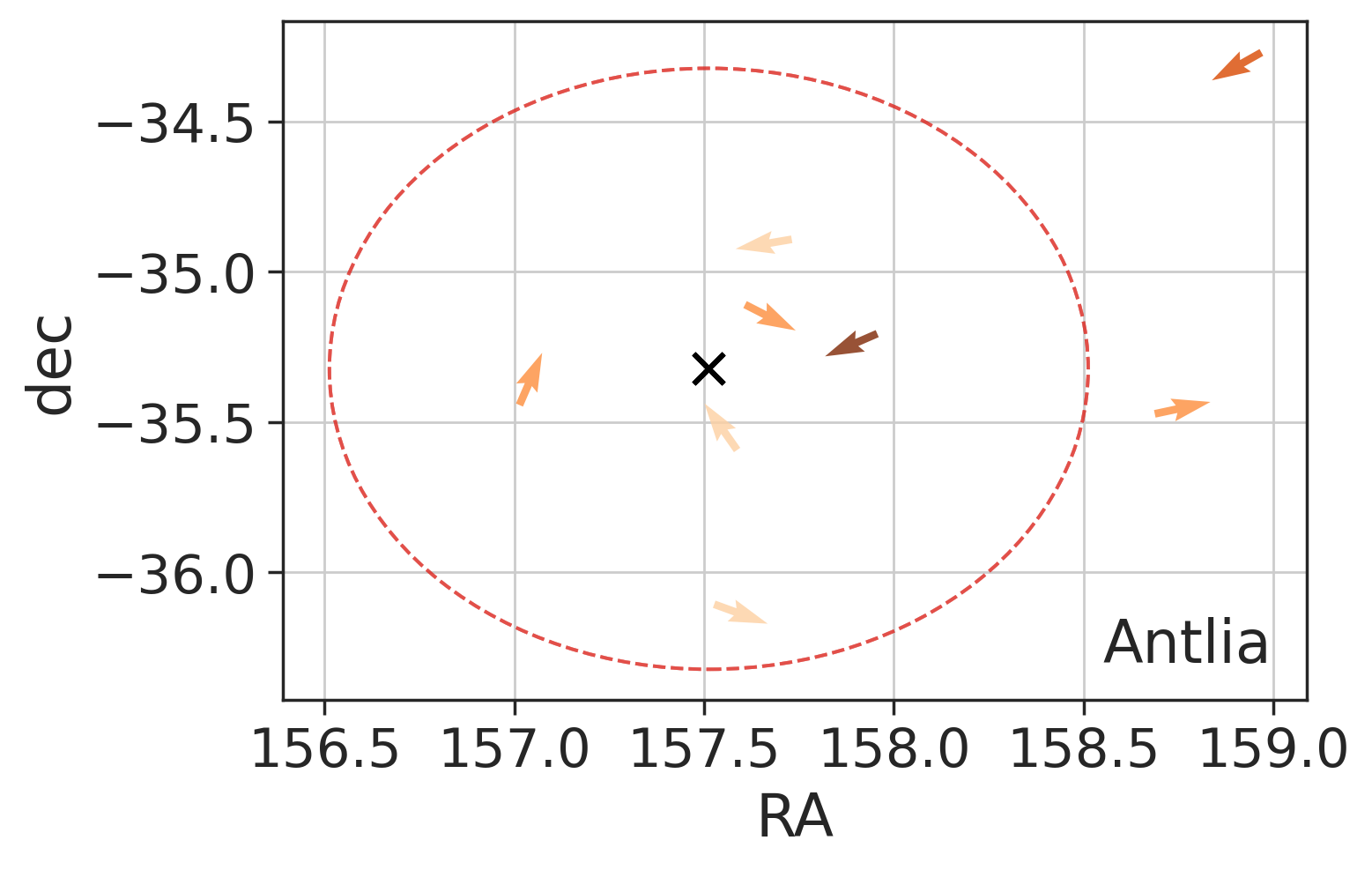}
\end{subfigure}\hfill
\begin{subfigure}{.31\linewidth}
\centering
\includegraphics[width=\textwidth]{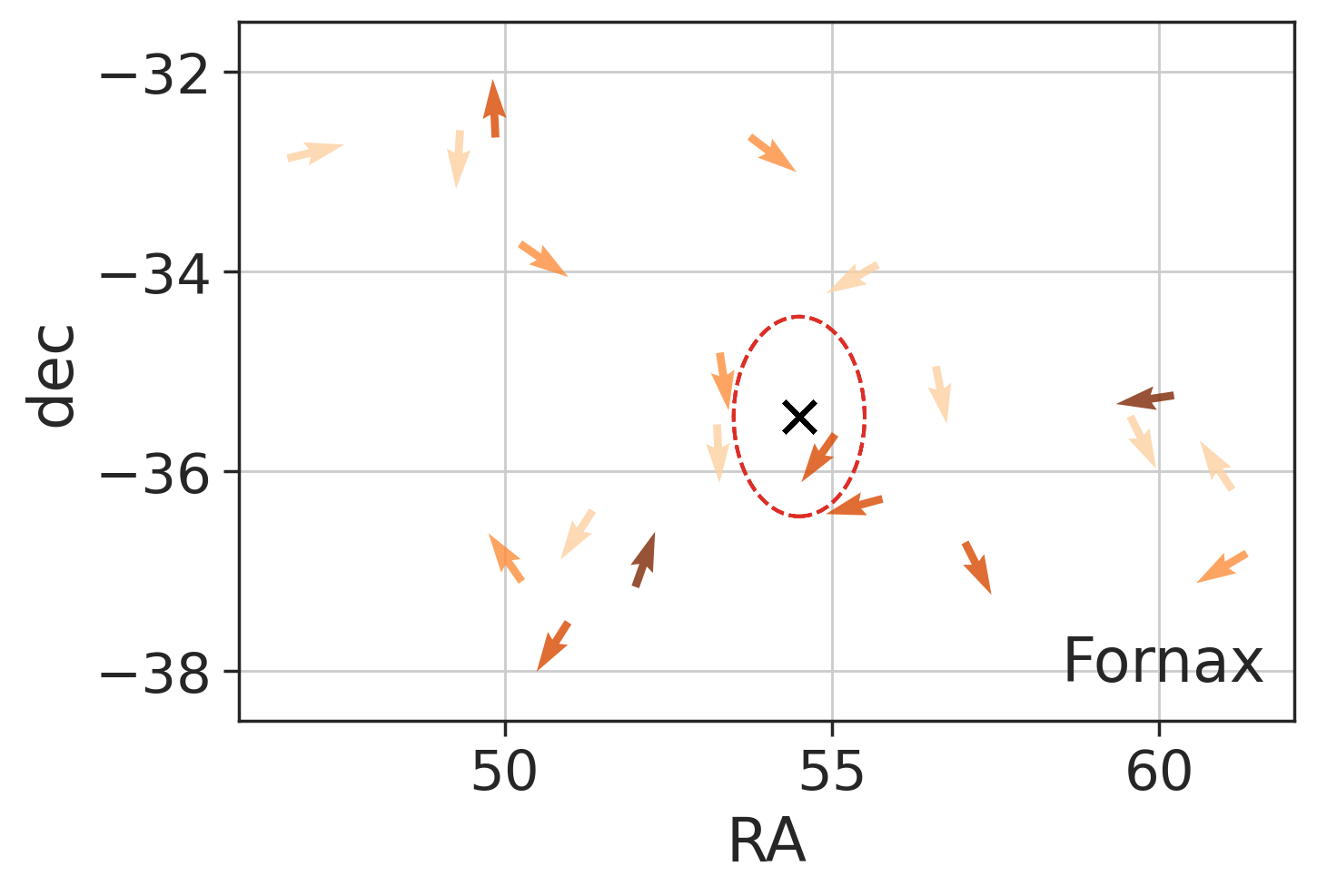}
\end{subfigure}\hfill
\begin{subfigure}{.318\linewidth}
\centering
\includegraphics[width=\textwidth]{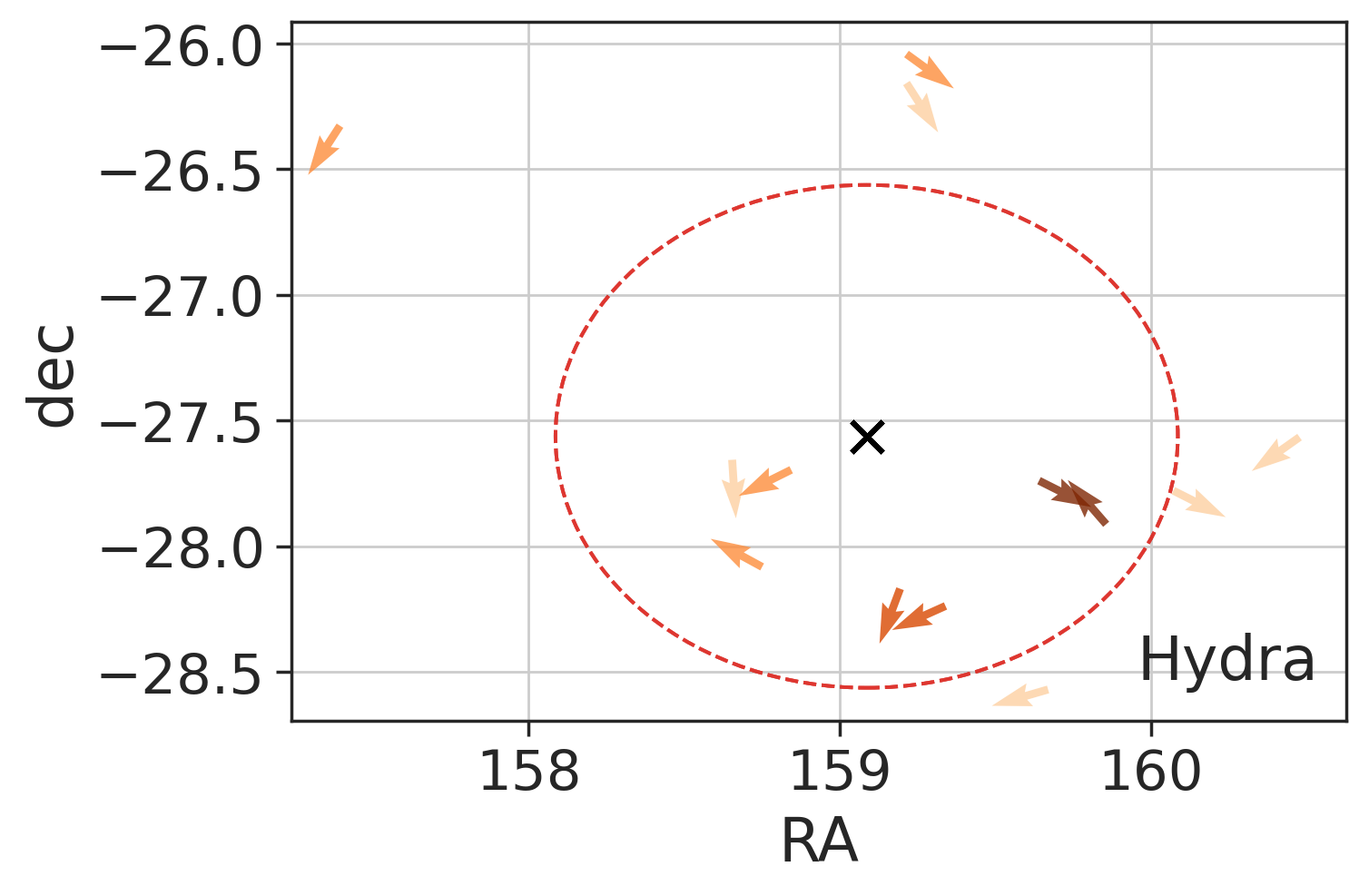}
\end{subfigure}
\begin{subfigure}{0.5\linewidth}
\centering
\includegraphics[width=\textwidth]{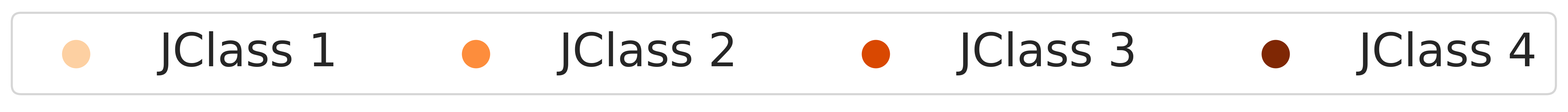}
\end{subfigure}
\caption{Spatial distribution of jellyfish candidates in each galaxy cluster: Antlia, Fornax, and Hydra, along with their trail vectors shown by arrows. Since only the direction is considered in this study, all trail vectors are shown with the same length. Galaxies with a failed trail vector calculation were excluded. The black cross marks the centre of the cluster. The red dashed circle indicates the virial radius.}
\label{fig:trail_vectors}
\end{figure*}

Fig.~\ref{fig:trail_vectors} shows the spatial distribution of the jellyfish candidates from each galaxy cluster, with the trail vectors shown by the arrows. Similar to \citet{roman2019}, we calculate the angle between the trailing vector and the line joining the galaxy to the cluster's centre to identify whether the galaxy moves towards or away from the cluster centre. The galaxy is considered infalling towards the centre if the angle is less than 90 degrees and outfalling if the angle is greater than 90 degrees. Table~\ref{tab:trailDirectionsNumbers} shows the distribution of the number of galaxies falling inwards or outwards from the respective cluster system. We consider galaxies in a cluster to preferentially fall toward the cluster when more than half of the galaxies are found to be infalling. Thus, galaxies from Antlia and Fornax preferentially fall toward the cluster, whereas there seems to be mild  or no specific preference for galaxies from Hydra to infall. However, we note that our analysis is affected by the limited sample size.

\begingroup

\begin{table}
\centering
\caption{ No. of infalling and outfalling galaxies, categorised by the JClass, in each galaxy cluster, as indicated by the trail vector directions. See Fig.~\ref{fig:trail_vectors} for the trail directions.}
\label{tab:trailDirectionsNumbers}
\begin{tabular}{@{}ccccccc@{}}
\toprule
Cluster & Direction  & JClass 1 & JClass 2 & JClass 3 & JClass 4 & Total \\ \midrule
Antlia  & Infalling  & 2        & 2        & 1        & 1 &        6       \\
        & Outfalling & 1        & 1        & 0        & 0 &        2       \\
Hydra   & Infalling  & 3        & 3        & 0        & 1 &        7       \\
        & Outfalling & 2        & 1        & 2        & 1 &        6       \\
Fornax  & Infalling  & 5        & 5        & 2        & 2 &        14       \\
        & Outfalling & 3        & 1        & 3        & 0 &        7        \\ \bottomrule
\end{tabular}
\end{table}

\endgroup

\subsection{Phase-space analysis} \label{subsec:phase_space}

The environment where a galaxy resides within a group or cluster may pose noteworthy morphological and physical transformations. In addition, the position and velocity of the galaxy with respect to the cluster centre are determinants for our understanding of the different dynamical effects at play. In particular, the phase-space diagram (\citealt{Jaffe15}) relates the peculiar line-of-sight (LOS) velocity $\Delta \mathrm{V_{los}}$ of each galaxy and their projected radial position $\mathrm{R_p}$ from the cluster centre. The line-of-sight velocity can be determined by
\begin{equation}
    \displaystyle{\frac{\Delta \mathrm{V_{los}}}{\sigma_{\mathrm{v}}} = \frac{c(z-z_{cl})}{(1+z_{cl})\sigma_{\mathrm{v}}}},
\label{eq:los_velocity}
\end{equation}
where $\sigma_{\mathrm{v}}$ is the velocity dispersion of the cluster, $c$ the speed of light, $z$ the spectroscopic redshift of a given galaxy, and $z_{cl}$ the redshift of the cluster. For the projected distance, we converted each angular distance in arcsec to a kpc scale based on the distance to the cluster (e.g., 1$\arcsec$\,=\,0.247\,kpc in Hydra, as discussed in \citealt{Arnaboldi12}). 

The spectroscopic properties of the jellyfish candidates were obtained from NED. However, for three candidates from Hydra, we did not find their spectroscopic properties in NED, for which we made use of the catalogue of ram pressure targets from Hydra published by the \texttt{WALLABY} survey \citep{Wang21}. The coordinates (RA, DEC) of these galaxies are: (159.854$^{\circ}$, -27.9125$^{\circ}$), (159.192$^{\circ}$, -28.1672$^{\circ}$) and (159.337$^{\circ}$, -28.2372$^{\circ}$). The properties of the three clusters are shown in Table \ref{tab:cluster_properties}. In the case of Fornax, we consider the cluster centre at NGC1399.

\begingroup
\begin{table}
       \centering
       \caption{Cluster properties: distance to the cluster, $D_{cl}$ (Mpc), virial radius, $\mathrm{R_{200}}$ (Mpc), virial mass, $\mathrm{M_{200}}$ ($\mathrm{M_{\odot}}$), velocity dispersion, $\sigma_{\mathrm{v}}$ (km\,s$^{-1}$), and spectroscopic redshift ($z_{cl}$). References: {\small
       (a) \citet{Wong16};
       (b) \citet{Sarkar22};
       (c) \citet{Ragusa2023}; (d) \citet{Hopp1985}; (e)\citet{Sarkar22}; (f) \citet{Tonry2001};  (g) \citet{Iodice2019}; (h) \citet{Drinkwater2001}; (i) \citet{Reiprich2002}; (j) \citet{Arnaboldi12}; (k) \citealt{Wang21};  (l) \citet{LimaDias21}.}
       } 
       \begin{tabular}{lccccc}
       \toprule
        Cluster & $D_{cl}$  & $\mathrm{R_{200}}$ & 
        $\mathrm{M_{200}}$ & $\sigma_{\mathrm{v}}$ & $z_{cl}$\\
         \midrule
        Antlia & 39.8$^{(a)}$ & 0.887$^{(b)}$ & $10^{14\,(c)}$ & 591$^{(c)}$ & 0.009$^{(a)}$\\
        Fornax & 19.9$^{(e)}$ & 0.7$^{(f)}$ & $7\times10^{13\,(g)}$  & 370$^{(g)}$ & 0.0046$^{(h)}$\\
        Hydra & 50$^{(i)}$ & 1.4$^{(j)}$ &  $3\times10^{14\,(k)}$   & 690$^{(l)}$ & 0.012$^{(l)}$\\
      \bottomrule
       \end{tabular}
       \label{tab:cluster_properties}
   \end{table}
\endgroup

\begin{figure*}
\begin{subfigure}{.33\linewidth}
\centering
\includegraphics[width=\textwidth]{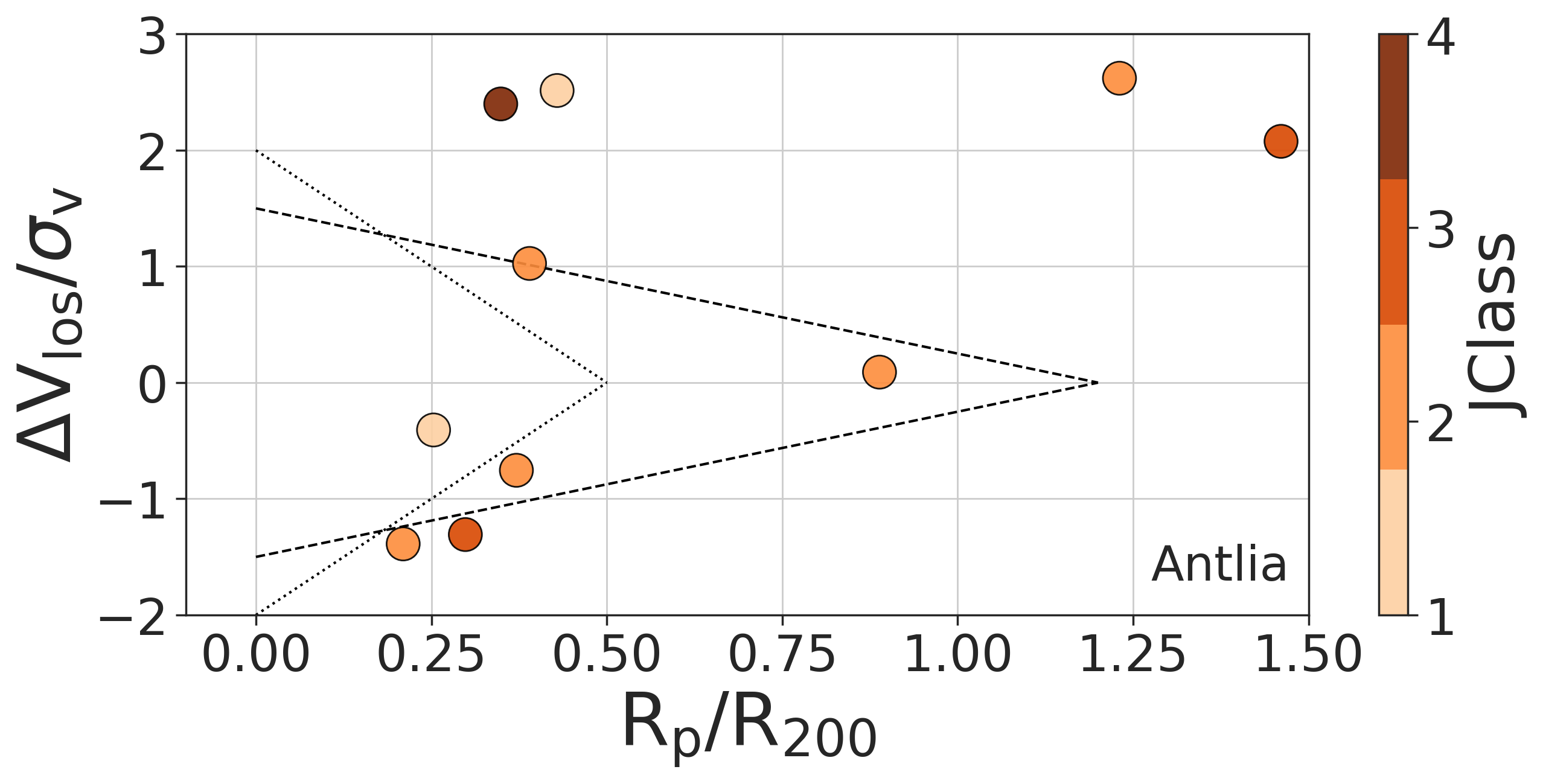}
\end{subfigure}\hfill
\begin{subfigure}{.33\linewidth}
\centering
\includegraphics[width=\textwidth]{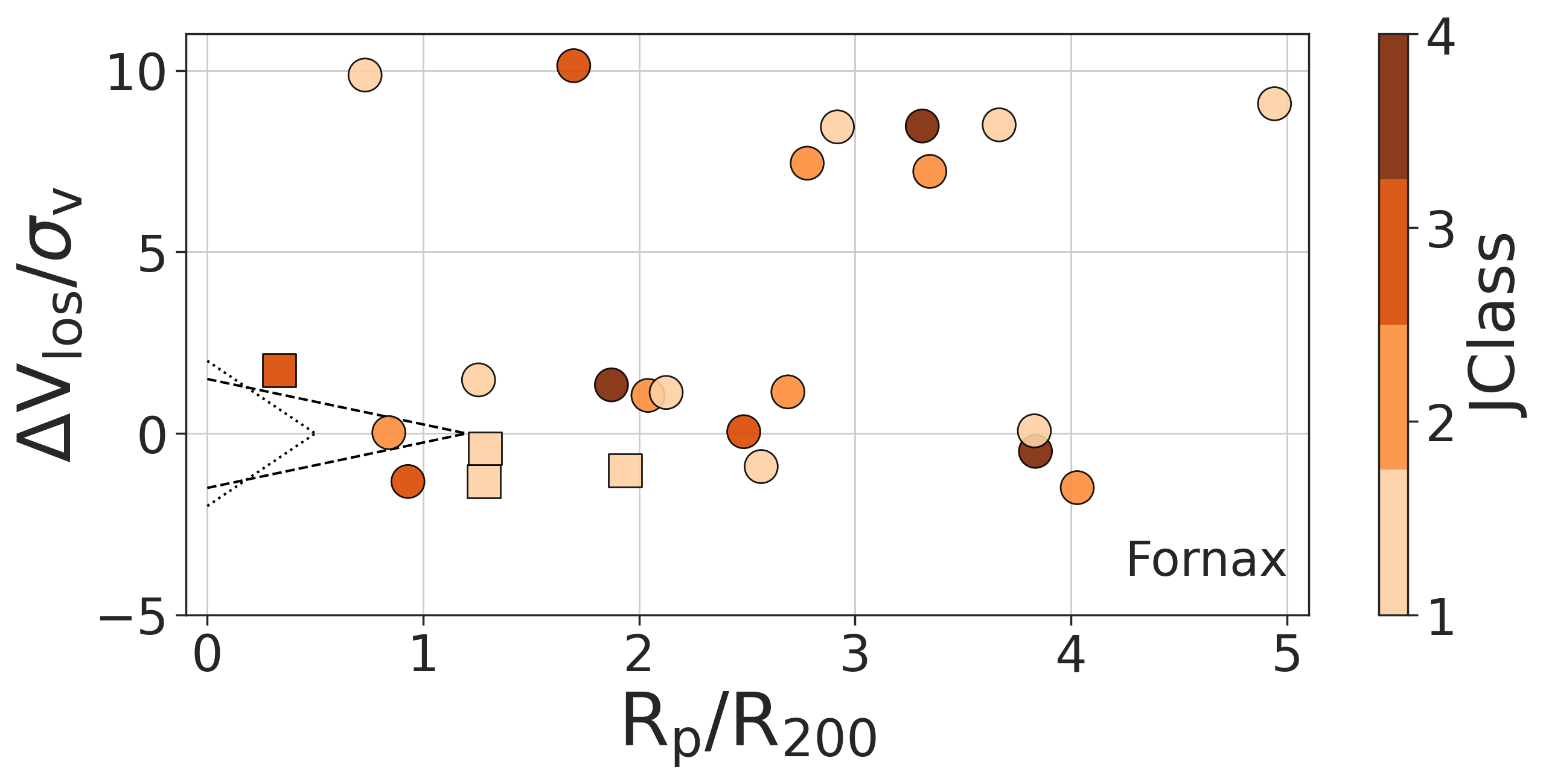}
\end{subfigure}\hfill
\begin{subfigure}{0.33\linewidth}
\centering
\includegraphics[width=\textwidth]{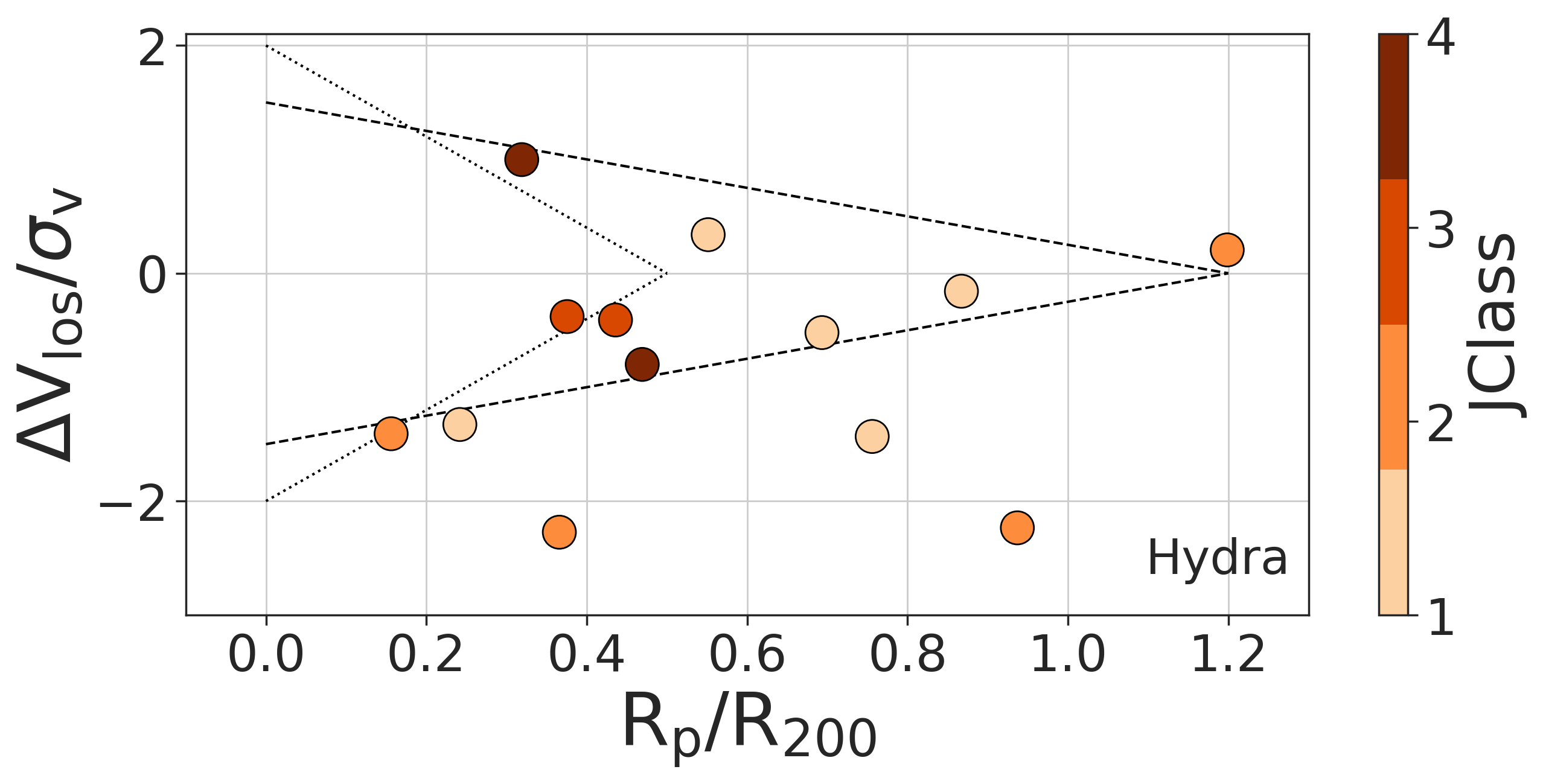}
\end{subfigure}
\caption{Projected phase-space diagram for the jellyfish candidates from Antlia, Fornax, and Hydra. The cluster centre for Fornax considered is NGC1399. The colour bar shows the JClass. In the case of Fornax, dots (squares) indicate candidates from the main (control) samples. The dotted and dashed lines indicate boundaries $B1$ and $B2$, respectively.}
\label{fig:phase_space}
\end{figure*}

In Fig.~\ref{fig:phase_space}, we show the distributions of the most secure jellyfish candidates from each cluster in the projected phase-space diagram. The x-axis shows the projected distance normalised by the virial radius $\mathrm{R_{200}}$. Following \citet{roman2019}, we define two boundaries: $(B1)$ $\displaystyle{|\Delta V_{los}/\sigma_{v}|\leq 1.5-(1.5/1.2)\times \mathrm{R_{p}/R_{200}}}$ (\citealt{Jaffe15}) and $(B2)$ $\displaystyle{|\Delta V_{los}/\sigma_{v}|\leq 2.0-(2.0/0.5)\times \mathrm{R_{p}/R_{200}}}$ \citep{Weinzirl17} (see also \citealt{Rhee2017,Pasquali2019}). 
The areas within the defined boundaries represent virialized galaxies. The purpose of segmenting the phase space into regions is to determine the most likely stage of the galaxy's orbit, such as recent infalling, backsplashing, or having already undergone virialization.

In the case of Antlia, some of the JClass 1 and 2 candidates are under the influence of the cluster, and have peculiar velocities of the order of the velocity dispersion of the cluster. Two candidates are close to boundary $B2$ (one of them being a JClass 3). However, the remaining candidates (including a JClass 4) are outside the influence of the cluster and exhibit relatively high LOS velocities. For Hydra, most of the candidates are found within the influence of the cluster. On the other hand, most of the Fornax candidates are located in the outskirts ($\mathrm{R_p > 2\times R_{200}}$), which is also confirmed by the spatial distribution of Fornax galaxies in Fig.~\ref{fig:trail_vectors}. This is not unexpected since Fornax's outskirts up to large radii are covered in this study. These candidates also exhibit high velocities with respect to the cluster velocity dispersion.  Two JClass 4 candidates are located at $R_p > 3R_{\textrm{200}}$. Further inspection of these candidates can reveal whether ram-pressure stripping is acting out to these large outskirts, which is a possible phenomenon \citep{Bahe2013}.

\begin{figure}
    \includegraphics[width=0.9\linewidth]{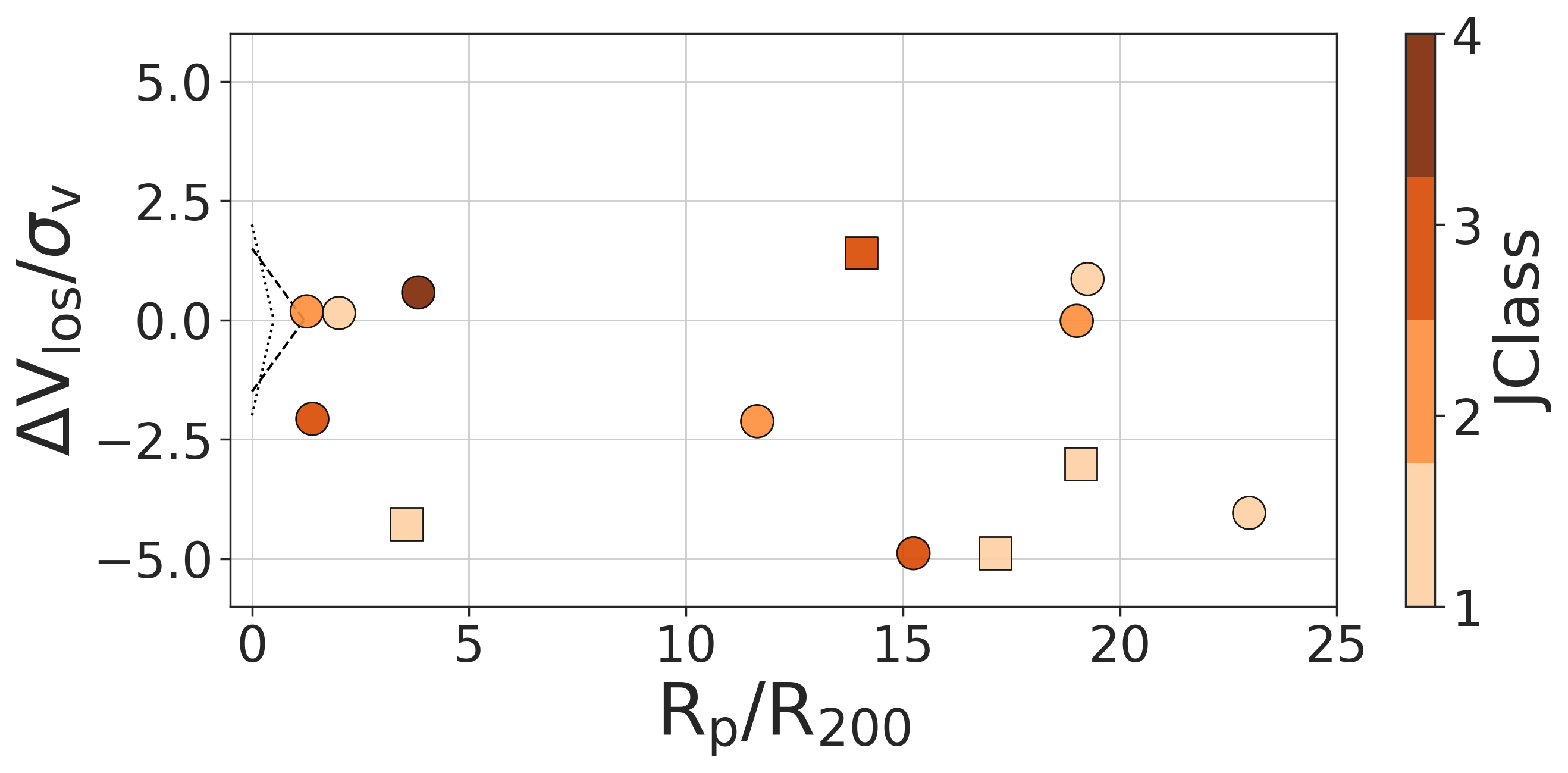}
    \caption{Projected phase-space diagram similar to the one for Fornax in Fig.~\ref{fig:phase_space}, but considering the centre at NGC1316.}
    \label{fig:phase_space_diagram_fornax2}
\end{figure}

\begin{figure}
    \includegraphics[width=0.9\linewidth]{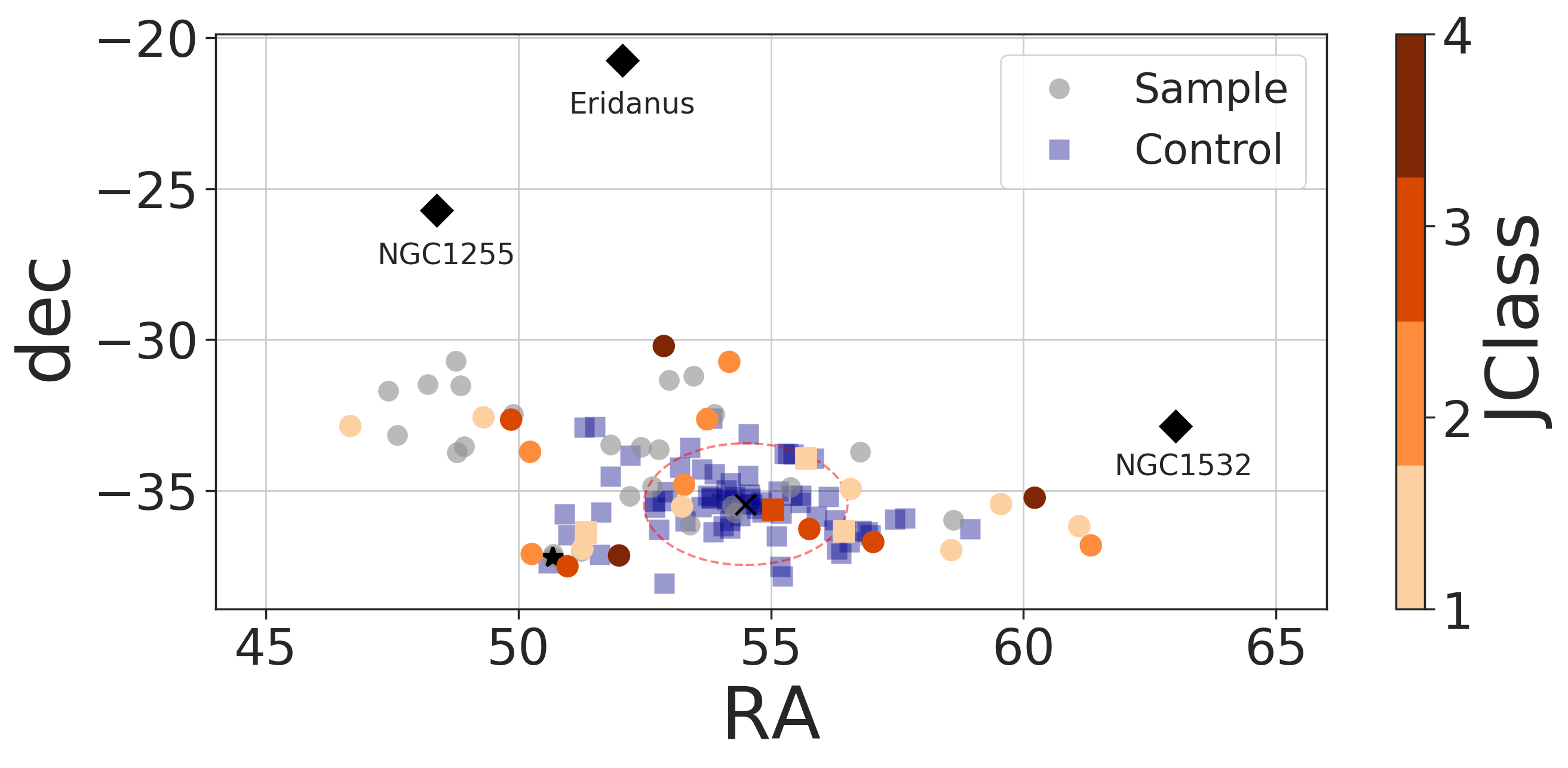}
    \caption{Spatial distribution of galaxies from Fornax. The dots (squares) indicate candidates from the main (control) samples. The colour bar indicates the JClass. The pink dashed circle indicates the virial radius. The black cross (star) shows the location of NGC1399 (NGC1316). The black diamonds indicate the locations of three nearby groups (NGC1225, Eridanus, and NGC1532).}
    \label{fig:sky_distribution_fornax_outliers}
\end{figure}

To better investigate the distribution of candidates in Fornax, we plotted the phase space diagram by considering the centre at NGC1316, a lenticular galaxy from an in-falling subgroup. However, as shown in Fig.~\ref{fig:phase_space_diagram_fornax2}, the jellyfish candidates are even further away from the influence of this subgroup. Finally, since Fornax is surrounded by three other groups (NGC1225, Eridanus, and NGC1532), another hypothesis is that they may influence these candidates. However, this conjecture is refuted by the spatial distribution map shown in Fig.~\ref{fig:sky_distribution_fornax_outliers}.Therefore, more investigation is needed to determine whether  this gravitational influence of Fornax is causing RPS in these candidates.
 We note that, another significant influence of the cluster environment on a galaxy, which reaches beyond the virialized region, is the extent of the virial shock surrounding the cluster. This shock boundary can extend several times beyond the virial radius \citep[e.g.][]{Bahe2013,zinger2018} and 
once a galaxy crosses the virial shock, the surrounding gas density increases and consequently there is a rise in ram pressure.

\section{Discussion} \label{sec:discussion}
This paper analyses 51 jellyfish galaxy candidates from three nearby galaxy clusters: Fornax, Antlia, and Hydra, observed in the S-PLUS survey. These candidates were derived using the traditional visual inspection approach, which produced a categorical RPS measure, the JClass, ranging from 1 to 4, representing the weakest to strongest RPS evidence. 
We have not recovered any JClass = 5 cases in our hunt for jellyfish candidates in these three clusters. It is possible that these clusters do not harbour extreme ram-pressure stripping galaxies, or that such stripped structures may be revealed by other observational methods that are beyond the parameters probed by this study; see \citet{serra23} for an example of a prominent tail in HI gas observed by MeerKAT in NGC1437A, which received a JClass = 3 in our study.

Following the identification of jellyfish candidates, we analysed their astrophysical properties. A morphological study revealed that moderate to extreme jellyfish candidates (JClass 2 and 3; JClass 4 was not studied for morphology due to scarcity of JClass 4 examples) are clumpier (higher entropy, $H$) and have more scattered flux (smaller Gini coefficient, $G$) than galaxies with weak or no evidence of RPS (JClass 0 and 1). The JClass 2 and 3 galaxies are more disc-like than JClass 0 and 1 galaxies, quantified by the lower S\'ersic indices of the former. The increasing ``disc-ness" as the jellyfish signatures become more prominent is expected since jellyfish galaxies are known to have a prominent galactic disc. The majority of the jellyfish candidates with JClass 2 and 3 are low stellar mass galaxies ($M_* < 10^{10}$ M$_{\odot}$), with most of them having $10^7$ M$_{\odot} < M_* < 10^9$ M$_{\odot}$ (see Fig. \ref{fig:SFR_mass}).
Low-mass star-forming galaxies are usually pure-disc systems with surface brightness profiles well described by an exponential law \citep[$n \sim 1$;][]{2006ApJS..162...49H,2015MNRAS.447.2603L,2015ApJS..219....4S}. These galaxies are the ones that are more easily perturbed in a dense environment \citep{2006PASP..118..517B,review_jellyfish,2023A&A...675A.108K}. 
Thus the jellyfish candidates with a higher JClass might be associated  with lower mass cluster members that are more sensitive to the surrounding environment compared to JClass 0 galaxies, characterised by a larger range of stellar masses that extends above $10^{10}$ M$_{\odot}$ (Fig. \ref{fig:SFR_mass}). 

An essential finding through this analysis is that high JClass galaxies ($\textrm{JClass} \geq 2$, are not entirely distinct from the others (JClass 0, 1). Instead, they prefer a specific sub-region of the morphological parameter space of the JClass 0 and 1 galaxies, as aptly demonstrated by Fig.~\ref{fig:morphJoint}. This overlap in space occupation could be because the employed morphological parameters produce degeneracies between extreme RPS galaxies and a specific type of non-jellyfish galaxies. These observations demonstrate that jellyfish candidates have complicated morphological characteristics that likely cannot be sufficiently described by a single morphological indicator. Thus, a combination of these morphological characteristics can be used to perform morphological cuts for candidate jellyfish sample selection in the future. Simulations studying the evolution of jellyfish galaxies (its different stages such as cluster infall, stripping of gas from its disc, and late-stage evolution) can allow tracking its position in the morphology space (e.g., the two-dimensional space of Fig.~\ref{fig:morphJoint}) to get direct insights into morphological evolution of jellyfish galaxies.  Also, interpreting weak jellyfish candidates (JClass 1 and 2) as examples of `disturbed' morphologies rather than jellyfish could also help explain why these candidates largely overlap with the non-jellyfish population.

The star formation activity analysis suggested that galaxies with $\textrm{JClass} \geq 1$ have a higher SFR than normal, star-forming galaxies (with JClass 0). This observation is expected since RPS is known to cause a temporary starburst in jellyfish galaxies before eventually quenching its cold gas \citep[e.g.,][]{Gullieuszik2017, enhancedSFR, Poggianti2019Starburst, roman2019, SFR_tails}.
The SFR vs. stellar mass plot trends suggest that this effect can become less pronounced for high-mass galaxies ($M_\star \gtrsim 10^9\,M_{\odot}$). However, it is not possible to make definitive conclusions since selection effects can be at play: our sample consists of few examples of low-mass normal, star-forming galaxies, and many jellyfish candidates have a low stellar mass, as seen in Fig.~\ref{fig:SFR_mass}. Such a disproportionate distribution of stellar mass of the galaxy candidates poses difficulties in comparing SFRs at any given mass in this study.

By analysing the tail direction of galaxies with $\textrm{JClass} > 1$ in their respective cluster systems, we find that galaxies from the Antlia and Fornax clusters have tails preferentially pointing away from the cluster centre, which suggests they are falling towards it probably for the first time. However, only a mild preference is observed for galaxies in the Hydra cluster since around half of the galaxies in Hydra indicate infall, whereas the other half indicate outfall. We find insufficient evidence that nearby groups could affect the ram pressure stripping in galaxies from Fornax.

Further insights might require more extensive analyses due to potential selection effects influenced by the galaxies' projected radii from the cluster centre. For instance, comparing our calculated trail vector directions with H{\sc i} tail directions from studies such as \citet{Wang21, 2023A&A...675A.108K} could be insightful but lies beyond the scope of this paper. 

The properties of jellyfish candidates are generally compared with non-jellyfish galaxies to gain insights into the differing physics between the two types of galaxies. Naturally, the astrophysics and the subsequent scientific implications depend on how the JClasses are assigned. The traditional visual classification used to separate jellyfish from non-jellyfish galaxies is highly dependent on human visual biases, which makes the classification subjective. In our visual classification, we often observed disagreement between different classifiers. It is also impractical to manually vet large datasets (comprising $\gtrsim 10^{6-7}$ galaxies) to identify rare jellyfish galaxies.

To mitigate these challenges, we proposed a semi-automated pipeline using SSL and demonstrated a proof-of-concept application of our pipeline. We pretrained an encoder network aimed at extracting generalised feature representations of galaxies without relying on human-based JClass for learning. Thus, SSL pretraining is ideal for learning morphologies from large, unlabelled galaxy datasets, a common theme in astronomy. The encoder is then used as a feature extractor on galaxies unseen during the pretraining. Noting the limitations of fully supervised learning or supervised learning on self-supervised representations in providing JClass estimates, we developed a novel downstream task based on the self-supervised representations. We assign a self-supervised-based JClass to refine the visually-assigned JClass using a weighted nearest-neighbor search on the self-supervised representations. For any given galaxy, we assign the refined JClass as the weighted mean of visual JClasses of the $k$ nearest neighbours in the self-supervised representation space. We demonstrated the application of this framework to a few galaxies having uncertain visual JClass classification, which in this study refers to the case where more than half of the visual classifiers predict different JClasses. Care was taken to ensure that the nearby galaxies used to assign the JClass to the concerned galaxy were not themselves uncertain for visual classification. The astrophysical analysis presented in Sect.~\ref{sec:astro_results} studied possible patterns in the changing astrophysics as the JClass increases; however, the JClass was obtained only by visual inspection. Our self-supervised pipeline can be integrated with visual classification to improve JClass estimates of uncertain visual classifications, which could reveal new patterns or improve the reliability of observed patterns.

Due to our limited dataset, it is difficult to reliably quantitatively assess how much disagreement exists between the visual and self-supervised JClasses. However, we have found many cases where SSL predicted weak and intermediate jellyfish candidates based on visual classification (JClass 1, 2) as non-jellyfish (JClass 0). This may help mitigate false-positive cases. We have also found cases where both approaches agree in their JClass estimate, particularly for visual non-jellyfish galaxies. Cases where the self-supervised predicted a stronger jellyfish signature were also present. Such cases could pave the way for detecting new jellyfish galaxies. We have validated our framework by finding that the self-supervised and visual JClasses agree well for visually confident JClass predictions, which was the expectation.

A current limitation of our self-supervised pipeline is based on the fact that jellyfish galaxies and non-jellyfish galaxies, even the most pronounced ones, have overlapping morphological features, such as extended emission or diffuse regions at their edges. Since the self-supervised representations are of high dimensionality (512 in this study), such similarity searches may highlight similarities based on features that do not help distinguish jellyfish from non-jellyfish galaxies. In other words, although the learned features are useful for classification, we can not guarantee that they directly correspond to fundamental astrophysical properties that are immediately interpretable.
As a result, we speculate that differentiating between merger and jellyfish galaxies will be difficult based on the current approach.

As seen in Sect.~\ref{sec:fixBad}, cases where the self-supervised JClass is greater than the visual JClass can be studied in more detail, such as their astrophysical and morphological characteristics, which is another potential application of SSL. Such studies can also differentiate true ram pressure stripping candidates from merger galaxies. Another research direction is identifying and disentangling features or learning a similarity metric (instead of fixing it to cosine similarity) that helps distinguish jellyfish from non-jellyfish galaxies, which could alleviate these issues. A larger pretraining dataset can also prove beneficial.

Although this study aimed to improve the JClass obtained from visual classification, there are several possible extensions of our study. The strongest jellyfish candidates based on visual classification can be used as queries for similarity search so that the resulting similar images output by the SSL pipeline can be further inspected for jellyfish identification. An advantage is that only the strongest jellyfish signatures need to be classified visually (which generally happens quickly and reliably), and our pipeline may automatically identify other strong, intermediate, or weak jellyfish signatures. Once identified by our pipeline, these jellyfish candidates can be inspected in more detail. Such an approach is helpful for swiftly identifying new jellyfish candidates from future galaxy surveys. Another research direction is to pre-train the self-supervised encoder on a large, unlabelled galaxy dataset and then perform fine-tuning of the model for the specific task of JClass assignment. The fine-tuning approach can significantly improve the generalisability of the learnt representations and has been shown to surpass fully supervised methods \citep[e.g.,][]{Liu2019, Hayat_2021, Hayat2021FineTune}. The loss function for fine-tuning could be based on a representation similarity metric.

\section{Conclusions} \label{sec:conclusions}

This study catalogues and analyses the astrophysical properties of 51 jellyfish candidates (possessing visual evidence of ram pressure stripping) within the Fornax, Antlia, and Hydra clusters from the S-PLUS survey data. Based on the Gini coefficient ($G$), entropy ($H$), and best-fit 2D S\'ersic index (\texttt{nFit2D}) morphological parameters, we find that galaxies possessing extreme ram pressure stripping prefer the following regions in the morphology space: low $G$, high $H$, and low nFit2D. However, the region where such galaxies are located overlaps with the overall span of normal, star-forming galaxies in the morphology space. We have found that galaxies with $\textrm{JClass} \geq 1$ possess an overall higher SFR compared to the sample of normal, star-forming galaxies. While we observe a strong preference for galaxies in the Antlia and Fornax clusters to infall toward the cluster centre, galaxies from Hydra possess a relatively weaker preference for infall. According to our study, the order of virialisation of the clusters is Hydra, Antlia, and Fornax, with Hydra being the most likely virialised system.

Another crucial contribution of our study is to present a semi-automated pipeline based on a branch of machine learning called self-supervised learning (SSL) to assist in visually classifying galaxies. The primary motivation to use our designed pipeline for identifying jellyfish galaxies is because, traditionally, their identification in optical wavelengths has predominantly depended on visual inspection, which is a time-consuming endeavour.

Our study analyses the capabilities of SSL to assist visual morphological classification of galaxies in the low-data regime ($\sim$200 galaxies only), which has been previously largely unexplored in an astronomical context. Despite the paucity of data, a similarity search using SSL revealed that the learnt representations of our galaxies are robust to orientation and noise. Our study thus shows that SSL can learn meaningful feature representations of galaxies even with limited data, likely due to its non-dependence on any labels during training. There are two immediate advantages of our self-supervised pipeline used to refine the visual JClasses. First, unlike laborious visual inspection, it is scalable to large datasets. Once the self-supervised encoder network is pretrained, it can be used for swift JClass assignment for new galaxies based on a simple, weighted nearest neighbour search. Second, although our pipeline uses the visually assigned JClasses for the final JClass prediction, the self-supervised encoder is trained agnostic to the visual JClass labels. Thus, the training is unaffected by the quality of visual JClass predictions. Furthermore, our pipeline relies only on confident visual classifications, significantly reducing misclassifications arising from human biases. Traditional approaches of supervised learning or training a supervised classifier on the extracted feature representations (i.e., the linear evaluation protocol) rely entirely on the quality of the visual JClass since these labels are used as ground truths in the learning process.

Our pipeline can also be used as a guide to train human classifiers to assist in their visual classification. Another application of our pipeline is identifying false positives and negatives during follow-up analysis after human classification. Our self-supervised strategy is designed to lay the groundwork for a more comprehensive search in the future. For example, with large astronomical datasets, more powerful semantic embeddings can be obtained, further improving the performance. It will then be possible to leverage our self-supervised pipeline to produce more reliable JClass estimates and thus pave the way for better constraining the properties of these rare jellyfish galaxies. Finally, the idea of a task-agnostic nearest neighbour search in the self-supervised representation space makes our pipeline highly adaptable for the seamless identification of any rare astronomical signatures within astronomical datasets of future astronomical surveys.

\section*{Acknowledgements}

The S-PLUS project, including the T80-South robotic telescope and the S-PLUS scientific survey, was founded as a partnership between the Funda\c{c}\~ao de Amparo \`a Pesquisa do Estado de S\~ao Paulo (FAPESP), the Observat\'orio Nacional (ON), the Federal University of Sergipe (UFS), and the Federal University of Santa Catarina (UFSC), with important financial and practical contributions from other collaborating institutes in Brazil, Chile (Universidad de La Serena), and Spain (Centro de Estudios de F\'isica del Cosmos de Arag\'on, CEFCA). We further acknowledge financial support from the S\~ao Paulo Research Foundation (FAPESP), the Brazilian National Research Council (CNPq), the Coordination for the Improvement of Higher Education Personnel (CAPES), the Carlos Chagas Filho Rio de Janeiro State Research Foundation (FAPERJ), and the Brazilian Innovation Agency (FINEP).
We acknowledge useful discussions with Angela Krabbe, Cristina Furlanetto, Karín Menéndez-Delmestre and Alvaro Alvarez-Candal.
ACS acknowledges funding from CNPq and the Rio Grande do Sul Research Foundation (FAPERGS) through grants CNPq-403580/2016-1, CNPq-11153/2018-6, PqG/FAPERGS-17/2551-0001, FAPERGS/CAPES 19/2551-0000696-9. This work was partially supported by the CAS PIFI programme 2021VMC0005. 
RSS and SS acknowledge the support from the China Manned Space Project with NO. CMS-CSST-2021-A07.
Y.J. acknowledges financial support from ANID BASAL project No. FB210003 and FONDECYT Regular No. 1230441.
RR acknowledges support from the Fundaci\'on Jes\'us Serra and the Instituto de Astrof{\'{i}}sica de Canarias under the Visiting Researcher Programme 2023-2025 agreed between both institutions. RR, also acknowledges support from the ACIISI, Consejer{\'{i}}a de Econom{\'{i}}a, Conocimiento y Empleo del Gobierno de Canarias and the European Regional Development Fund (ERDF) under grant with reference ProID2021010079, and the support through the RAVET project by the grant PID2019-107427GB-C32 from the Spanish Ministry of Science, Innovation and Universities MCIU. This work has also been supported through the IAC project TRACES, which is partially supported through the state budget and the regional budget of the Consejer{\'{i}}a de Econom{\'{i}}a, Industria, Comercio y Conocimiento of the Canary Islands Autonomous Community. RR also thanks to CNPq 311223/2020-6,  304927/2017-1 and  400352/2016-8, FAPERGS 16/2551-0000251-7 and 19/1750-2, CAPES/0001.
This work was partly done using GNU Astronomy Utilities (Gnuastro, ascl.net/1801.009) version 0.19. Work on Gnuastro has been funded by the Japanese Ministry of Education, Culture, Sports, Science, and Technology (MEXT) scholarship and its Grant-in-Aid for Scientific Research (21244012, 24253003), the European Research Council (ERC) advanced grant 339659-MUSICOS, the Spanish Ministry of Economy and Competitiveness (MINECO, grant number AYA2016-76219-P) and the NextGenerationEU grant through the Recovery and Resilience Facility project ICTS-MRR-2021-03-CEFCA.
This research has made use of the NASA/IPAC Extragalactic Database (NED), which is funded by the National Aeronautics and Space Administration and operated by the California Institute of Technology.
This research made use of the following python packages: \texttt{Astropy} (\citealt{Astropy13}, \citeyear{Astropy18}), \texttt{Matplotlib} (\citealt{Matplotlib}), \texttt{NumPy} (\citealt{NumPy}) and \texttt{SciPy} (\citealt{SciPy}).

\section*{Data Availability}
Part of the data can be found at the S-PLUS cloud at \url{https://splus.cloud/}.
The complete dataset underlying this paper will be shared upon a reasonable request to the corresponding authors.
We make the codes used in this study publicly available at \url{https://github.com/Yash-10/jellyfish_self_supervised}.


\bibliographystyle{mnras}
\bibliography{biblio}


\appendix

\section{Model training details}
\label{appn:moreTrainingDetails}

\subsection{Implementation details}\label{sec:moreTrainingDetails}

All experiments are conducted using the \texttt{pytorch-lightning} library (\citealt{falcon2019pytorch}; version 1.6.0), and our SimCLR implementation is motivated by the SimCLR tutorial\footnote{\url{https://uvadlc-notebooks.readthedocs.io/en/latest/tutorial_notebooks/tutorial17/SimCLR.html}} in \citet{lippe2022uvadlc}. We use ResNet-34 as our base encoder instead of the Resnet-50 used in the original SimCLR approach \citep{7780459} since larger models tend to overfit on small datasets \citep{Cao2021}. The ResNet-34 architecture is modified to accept our 12-channel input. To handle the relatively low-resolution images in our case as compared to typical images used for ResNet, such as those from ImageNet \citep{5206848}, we also change the stride from 2 to 1 in the first convolutional layer and reduce the amount of pooling by removing the first max pooling layer (\citealt{newell2020useful}; \citealt{Hayat_2021}). We use the default weight initialisation in PyTorch for all the models considered in our study. This choice of architecture yields a 512-dimensional representation vector for each image. Although increasing the dimensionality of the representations enhances their quality, higher dimensional representations may degrade the quality of the representations for smaller datasets \citep{Kolesnikov2019}. Hence, we refrain from experimenting with different representation sizes. Further studies can investigate the potential benefits of increasing representation dimensions for smaller datasets.

The projection head in our self-supervised model is a two-layer MLP with a ReLU activation function, mapping the representation vector onto a 128-dimensional space. Taking a cue from the observation by \citet{chen2020big} that a wider projection head improves performance, we opt for a four-fold wider hidden layer. We do not, however, increase the depth of the MLP, given that the benefits of deeper MLPs tend to saturate for already-wide projection heads.

\subsection{Hyperparameter tuning}\label{appn:hyperTuning}

\subsubsection{For self-supervised pretraining}\label{appn:hyperTuning-ss}

For pretraining, the following hyperparameters are tuned: learning rate, weight decay, temperature, and number of epochs. As mentioned in the main text, the batch size is fixed to 128. $K$-fold cross-validation is used to optimise the hyperparameters by dividing the training dataset into $K$ folds. We note that such an approach is less commonly used in the SSL context since pretraining datasets are generally sufficiently large, unlike the case here. The study by \citet[e.g.,][]{sdmis22}, for example, utilised $K$-fold cross-validation in SSL. We adopt such a validation procedure for two main reasons. First, our dataset size is too small to assume that a simple train-validation-test split would estimate the model performance reliably. Second, it would mean a part of the training data is set aside for validation, thus reducing the amount of training data.

The hyperparameters are tuned using a combination of contrastive loss (defined in Section~\ref{sec:ssl}) and top-5 accuracy (the number of times the desired patch is within the top five most similar examples to the original image in the sampled batch) on the validation split. For computational reasons, $K = 3$ is used instead of the common choices ($K = 5$ or $10$), which means the scores are averaged across three folds. We remind the reader that the top-5 accuracy is not the accuracy in the downstream classification task. The similarity is computed using the cosine similarity metric. The top-5 accuracy is used instead of top-1 since the former is less noisy. Even though there are two classes, jellyfish and non-jellyfish, here, the use of top-5 accuracy is valid, unlike traditional supervised classification, since the similarity is compared across all images from a given batch. The hyperparameters could be selected based on the downstream classification performance. However, such an approach is not used here since we are more concerned with image similarity than the final classification performance.

A progressive grid-search approach is used to select the optimal set of hyperparameters. First, a coarse search is performed on a wide range of hyperparameter values: three learning rate values uniformly selected from logarithmically spaced values in the range $10^{-5}$ to $10^{-2}$, weight decay uniformly selected from three logarithmically spaced values in the range $10^{-6}$ to $10^{-3}$, and temperature selected from {0.1, 0.5, 1.0}. Such a wide search informs of each hyperparameter's approximate optimal hyperparameter intervals. The optimal region found was around $3 \times 10^{-4}$ for learning rate, $3 \times 10^{-5}$ for weight decay, and 0.1 for temperature. A finer search is then performed around these values. In all cases till now, training was performed for 100 epochs to keep the computational costs low. We separately list the top five approaches based on top-5 accuracy and contrastive loss and only select the hyperparameter sets that appeared in both lists. This makes the hyperparameter selection less noisy. Six sets of hyperparameters appeared on both lists. Each case was then trained for longer, i.e., 300 and 500 epochs, to yield the optimal hyperparameter set: learning rate = $10^{-4}$, weight decay = $10^{-4}$, and temperature = 0.05, again decided based on the combination of top-5 accuracy and contrastive loss. As mentioned in the main text, training the model longer often helps contrastive learning. To test this, we trained the model using the optimal learning rate, weight decay, and temperature defined above for longer epochs, i.e. 700 and 1000 epochs. We found the performance to improve by training longer, which corroborates the fact. Hence, we use 1000 epochs for pretraining.

\begin{figure}
    \centering
    \includegraphics[keepaspectratio,width=0.48\textwidth]{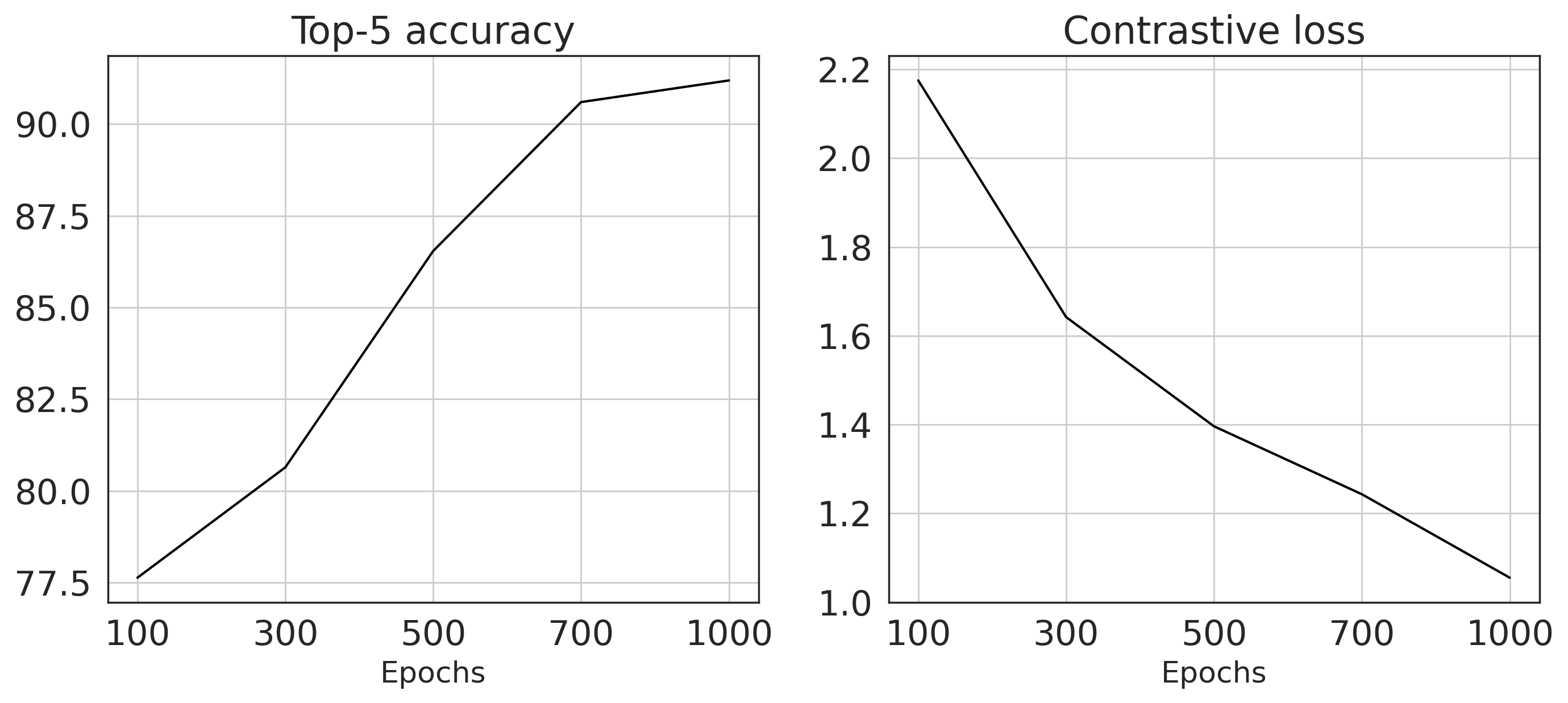}
    \caption{Averaged top-5 accuracy and contrastive loss, across the three folds, for the optimal hyperparameters (learning rate = $10^{-4}$, weight decay = $10^{-4}$, and temperature = 0.05), obtained during hyperparameter tuning of the self-supervised pretraining, as a function of the number of epochs.}
    \label{fig:hyperParamTuneEpoch}
\end{figure}

Fig.~\ref{fig:hyperParamTuneEpoch} compares the top-5 accuracy and contrastive loss, averaged across three folds, corresponding to the optimal hyperparameters obtained using the above procedure. The figure shows that training for 1000 epochs yields the highest accuracy and lowest contrastive loss. We have observed the standard deviation across the folds to be $< 2.5\%$. This difference is non-trivial, which may be due to our extremely small dataset that makes the individual folds not fully representative of the entire dataset.

\subsubsection{For supervised learning}\label{appn:hyperTuning-s}

The following hyperparameters are tuned: learning rate, weight decay, and number of epochs. As in Section~\ref{appn:hyperTuning-ss}, $K = 3$ is used for the $K$-fold cross-validation. The learning rate is selected from \{$10^{-5}$, $10^{-4}$, $10^{-3}$\}, weight decay from \{$10^{-5}$, $10^{-4}$, $10^{-3}$\}, and number of epochs from \{50, 70, 90\}. The macro-averaged precision and recall scores are used for selecting the optimal hyperparameter set. A similar grid search approach is used as in Section~\ref{appn:hyperTuning-ss}. This results in 27 different sets of hyperparameters. $K$-fold cross-validation is performed on each hyperparameter set, and the corresponding precision, recall, and accuracy scores are averaged across the three folds. The optimal hyperparameter set is chosen by selecting those that appear in the top three positions for all three metrics. In our case, two hyperparameter sets appeared in the top three positions. The tie was broken by selecting the set that ranked better across all three metrics. This procedure yielded the optimal hyperparameter set: learning rate = $10^{-5}$, weight decay = $10^{-3}$, and number of epochs = 90.

\begin{figure}
    \centering
    \includegraphics[keepaspectratio,width=0.48\textwidth]{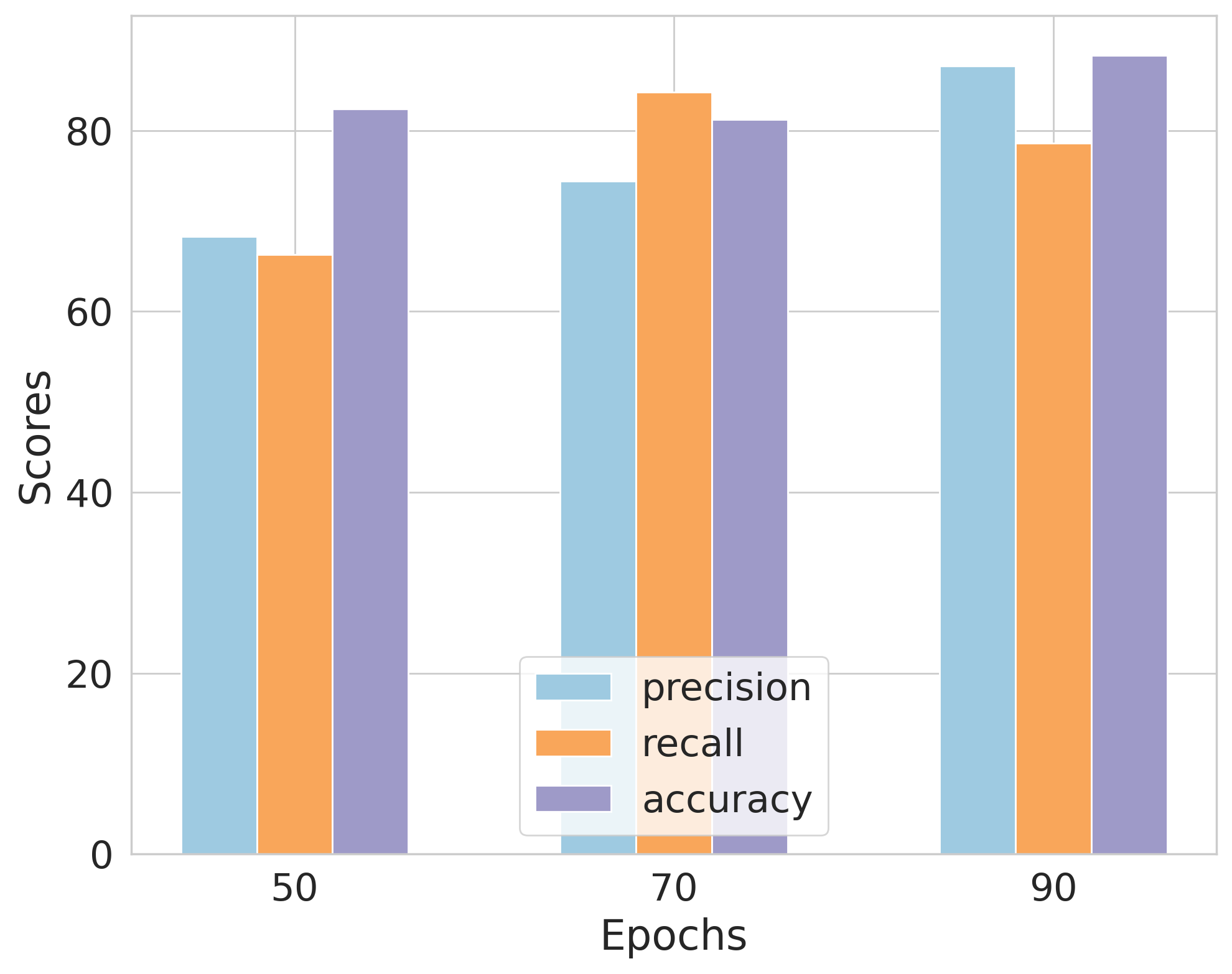}
    \caption{Averaged precision, recall, and top-1 accuracy, across the three folds, for the supervised classification using the optimal hyperparameters (learning rate = $10^{-5}$, weight decay = $10^{-3}$), obtained during hyperparameter tuning, as a function of the number of epochs.}
    \label{fig:supervised-hyperParamTuneEpoch}
\end{figure}

Fig.~\ref{fig:supervised-hyperParamTuneEpoch} shows the classification metrics averaged across the three folds as a function of the number of epochs. Overall, 90 epochs training performs better than 50 or 70 epochs training. Hence, we choose 90 epochs for comparison with the self-supervised approach.

\subsubsection{For linear evaluation}\label{appn:linearEvalHyperTune}

Linear evaluation results are discussed in Appendix.~\ref{sec:classificationCompare}. Here, the tuning is performed on the following hyperparameters: learning rate, batch size, and number of epochs. The representations from the training set (obtained during the train-test split during the linear evaluation) were divided into $K$ folds. Since the 512-dimensional representations are dealt with here rather than images, the hyperparameter tuning poses relatively less computational burden than the experiments in Section~\ref{appn:hyperTuning-ss} and \ref{appn:hyperTuning-s}. Hence, $K = 10$ is used here.

\begin{figure}
    \centering
    \includegraphics[keepaspectratio,width=0.48\textwidth]{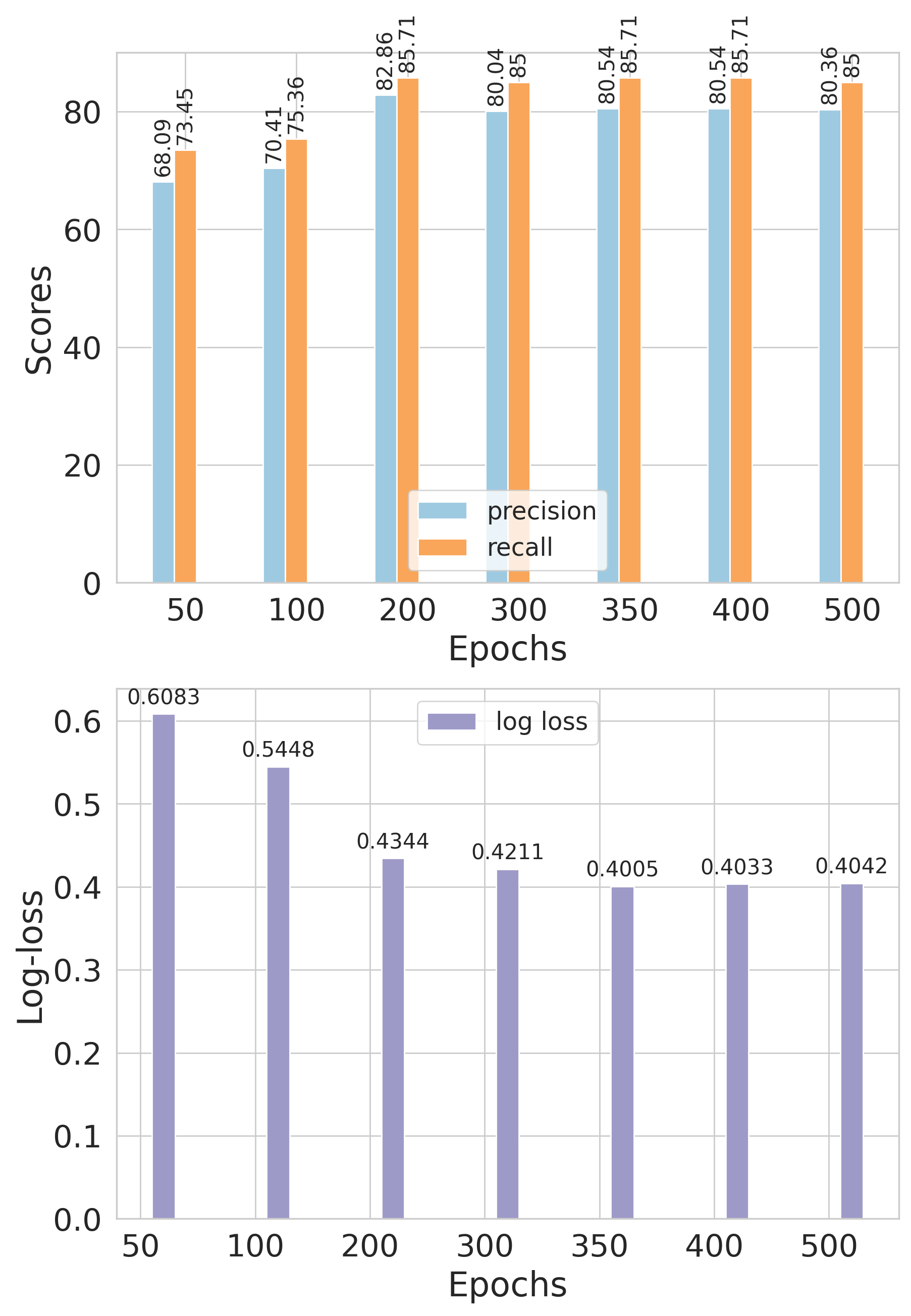}
    \caption{Averaged precision, recall, and logistic loss, across ten folds, for the linear evaluation using the optimal hyperparameters (learning rate = $5 \times 10^{-3}$ and batch size = 16) as a function of the number of epochs.}
    \label{fig:linearEval-hyperParamTuneEpoch}
\end{figure}

Similar to the hyperparameter tuning procedure of Sect.~\ref{appn:hyperTuning-s}, the macro-averaged precision and recall over the folds are compared. However, several ties were observed in this case while ranking different hyperparameter sets based on precision and recall scores. Hence, the logistic loss (also called the cross-entropy loss) was used to break the ties. The set that yielded the least loss was then selected. The optimal hyperparameter set we obtained is: batch size = 16, learning rate = $5 \times 10^{-3}$, and number of epochs = 350. As seen from Fig.~\ref{fig:linearEval-hyperParamTuneEpoch}, the number of epochs = 350 is optimal.

\section{Self-supervised vs. supervised learning}
\label{sec:classificationCompare}

We use the Resnet-34 architecture for the supervised CNN to match the encoder architecture of SSL. Since supervised approaches benefit from having the entire object within the image, we use all augmentations as in the self-supervised case except random-resize-and-crop. Moreover, to ensure the learning is not affected by class imbalance, we use weighted random sampling wherein images from the minority class (i.e., jellyfish) are oversampled in each sampled batch. We also reduce the learning rate by a factor of 0.1 after 70\% and 90\% of the total epochs during training for both approaches. For linear evaluation, we train the logistic regression classifier using the SGD optimiser with initial learning rate = $5 \times 10^{-3}$, batch size = 16, and the number of epochs = 350 but do not use weight decay.

It is widely acknowledged that supervised learning methods tend to exhibit subpar performance as the size of the dataset decreases. This is mainly due to their vulnerability to overfitting \citep{Ying_2019}, unless certain regularisation techniques, such as dropout \citep{srivastava2014dropout}, are employed. On the contrary, SSL demonstrates relative resilience to fluctuations in the training data size, as it does not rely on labels during the learning process but instead self-generates supervision from the data itself. We assess these claims by contrasting (a) the classification performance of a supervised logistic regression classifier, trained on self-supervised representations, with (b) vanilla supervised CNN. The former strategy is often called the `linear evaluation protocol' and is one of the common ways to evaluate self-supervised representation quality. Both methods use the ground-truth labels derived from visual classification results, with the only difference being that the former only uses labels during the supervised logistic regression classifier training but not for pretraining the encoder. The hyperparameter tuning details are described in Appendix~\ref{appn:hyperTuning}. For comparison purposes, we framed the problem as a binary classification. As stated in the main text, such a binary classification task greatly simplifies the comparison since a multi-label classification scheme (JClass 0, 1, 2, 3, and 4) exacerbates the class-imbalance issue due to the rareness of extreme jellyfish candidates. Moreover, to ascertain the robustness of the self-supervised representations to variations in seeing conditions, celestial locations, and other systematic factors, training is performed on galaxies from the Antlia and Hydra clusters, whereas the testing set includes images only from the Fornax cluster. For completeness, we discuss the other two cases (testing the model on the Hydra and Antlia cluster galaxies) in Appendix~\ref{appn:trainTestTwoCases}

Fig.~\ref{fig:classificationCompare} illustrates the comparison between the two approaches. Figs.~\ref{fig:sub1} and \ref{fig:sub2} suggest that the supervised approach leans more towards classifying galaxies as jellyfish than the self-supervised approach, as evidenced by the misclassification of eight non-jellyfish galaxies as jellyfish. This bias occurs despite adjustments made for class imbalance during the supervised learning process (Appendix~\ref{sec:moreTrainingDetails}). Conversely, when trained on self-supervised representations, a logistic regression classifier demonstrates enhanced robustness to imbalance, as it does not depend on labels for learning the representations. This finding indicates that self-supervised representations have successfully captured meaningful information regarding the galaxies' jellyfish-ness. This, in turn, enables the downstream classifier to differentiate more accurately between jellyfish and non-jellyfish galaxies. Overall, SSL demonstrates an improvement in classification performance over supervised learning.

In the context of this study, an effective classification scheme should generate a robust recall rate ($\dfrac{\textrm{TP}}{\textrm{TP} + \textrm{FN}}$; TP: True positive, FN: False negative; positive denotes the jellyfish category) for jellyfish candidates while maintaining high precision ($\dfrac{\textrm{TP}}{\textrm{TP} + \textrm{FP}}$) for non-jellyfish galaxies. Considering the rarity of jellyfish candidates, a certain level of false positives can be tolerated. Provided these inaccuracies are infrequent, manual re-inspection and refinement of these galaxies by human classifiers is feasible. In contrast, overlooking jellyfish candidates can compromise the utility of such a model. Figs.~\ref{fig:sub1} and \ref{fig:sub2} indicate that both self-supervised and supervised methodologies achieve identical recall scores for the jellyfish category (83\%). The precision scores for the non-jellyfish category are also identical (94\% and 93\% for self-supervised and supervised learning, respectively). Thus, self-supervised and supervised learning provide similar performances in such a context.

We note that in our study, a threshold probability of 0.5 is used for classification in both the self-supervised and supervised methods. Alternative threshold choices were not experimented with. Figs.~\ref{fig:sub4} and \ref{fig:sub5}, which display the predicted probabilities, can serve as a general guide for hypothesising how results might shift with the application of different thresholds. Given the rarity of jellyfish candidates, one might consider a higher probability threshold (analogous to the 0.8 threshold used for the visually classified raw score in \citealt{2023arXiv230409202Z}). If a higher threshold is used, even more significant improvements can be obtained using the self-supervised approach, as suggested by the probabilities observed in Figs.~\ref{fig:sub4} and \ref{fig:sub5}.

\begin{figure*}
\begin{subfigure}{.23\linewidth}
\centering
\includegraphics[width=1\textwidth,keepaspectratio]{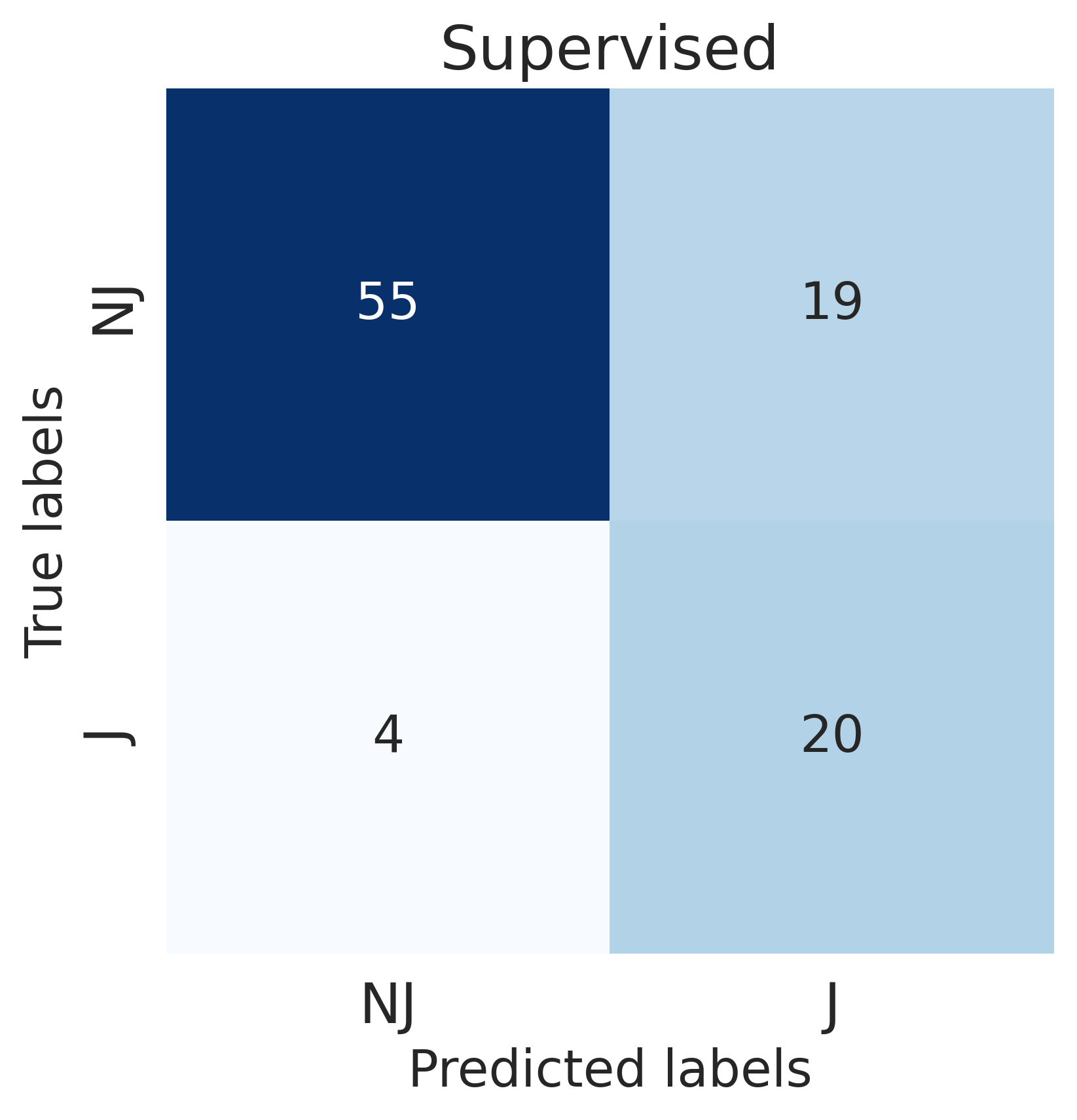}
\caption{}
\label{fig:sub1}
\end{subfigure}%
\begin{subfigure}{.23\linewidth}
\centering
\includegraphics[width=1\textwidth,keepaspectratio]{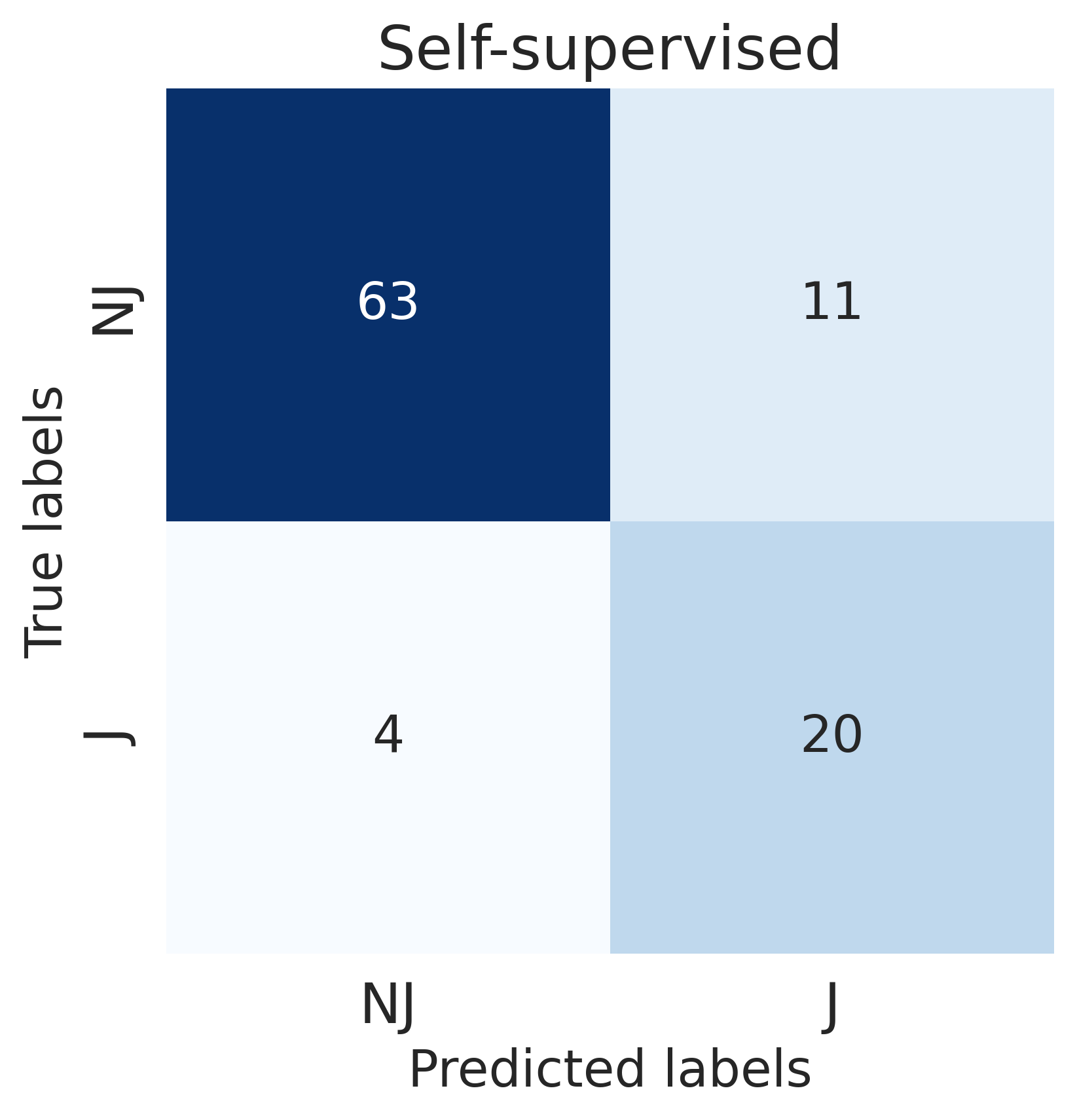}
\caption{}
\label{fig:sub2}
\end{subfigure}
\begin{subfigure}{.48\linewidth}
\centering
\includegraphics[width=1\textwidth,keepaspectratio]{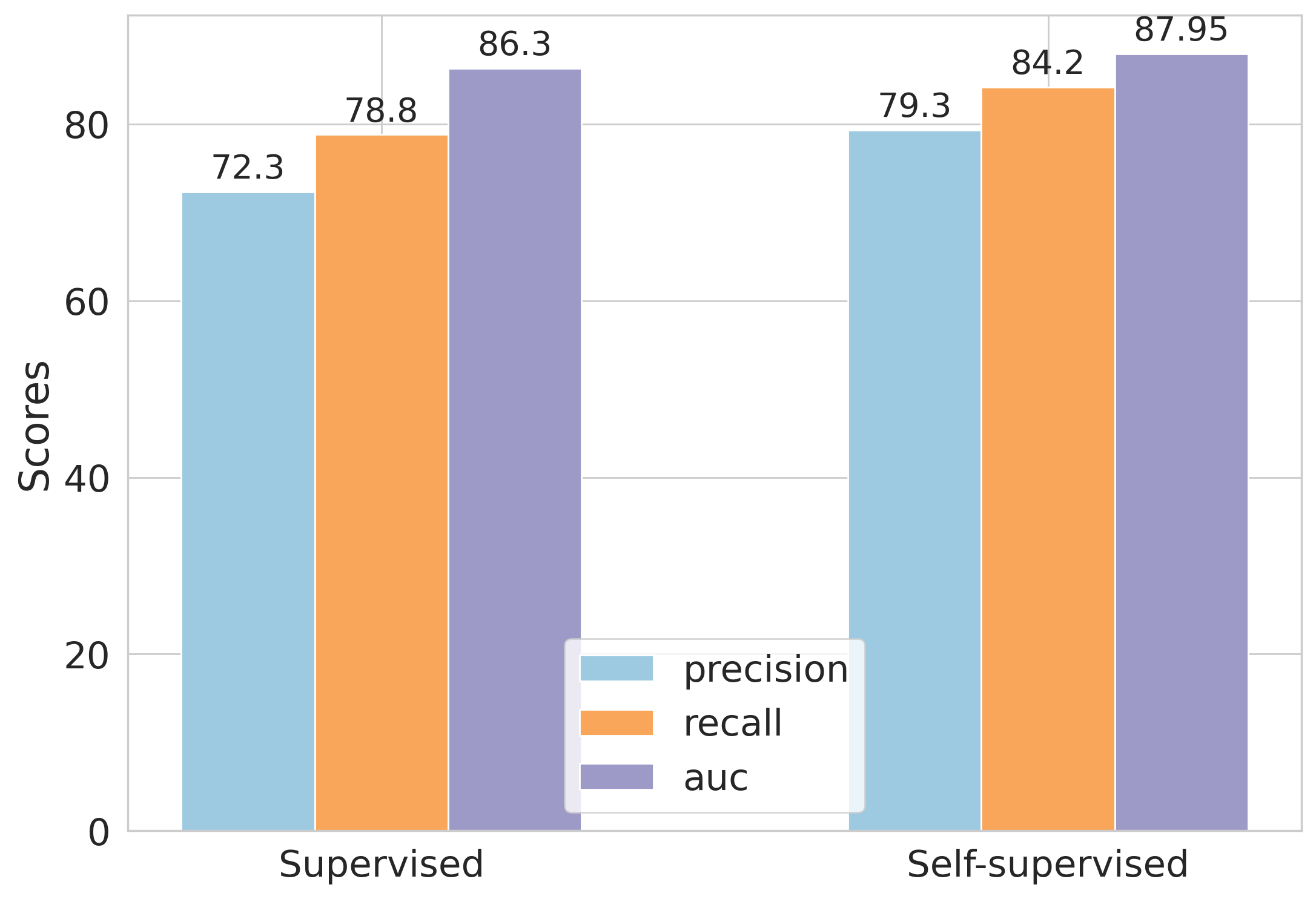}
\caption{}
\label{fig:sub3}
\end{subfigure}
\begin{subfigure}{.48\linewidth}
\centering
\includegraphics[width=1\textwidth,keepaspectratio]{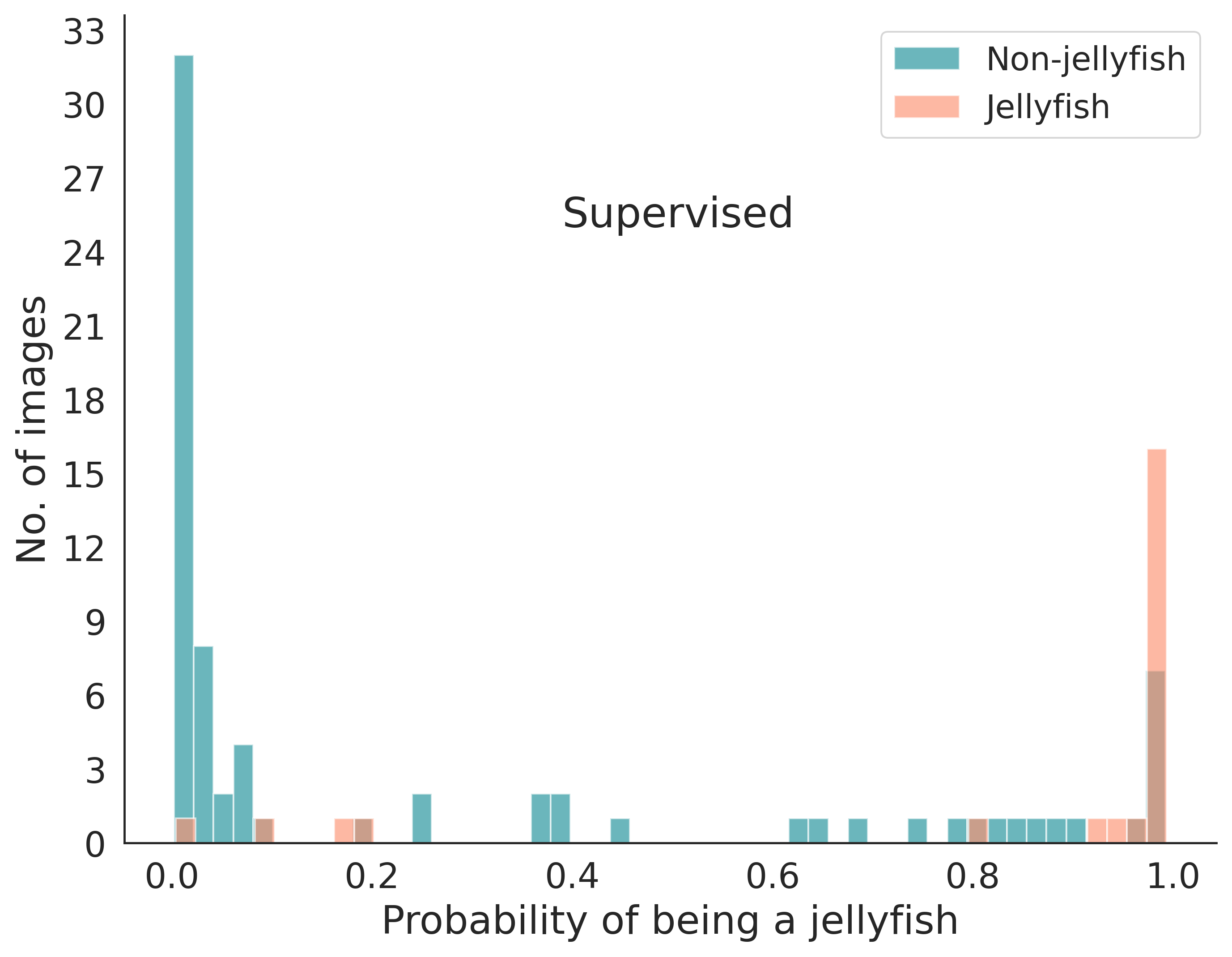}
\caption{}
\label{fig:sub4}
\end{subfigure}%
\begin{subfigure}{.48\linewidth}
\centering
\includegraphics[width=1\textwidth,keepaspectratio]{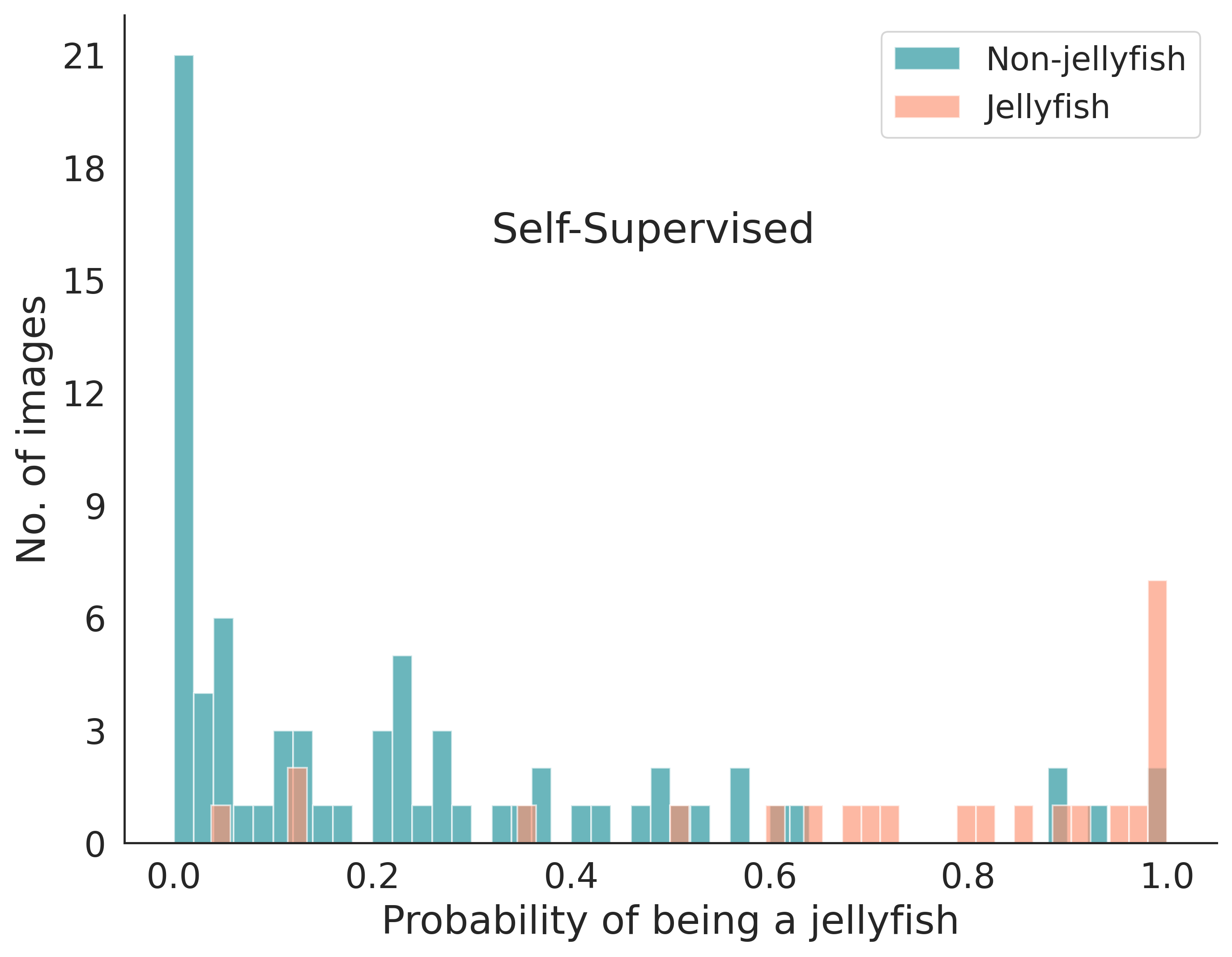}
\caption{}
\label{fig:sub5}
\end{subfigure}%
\caption{Classification performance comparison between the supervised and self-supervised approaches. Panels (a) and (b) show the confusion matrices, (c) compares the macro-averaged precision, recall, and area under the Receiver operating characteristic (ROC) curve (AUC), and (d) and (e) show the prediction probabilities output by the supervised ResNet and the logistic regression classifier on the self-supervised representations. By definition, AUC lies in the $[0, \,1]$ range, but here we show it as a percentage only for visualisation purposes.}\label{fig:classificationCompare}
\end{figure*}

\section{Data augmentation ablation study}\label{appn:dataAugMoreDetails}

Data augmentation is critical in learning good-quality representations for the contrastive learning framework used here \citep{Cheng2020Simclr}. Ablation studies are an indirect way to understand which augmentations (or combinations of augmentations) are critical or relatively unhelpful for learning good representations. Such ablation studies are commonplace in SSL approaches and have been studied by many works (see, e.g., \citealt{Hayat_2021}; \citealt{Kinakh2021}).

The procedure is as follows: certain augmentation(s) from the pipeline are turned off, pretraining is performed by learning representations of images using images from the training dataset, and the linear evaluation protocol from Appendix.~\ref{sec:classificationCompare} is employed on the test dataset. The `baseline' denotes the case where all augmentations described in the main text are used. An augmentation or a set of augmentations is considered important for the downstream classification task if the test performance decreases compared to the baseline after removing that augmentation or set of augmentations. Similar to the main text, training is performed on galaxy images from the Antlia and Hydra galaxy clusters, and testing is done on images from the Fornax cluster.

\begin{figure}
    \centering
    \includegraphics[keepaspectratio,width=0.5\textwidth]{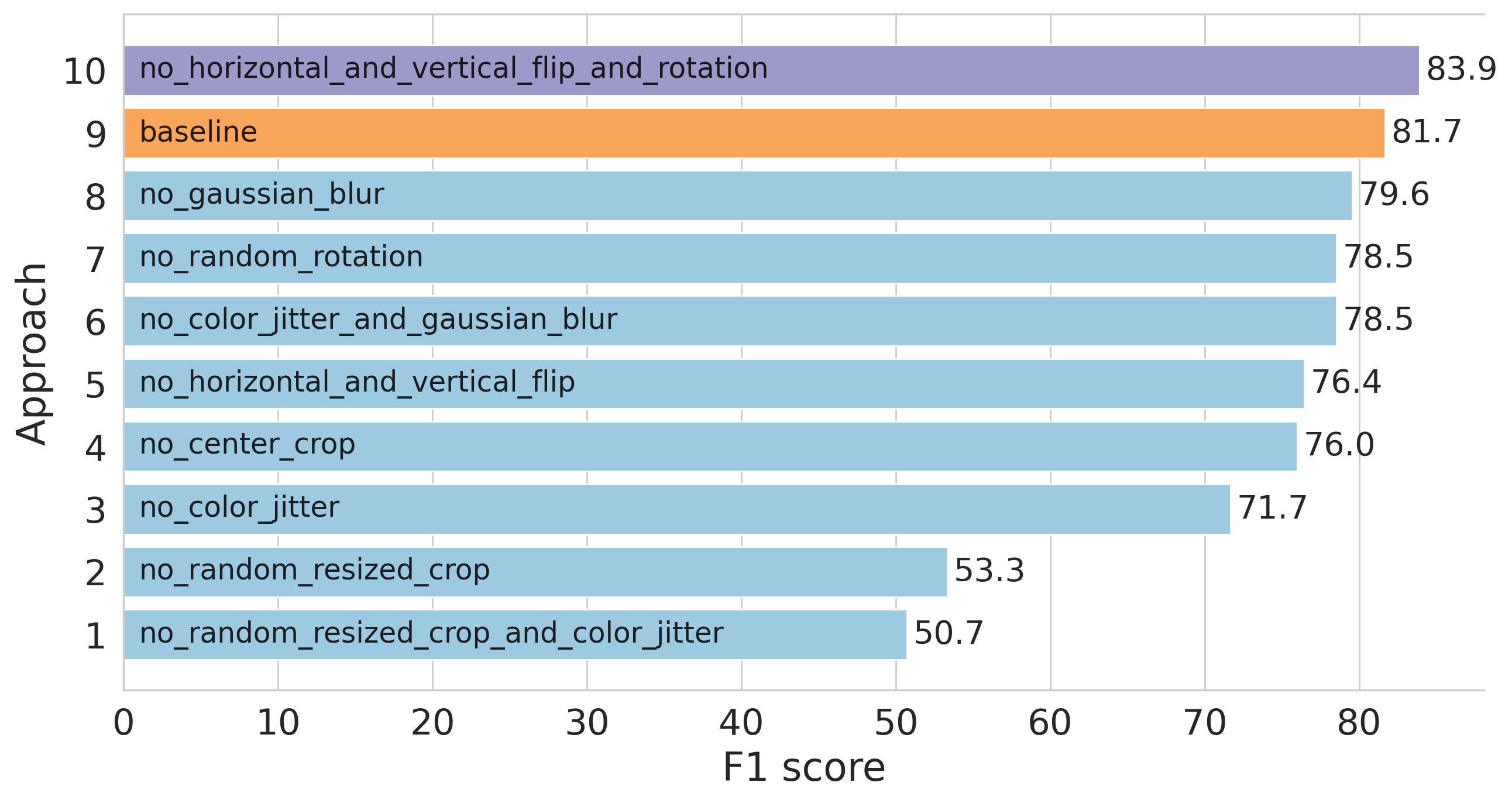}
    \caption{Classification F1 scores under various omission combinations of data augmentations. The labels on each bar describe which augmentations were removed from the pipeline. For example, no\_horizontal\_and\_vertical\_flip\_and\_rotation means that horizontal flip, vertical flip, and random rotation augmentations were removed. The `baseline' approach is marked by orange colour.}
    \label{fig:dataAugAblation}
\end{figure}

Fig.~\ref{fig:dataAugAblation} shows the F1 scores of the classification under different sets of augmentations. Accuracy is not used due to the class imbalance. It can be observed that the F1 score decreases from the baseline score, 81.7, to 50.7 when random resized crop and colour jitter are omitted. Thus, this combination of augmentations is vital for classification performance. Since the F1 score is the least among all F1 scores in the plot, we conclude that random resized crop and colour jitter is the most important set of augmentations. This also confirms the observation in \citet{Cheng2020Simclr}. Even though our colour jitter implementation differs from theirs, it is noteworthy that the result still holds. The figure also shows that random resized crop is the most important augmentation in our case, followed by colour jitter, following a similar logic. All other augmentations are also important for the classification task, except when horizontal flip, vertical flip, and random rotation augmentations are removed (Approach 10 in the figure), since the performance improves when these three augmentations are removed from the augmentation pipeline. A possible reason for this is that, since flips are a special case of rotation augmentation, incorporating both kinds of augmentations leads to redundancy, which could reduce the quality of the learnt representations. For approaches 1 and 2 in Fig.~\ref{fig:dataAugAblation}, we used a smaller batch size of 16 instead of 128 for self-supervised training due to memory constraints when not using cropping.

\section{Importance of accounting for background/nearby sources for effective similarity search}\label{appn:galmaskMotivation}

We use the Grad-CAM pixel attribution method on the self-supervised representations extracted by the trained model to visualise how background or nearby astronomical sources can affect the similarity search. Implementation is taken from the \texttt{grad-cam} Python library \citep{jacobgilpytorchcam}\footnote{The tutorial can be found here: \url{https://jacobgil.github.io/pytorch-gradcam-book/Pixel\%20Attribution\%20for\%20embeddings.html}}.

To demonstrate the effects due to background sources, we select a few cases in which \texttt{galmask} could not effectively remove sources near the galaxy of interest. We seek to find whether the similarity search can inadvertently find similar images based on nearby sources instead of the central galaxy. A query-by-example is run as described in Sect.~\ref{sec:query}. The embedding of the query image is the concept embedding. Grad-CAM is then used to highlight regions in the images closest to the query image. This test is valid since our self-supervised approach does not account for background sources during training, such as using data augmentations. Thus, if background sources are indeed dominating similarity decisions, it can be attributed to the model's incapability to be robust to background sources rather than inappropriate training.

\begin{figure*}

    \begin{subfigure}[b]{0.75\textwidth}
    \includegraphics[width=1\linewidth,keepaspectratio]{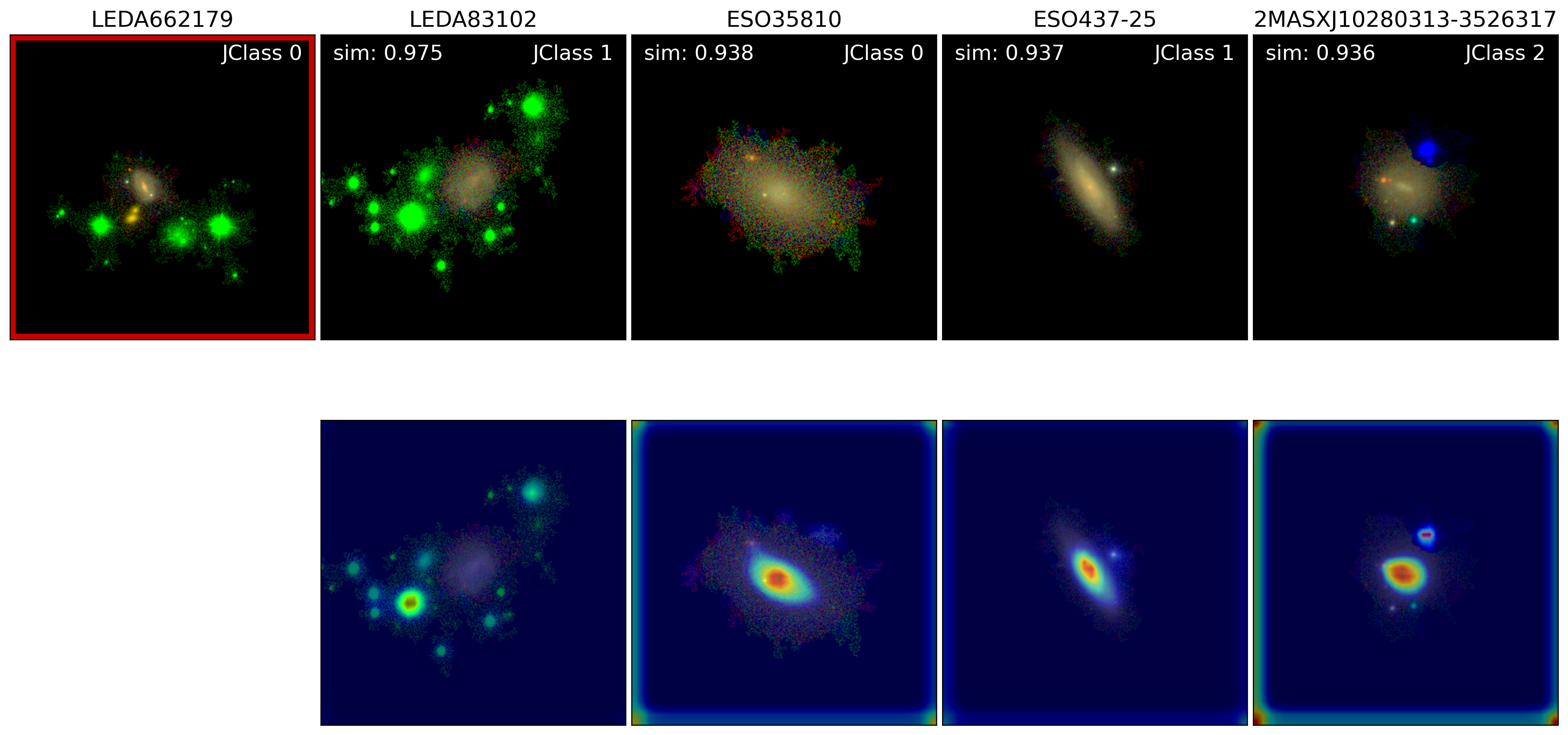}
    \end{subfigure}

    \begin{subfigure}[b]{0.75\textwidth}
    \includegraphics[width=1\linewidth,keepaspectratio]{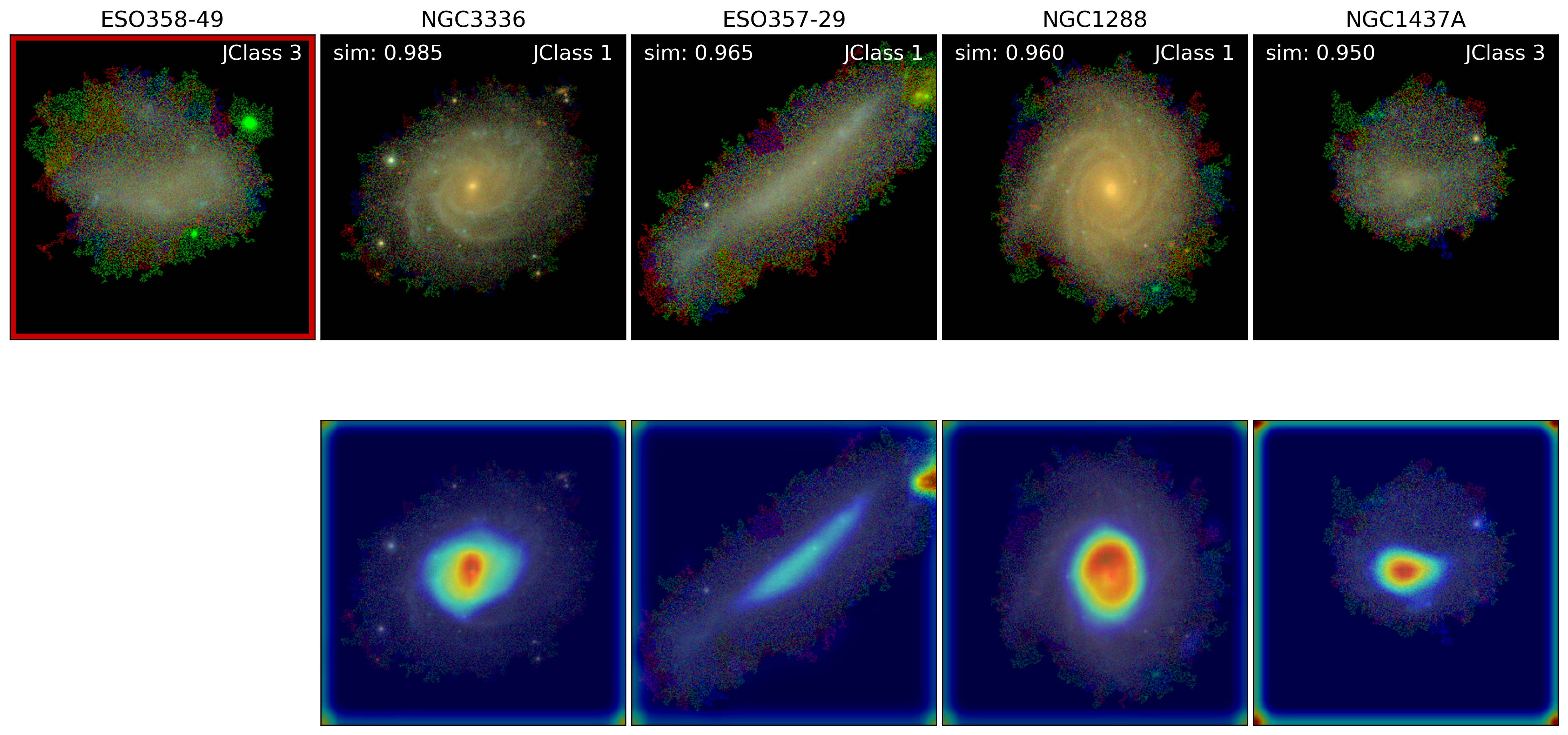}
    \end{subfigure}

    \begin{subfigure}[b]{0.75\textwidth}
    \includegraphics[width=1\linewidth,keepaspectratio]{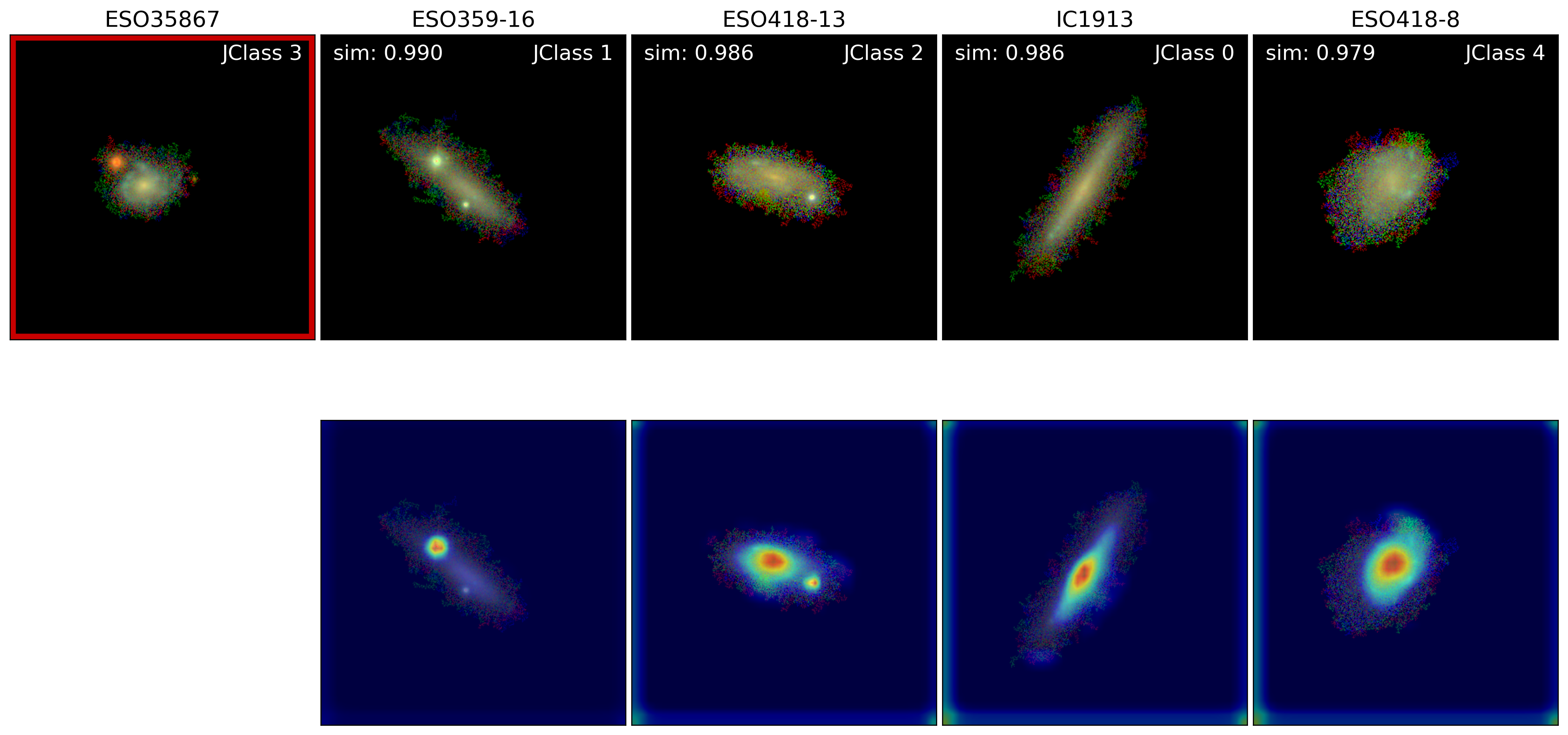}
    \end{subfigure}

    \caption{Demonstration using the Grad-CAM method of how background sources in the galaxy images can affect similarity search. Three examples are shown. The similarity search results are shown for each example, and the corresponding Grad-CAM heatmaps overlayed on the galaxy images are shown in the bottom row. This figure motivates handling background sources before or during model training, especially when the background sources have a size or brightness comparable to the central galaxy under consideration.}
    \label{fig:galmaskMotivation}
\end{figure*}

We now discuss Fig.~\ref{fig:galmaskMotivation}. The LEDA662179 query image in the first example shows bright green-coloured sources surrounding the central galaxy. The closest image (LEDA83102), using the similarity search, also contains similar background sources. The Grad-CAM heatmap for LEDA83102 highlights the bright green-coloured source instead of the central galaxy, which suggests that the similarity search was affected by the neighbouring sources.

A similar phenomenon is observed in the second example (ESO358-49 query image). For example, the heatmap for the second-closest image (ESO357-29) highlights the two closely-spaced sources in its bottom right more prominently than the central galaxy. This can be attributed to the fact that the query image (ESO358-49) also contained two similar sources, so the similarity search focused on the surrounding sources instead of the central galaxy. However, the subsequent similar images were not severely affected by surrounding sources. This could be because the surrounding sources in these similar images did not dominate the image in spatial size or brightness. The third example also shows that the nearby sources in the top two similar images to ESO35867 (ESO359-16 and ESO418-13) significantly contributed to the high similarity to ESO35867. However, ESO359-16 and ESO418-13 were considered visual mergers (not labelled in the figure). Thus, interacting galaxies could pose difficulties in similarity search, especially if there is a considerable difference in the spatial size or brightness of the interacting galaxies.

These examples suggest the importance of removing background sources, as deep learning models are not inherently robust to background sources. Thus, we have used the \texttt{galmask} package to explicitly remove background sources for the machine learning application in the main text. Cases where \texttt{galmask} could not effectively remove background sources were not considered part of our machine learning dataset. An alternative possibility to solve this issue is to account for background sources in the data augmentation pipeline, but this was not experimented in our study.

\section{Classification performance on training and testing on different galaxy clusters}\label{appn:trainTestTwoCases}

Appendix.~\ref{sec:classificationCompare} discussed the case where testing was performed on galaxies from the Fornax galaxy cluster. For completeness, we extend the comparison of supervised and self-supervised classification results, where testing is performed on the Antlia and Hydra galaxy clusters.

\begin{figure}
    \centering
    \includegraphics[width=0.48\textwidth]{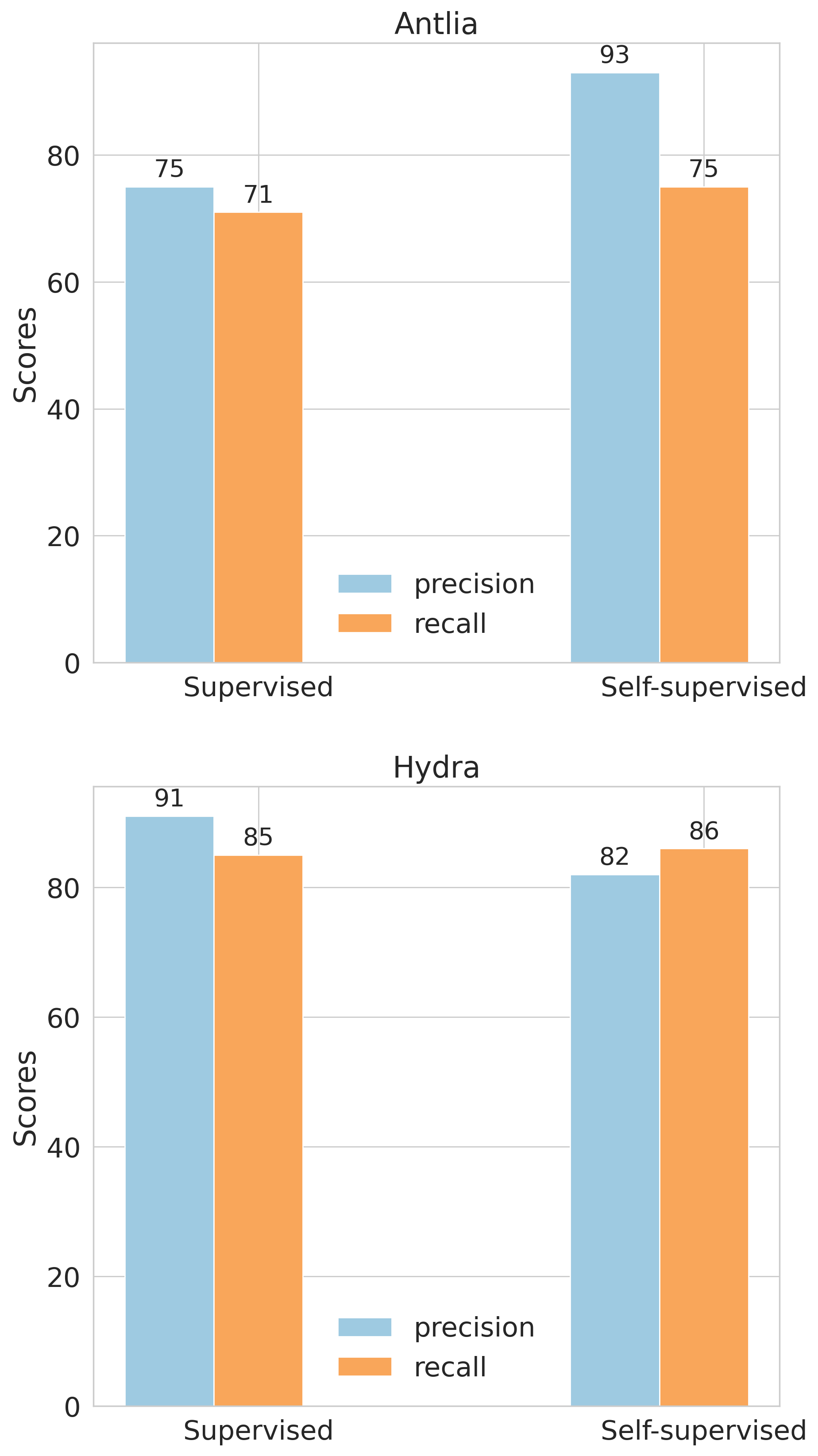}
    \caption{Macro-averaged precision and recall scores for supervised and self-supervised approaches when testing on galaxy images from the Antlia and Hydra clusters.}
    \label{fig:antliaHydra-S-SS-Comparison}
\end{figure}

Fig.~\ref{fig:antliaHydra-S-SS-Comparison} shows the macro-averaged precision and recall of the supervised and self-supervised classifications. In the case of the Antlia galaxy cluster, self-supervised classification outperforms the supervised case due to higher precision and recall scores. However, in the Hydra cluster case, the supervised approach has a higher precision (91\%) than the self-supervised approach (82\%). On the other hand, the recall scores are similar for the self-supervised approach (86\%) and the supervised approach (85\%). These two experiments reinforce the fact that SSL provides results competitive to supervised learning and can even improve classification results compared to supervised classification.

\label{lastpage}
\end{document}